\documentclass[%
 aip,
 amsmath,amssymb,
reprint, 
]{revtex4-2}

\usepackage{graphicx}
\usepackage{dcolumn}
\usepackage{bm}
\usepackage{mathtools}
\usepackage{amssymb}
\let\vec\mathbf
\usepackage{physics}

\usepackage[utf8]{inputenc}
\usepackage[T1]{fontenc}
\usepackage{mathptmx}
\usepackage{etoolbox}
\usepackage{placeins}
\makeatletter
\def\@email#1#2{%
 \endgroup
 \patchcmd{\titleblock@produce}
  {\frontmatter@RRAPformat}
  {\frontmatter@RRAPformat{\produce@RRAP{*#1\href{mailto:#2}{#2}}}\frontmatter@RRAPformat}
  {}{}
}%
\makeatother
\draft 

\begin{document}


\title{Transferability of atomic energies from alchemical decomposition}
 
\author{Michael J. Sahre}

\affiliation{Vienna Doctoral School in Chemistry (DoSChem) and Institute  of  Theoretical  Chemistry and Faculty of Physics, University of Vienna, 1090 Vienna, Austria}

\author{Guido Falk von Rudorff}
\affiliation{University Kassel, Department of Chemistry, Heinrich-Plett-Str.40, 34132 Kassel, Germany}
\affiliation{Center for Interdisciplinary Nanostructure Science and Technology (CINSaT), Heinrich-Plett-Straße 40, 34132 Kassel}

\author{Philipp Marquetand}
\affiliation{Institute  of  Theoretical  Chemistry, Faculty of Chemistry, University of Vienna, W\"ahringer Str. 17, 1090 Vienna, Austria}

\author{O. Anatole von Lilienfeld}
\affiliation{Chemical Physics Theory Group, Department of Chemistry, University of Toronto, St. George Campus, Toronto, ON, Canada}
\affiliation{Department of Materials Science and Engineering, University of Toronto, St. George Campus, Toronto, ON, Canada}
\affiliation{Vector Institute for Artificial Intelligence, Toronto, ON, M5S 1M1, Canada}
\affiliation{ML Group, Technische Universit\"at Berlin and Institute for the Foundations of Learning and Data, 10587 Berlin, Germany}
\affiliation{Berlin Institute for the Foundations of Learning and Data, 10587 Berlin, Germany}
\affiliation{Department of Physics, University of Toronto, St. George Campus, Toronto, ON, Canada}
\affiliation{Acceleration Consortium, University of Toronto, Toronto, ON}

\begin{abstract}
We study alchemical atomic energy partitioning  as a method to estimate atomisation energies from atomic contributions which are defined in physically rigorous and general ways through use of the uniform electron gas as a joint reference. We analyze quantitatively the relation between atomic energies and their local environment using a dataset of 1325 organic molecules. The atomic energies are transferable across various molecules, enabling the prediction of atomisation energies with a mean absolute error of 20~kcal/mol - comparable to simple statistical estimates but potentially more robust given their grounding in the physics-based decomposition scheme. A comparative analysis with other decomposition methods highlights its sensitivity to electrostatic variations, underlining its potential as representation of the environment as well as in studying processes like diffusion in solids characterized by significant electrostatic shifts. 
\end{abstract}


\maketitle 


\section{Introduction}
Understanding structure-property relationships is essential for the rational design of chemical compounds.
In this regard, the decomposition of molecules into smaller fragments has emerged as a valuable concept in chemistry.
The impact of a specific fragment on the molecular properties is often transferable,\cite{Fias2017}
enabling the estimation of molecular properties from fragment-based coarse grained models. Due to their reduced dimensionality compared to models explicitly considering the interactions between all electron and nuclei, such fragment based approaches can for example be used to study very large molecules.\cite{Herbert2019,Liu2020}
Furthermore, the lower complexity can be helpful to construct simplified models that encompass only the essential molecular characteristics for explaining trends among various molecules.
This is for example exploited in the Hammett model, which assumes the effect of a specific functional group on a molecular property to be a scalar value that is transferable across different molecules.\cite{hammett_35, hammett_37} 
More recently, the decomposition of molecules into smaller fragments has also been used in combination with machine learning.\cite{ccs_review1,Musil2021} For instance, so-called amon models infer the energy of a query compound based on a linear combination of reference results for its constituting building blocks\cite{amons} and many artificical neural networks predict total energies as a sum of atomic energies.\cite{behler_review,10.1063/1.5017898,schnet}
These models decompose properties based on statistics and for a given set of training molecules but accurate out-of-sample predictions are generally not guaranteed.

Molecular properties can also be partitioned via rigorous physics based decomposition schemes.
One of the earliest definitions of atoms in molecules dates back to Politzer and Parr who showed that the energy of a molecule can be expressed exactly as a function of the electrostatic potentials $V_{\text{A},0}$ at the positions of the nuclei as
\begin{equation}\label{eq:PolPArr}
    E^\text{mol} = \frac{1}{2} \sum_\text{A} Z_\text{A} V_{\text{A},0} -\frac{1}{2} \sum_\text{A} \int_0^{Z_\text{A}} Z'_\text{A} \left( \pdv{V_{\text{A},0}}{Z'_\text{A}} -V_{\text{A},0} \right) dZ'_\text{A}
\end{equation}
with the electrostatic potential
\begin{equation}\label{eq:elstat_politzer}
    V_{\text{A},0} = \sum_{\text{B}\neq \text{A}} \frac{Z_\text{B}}{|\vec{A} - \vec{B}|} - \int d\vec{r} \frac{\rho(\vec{r}, \{Z'_\text{A}\})}{|\vec{r}-\vec{R}_\text{A}|}
\end{equation}
and the nuclear charges and positions $Z_\text{A/B}, \vec{R}_\text{A/B}$ for all nuclei in the molecule.\cite{Politzer1974,Politzer2002,Politzer2021}
In the second term of Eq.~\eqref{eq:PolPArr}, the nuclear charges $\{Z'_\text{A}\}$ are scaled up linearly from 0 to their respective value in the molecule. The electron density $\rho(\vec{r}, \{Z'_\text{A}\})$ is a function of the varying nuclear charges 
and transforms into the electron density of the molecule along the integration path starting from an infinitely spread out charge distribution.
This idea built up on work by Wilson who employed the Hellman-Feynman theorem to demonstrate that the energy of a molecule can be expressed as a function of only the electron density and the external potential by scaling the nuclear charges as described above.\cite{Wilson1962} 
Similar expressions for the energy of single atoms were derived even earlier\cite{Kohn,foldy,LOWDIN195946} and can be traced back to Hylleraas who calculated energies of Helium-like atoms by treating the energy dependence on the nuclear charge with perturbation theory.\cite{Hylleraas1930-ww}
Since the scaling of nuclear charges corresponds to the generation or transformation of elements the term quantum alchemy was introduced to describe these processes.\cite{PhysRevLett.72.4001,VonLilienfeld2006,giorgio}

Eq.~\eqref{eq:PolPArr} is an exact decomposition of the energy into atomic contributions but the calculation of these atomic energies is challenging since the isoelectronic density must be known as a function of the variable nuclear charges along the scaling path.
Additionally, it was demonstrated that atomic energies are path dependent such that a meaningful comparison of atomic energies requires a common reference and scaling path.\cite{rudorff_atomic_energies}
To address these issues, the uniform electron gas was proposed as reference from which arbitrary isoelectronic compounds can be generated by linear up-scaling of the nuclear charges in the respective compound.\cite{rudorff_atomic_energies}
In this reference system, a positive background charge density equal to the negative uniform electron density is present to ensure electric charge neutrality. As the positive background charge density approaches zero the reference becomes equal to the infinitely spread out electron density originally proposed by Wilson.
For this limiting case, an alternative atomic energy expression to Eq.~\eqref{eq:PolPArr}, decomposing the energy of an arbitrary compound $E^\text{mol}$ with respect to an isoelectronic uniform electron gas reference $E^\text{UEG}$, was derived by application of the Hellmann-Feynman theorem and the chain rule
as
\begin{equation}\label{eq:guido_intro}
E^\text{mol} - E^\text{UEG} = \sum_I \underbrace{Z_I\left(  -\int d\vec{r} \frac{\int_0^1 d\lambda \rho(\lambda, \vec{r})}{|\vec{r}-\vec{R}_I|}+\frac{1}{2}\sum_{J \neq I} \frac{Z_J}{|\vec{R}_J-\vec{R}_I|}\right)}_{:= \Delta E_I}.
\end{equation}
$Z_I$ and $\vec{R}_I$ indicate nuclear charge and position, respectively. $\lambda$ is a coupling variable linearly connecting the Hamiltonian of the uniform electron gas at $\lambda =0$ with the Hamiltonian of the molecule at $\lambda = 1$ and $\rho(\lambda, \vec{r})$ is the electron density as a function of spatial coordinate $\vec{r}$ and $\lambda$. 

This is not the only possibility to define atomic energies. Another approach is for instance the interacting quantum atom (IQA),\cite{IQA,IQA_method} which is based upon the quantum atom in molecule introduced by Bader.\cite{bader} This scheme relies on the decomposition of the electron density into well-defined fragments based on the topology of the density.
Furthermore, it was recently proposed to calculate atomic energies by partitioning the energy among localized molecular orbitals, combined with an atomic charge partitioning scheme (in the following abbreviated as MO/AC).\cite{Eriksen2020}

Due to their well defined connection to the underlying physics such decomposition schemes can be useful to gain insights into the nature of chemical bonding.
The interacting quantum atom approach has for example been employed to examine binding energy trends of first row diatomics,\cite{binding_IQA} the role of electrostatics in torsional energy potentials\cite{Darley2008} or the development of machine learning force fields\cite{Popelier2015,fflux}. MO/AC was used to classify different types of electronic excitations\cite{Eriksen2022} or the development of machine learning models predicting atomisation energies.\cite{Kjeldal2023}

Recently, also the alchemical partitioning scheme in Eq.\eqref{eq:guido_intro} was utilized to construct a simple expression for binding energies between $p$-block elements with only the nuclear charges of the binding partners as variables.\cite{qaa_bde} However, Eq.~\eqref{eq:guido_intro} is only valid in the limit of the density of the uniform electron gas approaching zero, e.g. in the limit of an infinitely large supercell. This cannot be realized in practical calculations and consequently the sum of atomic energies $\sum_I \Delta E_I$ in Eq.~\eqref{eq:guido_intro} does not add up to $E^\text{mol}-E^\text{UEG}$. 

Here we introduce and benchmark an alternative approach to Eq.~\eqref{eq:guido_intro} rectifying this deviation by incorporation of a positive uniform background potential and periodic boundary conditions for finite supercells.
The modified alchemical decomposition scheme is then applied to a diverse set of 1325 organic isoelectronic molecules with six to eight heavy atoms, extracted from the QM9 dataset. We analyse the relationship between the local environment and alchemical atomic energies and develop a model to estimate the atomisation energies of the respective molecules.
The alchemical decomposition scheme is also compared to the IQA and MO/AC methods by calculating the respective atomic energies for the same dataset. 
By analysing the different relations between local structure and atomic energies in these methods we evaluate which kind of problems, e.g. estimation of energy or fingerprint for molecules in machine learning and which kind of systems, e.g. molecules or materials, might be treated best by which decomposition scheme.

\section{Methods}

\subsection{Atomic energies with periodic boundary conditions}

\begin{figure*}
\includegraphics[width=1.0\textwidth]{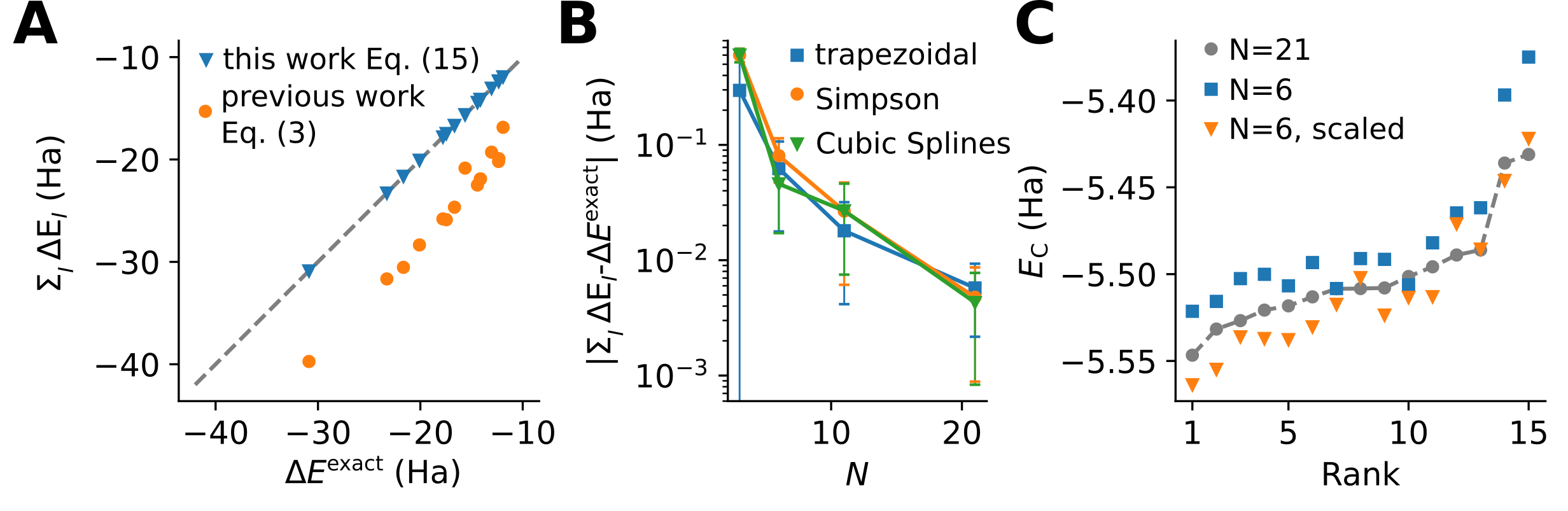}
\caption{Panel A: The sum of atomic energies for 15 different molecules vs the exact molecular energy with respect to the uniform electron gas $\Delta E^\text{exact}$ computed including periodic boundary conditions (pbc) and background potential via Eq.~\eqref{eq:dE_pbc} or without pbc via Eq.~\eqref{eq:guido_intro}. Panel B: MAE of $\sum_I \Delta E_I$ vs $\Delta E$ for the 15 molecules for an increasing number of integration points along the $\lambda$-path using different integration methods. Panel C: Comparison of atomic energy rankings for carbon atoms in 15 different molecular environments with varying number of integration points $N$ and rescaling (Eq.~\eqref{eq:rescaling}).}
\label{fig:methods}
\end{figure*}

We calculated atomic energies with Eq.~\eqref{eq:guido_intro}
for twelve molecules of structure CH$_3$-YH$_y$ with Y = C, N, O, F, Si, P, S, Cl, Ge, As, Se, Br and $y$ such that all valencies of Y are saturated and the molecules HCN, ethene and ethyne (15 molecules in total).
According to Eq.~\eqref{eq:guido_intro} the sum of atomic energies for any compound should add up to the energy difference between the energy of the compound and the uniform electron gas.
However, for the 15 molecules the sum of atomic energies was on average about 8~Ha lower than the exact energy difference as illustrated in Figure~\ref{fig:methods}A.

The reason for this deviation is that the Hamiltonian of the uniform electron gas contains a positive, uniform background potential. In a practical calculation this potential is modeled via periodic boundary conditions and the energies of the molecules are evaluated in periodic supercells. The periodicity and the background potential were however not considered in the derivation of Eq.~\eqref{eq:guido_intro}. 
Furthermore, the external potential is not coulombic (as assumed in Eq.~\eqref{eq:guido_intro}) if pseudopotentials are used.

If the Hamiltonians of the uniform electron gas $H^\text{UEG}$ and an isoelectronic compound $H^\text{mol}$ are linearly coupled as
\begin{equation}\label{eq:Hamiltonian}
    H(\lambda) = H^\text{mol} \lambda + (1-\lambda) H^\text{UEG}
\end{equation}
with $\lambda$ being a dimensionless variable defined over the interval $[0,1]$, the periodic external potential is given by
\begin{equation}
  V_\text{ext}^\text{pbc}(\vec{r}, \lambda) = \sum_\vec{T} \sum_I V_I^\text{PP}(\lambda, \vec{r}-\vec{R}_I-\vec{T}) + \rho^\text{bg}(\lambda) V^\text{bg}(\vec{r}-\vec{T}).
\end{equation}
$\sum_\vec{T}$ is a sum over the lattice vectors of the periodic system, $V_I^\text{PP}$ is the pseudopotential of the $I$-th nucleus, $\rho^\text{bg}$ is the uniform background charge and $V^\text{bg}$ is the associated potential. The background charge must be introduced to obtain a neutral system for $\lambda < 1$ since the energy would diverge otherwise.
In this work, we use pseudopotentials from Goedecker, Teter and Hutter\cite{Hartwigsen1998,Krack2005} which can be expressed analytically as
\begin{equation}
    V_I^\text{PP}(\lambda, r) = V_{I, \text{loc}}(\lambda, r) + V_{I,\text{nl}}(\lambda, \vec{r}, \vec{r}')
\end{equation}
with 
\begin{equation}
\begin{split}
    V_{I, \text{loc}}(\lambda, r) &= -\frac{Z_\text{V}(\lambda)}{r} \text{erf}(\alpha^\text{PP}r) \\
    &+ \sum_i C_i^\text{PP}(\lambda)\left(\sqrt{2} \alpha^\text{PP}r\right)
    \exp{-\left(\alpha^\text{PP}r\right)^2}
\end{split}
\end{equation}
and
\begin{equation}
    V_{I,\text{nl}}(\lambda, \vec{r}, \vec{r}') = \sum_{lm}\sum_{ij}\braket{\vec{r}}{p_i^{lm}}h_{ij}^l(\lambda)\braket{p_j^{lm}}{\vec{r}'}.
\end{equation}
The $\lambda$-dependence of the parameters follows from the linear coupling of the Hamiltonians in Eq.~\eqref{eq:Hamiltonian} as $Z_\text{V}(\lambda) = \bar{Z}_\text{V} \lambda$, $C_i^\text{PP}(\lambda) = \bar{C}_i^\text{PP}\lambda$ and $h_{ij}^l(\lambda) = \bar{h}_{ij}^l \lambda $. $\bar{Z}_\text{V}, \bar{C}_i^\text{PP}$ and $\bar{h}_{ij}^l $ denote the optimized parameters reported in Ref.~\citenum{Hartwigsen1998,Krack2005} such that the pseudopotential is zero for $\lambda = 0$ and equal to the potentials in Ref.~\citenum{Hartwigsen1998,Krack2005}  at $\lambda = 1$. A detailed description of the other paramters can be found in Ref.~\citenum{Krack2005}.

The uniform background charge density is given by
\begin{equation}
    \rho^\text{bg}(\lambda) = \frac{N_e - \sum_I Z_{I, \text{V}}(\lambda)}{\Omega_\text{cell}}
\end{equation}
with $N_e$ the number of valence electrons, $Z_{I, \text{V}}(\lambda)$ the valence charge of nucleus $I$ at a given $\lambda$-value and $\Omega_\text{cell}$ the volume of the supercell.
The uniform background potential is defined as
\begin{equation}
    V^\text{bg}(\vec{r}) = -\int_{\Omega_\text{cell}} d\vec{r}' \frac{1}{|\vec{r} -\vec{r}'|}
\end{equation}
where $\int_{\Omega_\text{cell}}$ indicates integration over a single supercell.

Furthermore, the nuclear repulsion energy $E^\text{nuc}$ depends on the interaction between the nuclei $E^\text{NN}$, between nuclei and background potential $E^\text{Nbg}$ and the self-interaction of the background $E^\text{bg}$ such that
\begin{equation}
    E^\text{nuc} = E^\text{NN} + E^\text{Nbg} + E^\text{bg}
\end{equation}
with
\begin{equation}\label{eq:ENN_pbc}
    E^\text{NN} = \frac{1}{2} \sum_{I,J'} \sum_\vec{T}^{'}  \frac{Z_{I,\text{V}}(\lambda) Z_{J,\text{V}}(\lambda)}{|\vec{R}_I-\vec{R}_{J'}-\vec{T}|}
\end{equation}
where the $'$ indicates the exclusion of the divergent terms for $\vec{R}_I = \vec{R}_J$ at $\vec{T}=0$
\begin{equation}
    E^\text{Nbg} = \sum_{I,\vec{T}} -Z_{I,\text{V}}(\lambda) \rho_\text{bg}(\lambda) V^\text{bg}(\vec{R}_I-\vec{T})
\end{equation}
and
\begin{equation}    E^\text{bg} = \frac{1}{2}\rho^2_\text{bg}(\lambda) \sum_\vec{T} -\int_{\Omega_\text{cell}} d\vec{r} V^\text{bg}(\vec{r} - \vec{r}' - \vec{T}).
\end{equation}

In analogy to Ref.\citenum{rudorff_atomic_energies} (see also supplementary material section I), the atomic energy decomposition can be derived by application of Hellmann-Feynman theorem and chain rule as
\begin{equation}\label{eq:dE_pbc}
    E^\text{mol} - E^\text{UEG} = \sum_I \int_0^1 d\lambda \underbrace{\pdv{E^\text{nuc}}{Z_I} \pdv{Z_I}{\lambda} + \int d\vec{r} \Delta V_I \rho_\lambda(\vec{r})}_{:= \pdv{E_I}{\lambda}}
\end{equation}
with $\Delta V_I = \sum_\vec{T} V_I^\text{PP}(\lambda=1, \vec{r}-\vec{R}_I-\vec{T}) - \bar{Z}_I / \Omega_\text{cell} V^\text{bg}(\vec{r}-\vec{T})$.
The evaluation of this expression is non-trivial because because the individual terms, e.g. $\int d\vec{r} \sum_\vec{T} V_I^\text{PP}(\vec{r}-\vec{T}) \rho_\lambda (\vec{r})$, can be divergent.
Thus, we propose to approximate $\partial_\lambda E_I$ by central finite differences as
\begin{equation}\label{eq:cfd}
    \pdv{E_I}{\lambda} \approx \frac{E[V_I(\lambda+\Delta \lambda)]-E[V_I(\lambda-\Delta \lambda)]}{2 \Delta \lambda}.
\end{equation}
$E[V_I(\lambda \pm \Delta \lambda)]$ is the energy of a system where the parameters $Z_\text{V}, \{C_i\}$ and $\{h_{ij}\}$, of all pseudopotentials except for the $I$-th one are scaled by $\lambda$ and the same parameters for $V_I$ are scaled by $\lambda \pm \Delta \lambda$ with $\Delta \lambda$ being a small value (in this work $\Delta \lambda = 5\times 10^{-5}$).
The periodic boundary conditions and the background potential are then consistently considered in the evaluation of $E[V_I(\lambda \pm \Delta \lambda)]$ by the respective electronic structure code.
This improves the agreement between the sum of atomic energies and exact energy difference between compound and uniform electron gas, $\Delta E^\text{exact}$, by three orders of magnitude (MAE = 4~mHa) compared to Eq.~\eqref{eq:guido_intro} (MAE = 8~Ha) for the compounds mentioned above (see Figure~\ref{fig:methods}A).

\subsection{Technical aspects}
The calculation of atomic energies with Eq.~\eqref{eq:dE_pbc} is performed by numerical integration of $\partial_\lambda E_I$ along the $\lambda$-path. Inspection of the derivatives for different atoms $\partial_\lambda E_I$ as a function of $\lambda$ shows that the derivatives are zero at $\lambda = 0$ and increase monotonously with increasing $\lambda$ (Figure S1). Thus, the derivative at $\partial_\lambda E_I (\lambda = 0)$ was not calculated explicitly but always set to to 0. 
The error of $\sum_I \Delta E_I$ with respect to $\Delta E^\text{exact}$ for the test molecules mentioned above is shown in Figure~\ref{fig:methods}B for different numerical integration methods and increasing number of equally spaced integration points along the $\lambda$-path.
The MAE over all molecules decreases with increasing number of grid points. For 21 equally spaced integration points, the errors are 5.9, 4.5 and 4.4~mHa for integration with trapezoidal rule, Simpsons rule and cubic splines, respectively, which is still somewhat above chemical accuracy (1.6~mHa). 
Unfortunately, the use of an even denser integration grid is not feasible, in particular, if the decomposition scheme is applied to a larger set of molecules due to the slow convergence of the SCF-calculations for the unusual systems $\lambda < 0$.

While accurate absolute energies require a dense integration grid, qualitative trends in atomic energies can already be recovered for a coarse grid. This is illustrated in Figure~\ref{fig:methods}C, which shows atomic energies for carbon in 15 different molecular environments sorted by increasing energy value as predicted with 6 or 21 equally spaced integration points along $\lambda$.
The overall trend is already reproduced if atomic energies are calculated via integration with Simpson's rule over 6 points along the $\lambda$-path as evident from a Kendall's $\tau$ coefficient of 0.77.

Furthermore, the atomic energies can be rescaled to yield the exact energy difference as
\begin{equation}\label{eq:rescaling}
    \Delta \tilde{E}_I^N = \Delta E^\text{exact} \frac{\Delta E_I^N}{\sum_J \Delta E_J^N},
\end{equation}
where $\Delta E^\text{exact}$ is the exact energy difference between energy of the whole molecule and the uniform electron gas, $\Delta E_I^N$ is the atomic energy obtained after integration over $N$ points along $\lambda$ and $\sum_J \Delta E_J^N$ is the sum of all atomic energies in the molecule numerically integrated over $N$ points.
Rescaled atomic energies can be used if exact values are required, for example for the training of machine learning models.
For $N = 6$ and our test set, rescaling even improves the ordering of atomic energies (Kendall's $\tau = 0.85$) with respect to the dense integration grid with $N = 21$. Thus, we expect to recover atomic energy trends with varying local environment at least qualitatively at this grid size.

Furthermore, finite size effects can impact the atomic energies. For example the symmetrical hydrogen atoms in ethane have different interactions with periodic images because the molecular symmetry differs from the cell symmetry. As a consequence, their atomic energies differ by up to 6~mHa. It might be possible to account for this effect via finite size corrections as used in the calculation of point defect energies.\cite{PhysRevB.89.195205} However in this study, we simply average over the energies of symmetrical atoms because changes in atomic energy due to varying local enviromnent tend to be larger than the error due to finite size effects.
For example, in case of the symmetrical hydrogen atoms in ethane, the error due to the finite size effects (6~mHa) is significantly smaller than the change in atomic energy due to the different local environment of hydrogen in the amino group in CH$_3$NH$_2$ (15~mHa).

The computation of derivatives via central finite differences (Eq.~\eqref{eq:cfd}) entails an additional $2 M N$ self-consistent field (SCF) calculations, where $M$ is the number of atoms in the system and $N$ is the number of integration points along the $\lambda$ coordinate. Although these SCF calculations converge rapidly, they significantly increase the computational cost of this method. 
According to the Hellmann-Feynman theorem, a perturbative \emph{ansatz} can be employed instead,\cite{lili_2009,VonRudorff2020} where the energy $E[V(\lambda \pm \Delta \lambda), \rho_\lambda]$ is evaluated for the density $\rho_\lambda$ at $\lambda$ instead of the self-consistent density $\rho_{\lambda\pm\Delta \lambda}$ as
\begin{equation}\label{eq:pert}
     \pdv{E_I}{\lambda} = \frac{E[V(\lambda + \Delta \lambda), \rho_\lambda] - E[V(\lambda - \Delta \lambda), \rho_\lambda]}{2\Delta \lambda}
\end{equation}
Such a perturbative approach has already been used previously for the calculation of alchemical derivatives with periodic boundary conditions.\cite{VonLilienfeld2005,Chang2018} 
As shown in the supplementary material section II, Eq.~\eqref{eq:pert} should recover the exact derivative $\partial_\lambda E_I$ independent of the chosen $\Delta \lambda$. In practice, the agreement between Eq.~\eqref{eq:pert} and the finite difference approach Eq.~\eqref{eq:cfd} worsens with increasing values of $\Delta \lambda$. This might be due to the practical implementation of the energy expression with a uniform background potential in the used electronic structure code (CPMD\cite{CPMD}).
Nevertheless, perturbative atomic energies are nearly identical to those obtained via self-consistent calculations for a small value of $\Delta \lambda = 5 \times 10^{-5}$ with a mean absolute deviation of 0.03~mHa for the 91 atoms in our test set of 15 small molecules.
We assessed the same error for 361 fragments in 55 molecules of the QM9 database. In this case, the deviation was 0.02~mHa. Thus, atomic energies for the molecules in QM9 were evaluated via Eq.~\eqref{eq:pert}. 

Furthermore, the scaling of non-local parameters $h_{ij}$ of the pseudopotentials does not have a significant impact on the atomic energies (see Figure S2) as has been reported in a previous study.\cite{lili_2009} Therefore, only the parameters of $V_\text{loc}$ were scaled for the calculations of the QM9 molecules but rescaling of the $h_{ij}$ could be included straightforwardly if necessary.

The comparison of absolute atomic energies from the alchemical energy decomposition scheme with other decompositions is not meaningful because the alchemical energies are given with respect to the uniform electron gas and also differ significantly from energies obtained from an all electron calculation.
Thus, we define atomic contributions to the atomisation energy as 
\begin{equation}\label{eq:atomisation_energy}
    E_I^\text{at} = \Delta E_I + \frac{Z_{I,\text{V}}}{N_e} E^\text{UEG} - E_I^\text{free}
\end{equation}
with $\Delta E_I = \int_0^1 d\lambda \partial_\lambda E_I$ as defined in Eq.~\eqref{eq:dE_pbc}, the nuclear valence charge $Z_{I,\text{V}}$, the total number of valence electrons $N_e$, the energy of the uniform electron gas $E^\text{UEG}$ and the energy of the free pseudo atom (with nuclear charge $Z_{I,\text{V}}$) $E_I^\text{free}$. The second term removes the shift in energy due to the choice of the uniform electron gas as the reference.
The sum of atomic atomisation energies $\sum_I E_I^\text{at}$ is then equal to the atomisation energy of the respective molecule.

\section{Results and Discussion}
\subsection{Alchemical atomic energies of organic molecules}
The alchemical atomic energy decomposition scheme was applied to the subset of molecules in QM9 with 38 valence electrons (in total 1325 molecules). Lewis graphs of the corresponding molecular structures (coordinates directly taken from QM9\cite{qm9}) of the respective molecules are shown in the supplementary material (Figure~S5 - S17).
All alchemical atomic energies presented in this section correspond to atomisation energies as defined in Eq.~\eqref{eq:atomisation_energy}.
They were obtained by first calculating atomic energies with respect to the uniform electron gas by numerical integration of Eq.~\eqref{eq:dE_pbc} for each atom in a molecule. The required derivatives were evaluated with the perturbative approach Eq.~\eqref{eq:pert}. These atomic energies were further rescaled according to Eq.~\eqref{eq:rescaling} and finally atomic atomisation energies were obtained from Eq.~\eqref{eq:atomisation_energy}. 

\begin{figure}
\includegraphics[width=0.48\textwidth]{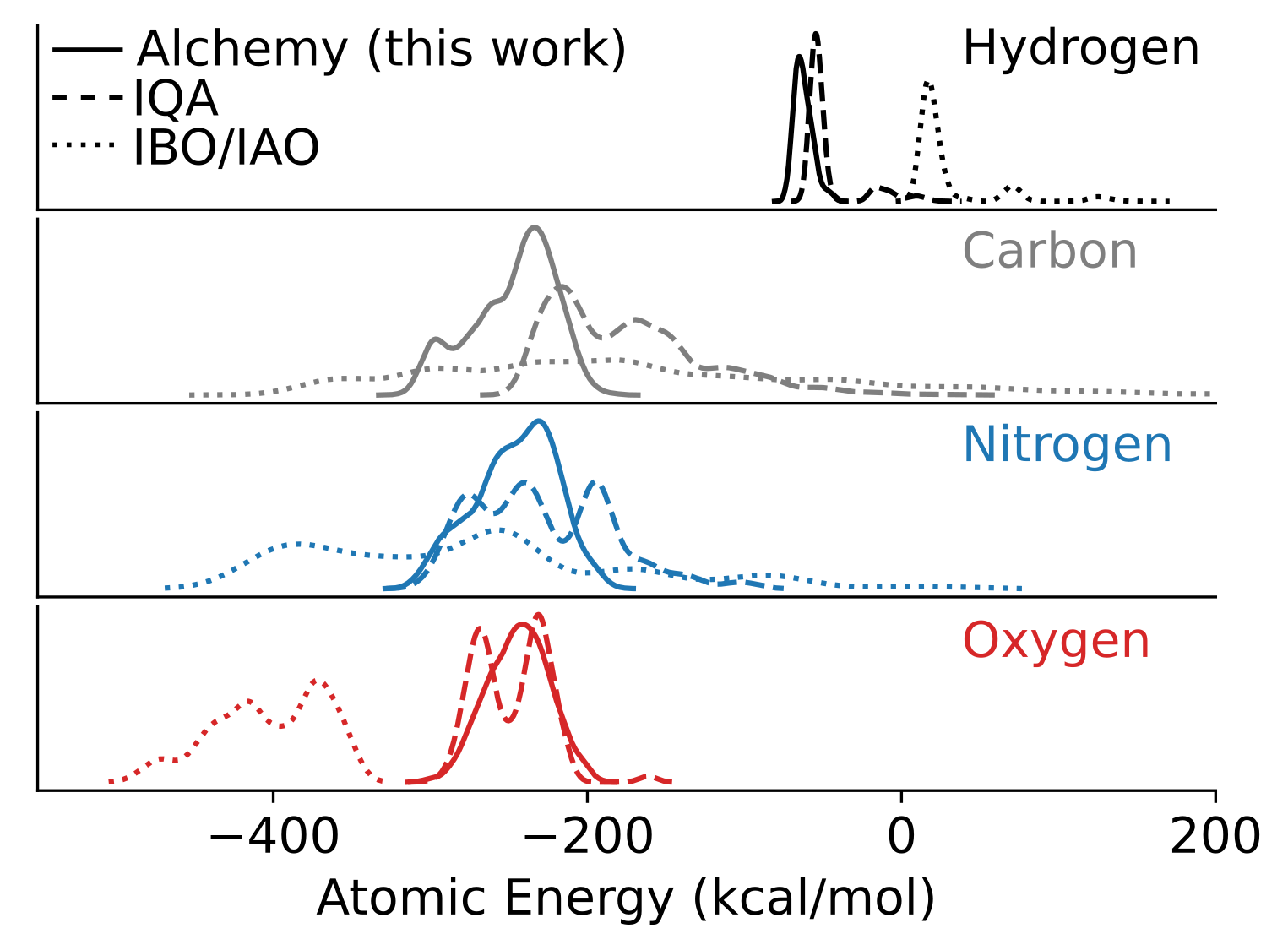}
\caption{Atomic contributions to the atomisation energy in different local environments for 1325 drug-like molecules ($\sim$18000 atoms) with 6-8 heavy atoms split by element from different decomposition schemes.}
\label{fig:ae_dist_diff_methods_modified}
\end{figure}

Figure \ref{fig:ae_dist_diff_methods_modified} shows the atomic energy distributions for hydrogen (8644 atoms), carbon (6323 atoms), nitrogen (1661 atoms) and oxygen (1341 atoms). The dataset contains also 9 fluorine atoms for which atomic energies are omitted in the Figure.

The magnitude of the atomic energies for carbon, nitrogen and oxygen is similar (mean values are $-247$, $-246$, $-244$~kcal/mol, respectively) and the distributions span similar ranges (135~kcal/mol for carbon and nitrogen and 110~kcal/mol for oxygen). The average atomic energy of hydrogen is $-62$~kcal/mol and the respective distribution spans a range of 37~kcal/mol. The narrower range of atomic energies for hydrogen could be explained by a lower diversity of molecular environments. Hydrogen can have only three different binding partners (carbon, nitrogen or oxygen). In contrast, in our dataset oxygen has 6, nitrogen 14 and carbon 34 different sets of binding partners.

\begin{figure*}
\includegraphics[width=1.0\textwidth]{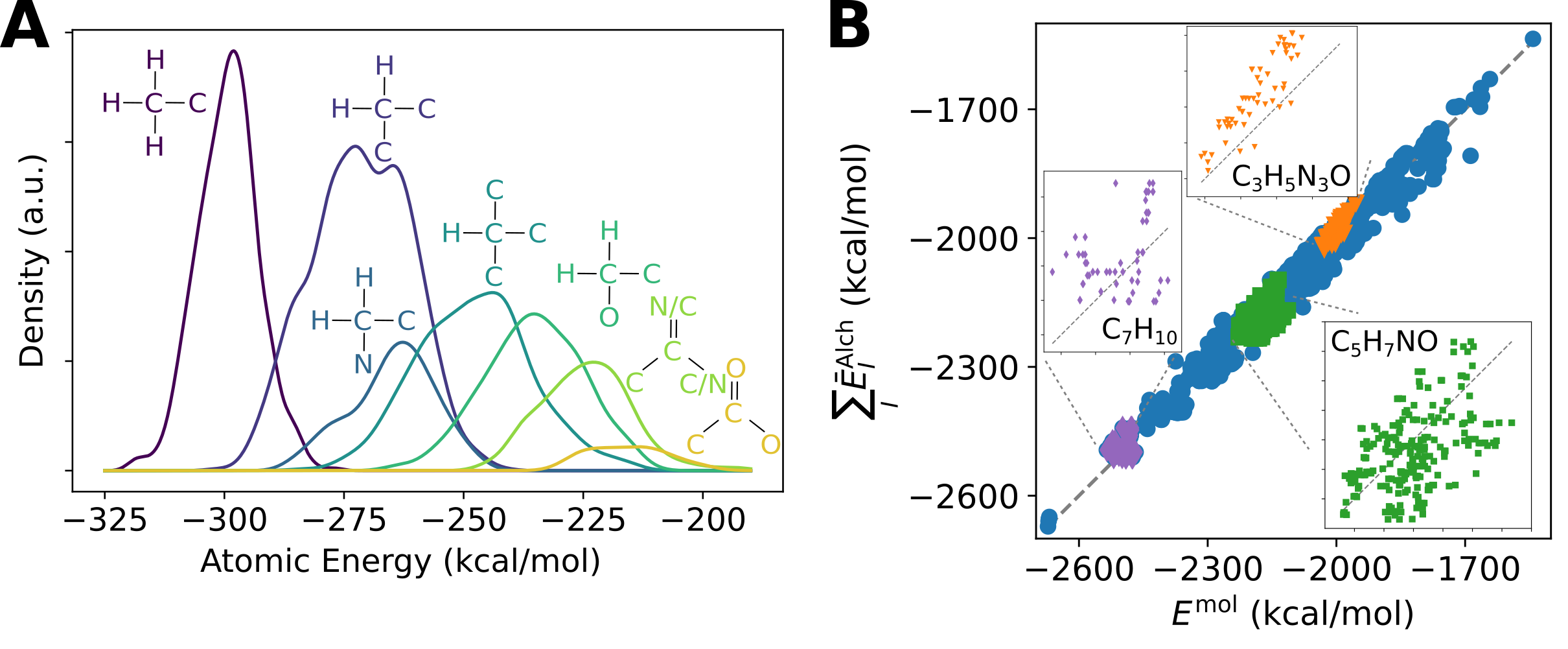}
\caption{Modelling the atomisation energy from atomic contributions obtained via the alchemical energy decomposition. Panel A: Atomic energy distributions for carbon in different local environments. Panel B: Estimated atomisation energies from the sum of average atomic energies for the local environments in a molecule. Insets show predictions for selected stoichiometries.}
\label{fig:local_envs_predictions}
\end{figure*}

We explore the impact of the local environment on the atomic energy in more detail by grouping the atomic energies by binding partners.
Exemplarily, selected atomic energy distributions for carbon with different binding partners are shown in Figure~\ref{fig:local_envs_predictions}A.
The atomic energies for the same set of binding partners span energy ranges of 50~kcal/mol or less indicating that the atomic energy is sensitive to the local environment. A similar relation can be observed for the elements H, N and O (Figure~S3).

This can be exploited to estimate atomisation energies by adding up the mean atomic energy for each local environment in a molecule as
\begin{equation}\label{eq:estimate}
    E^\text{mol} \approx \sum_I \bar{E}(Z_I, P_I),
\end{equation}
where $I$ is a sum over all atoms in the molecule, $Z_I$ is the nuclear charge of $I$, $P_I$ describes the set of binding partners of $I$ and $\bar{E}(Z_I, P_I)$ is the atomic energy averaged over all atoms with nuclear charge $Z_I$ and the binding partners $P_I$ in our dataset. The atomisation energies of the 1325 molecules can be estimated with a MAE of $29 \pm 23$~kcal/mol by Eq.~\eqref{eq:estimate}. Furthermore, energies are overestimated for molecules with a low atomisation energy and underestimated for molecules with a high atomisation energy. The MAE can be reduced to $23 \pm16$~kcal/mol by accounting for the systematic error by combining Eq.~\eqref{eq:estimate} with a linear fit as 
\begin{equation}\label{eq:estimate_lin}
    E^\text{mol} \approx m \sum_I \bar{E}(Z_I, P_I) + b,
\end{equation}
with slope $m$ and intercept $b$ chosen such that $\sum_J \left(m \sum_J \sum_{I(J)} \bar{E}(Z_I, P_I) + b -  E_J^\text{mol}\right)^2$ is minimal, where $J$ runs over all molecules in the dataset.

As displayed in Figure~\ref{fig:local_envs_predictions}B, Eq.~\eqref{eq:estimate_lin} can predict the stability trend for the molecules in our dataset. While the overall error is one order above chemical accuracy, the model might still be useful as a baseline for machine learning models of the atomisation energy.

To set the accuracy of our baseline into perspective, we compare it to other simple estimates of atomisation energies. For example bond-counting yields a MAE of 71~kcal/mol for the QM7-dataset.\cite{Rupp2012} While QM7 is a larger and likely more diverse dataset than the subset of 1325 QM9 molecules used in this study, this result indicates that the alchemical decomposition scheme can be competitive to conventional energy estimates.
The dressed atom is another, statistical estimate of atomisation energies.\cite{Hansen2015} 
In the dressed atom model, the atomisation energy is approximated as
\begin{equation}
    E^\text{mol} \approx \sum_\text{elements} N^\text{el} \times E^\text{mean,el},
\end{equation}
where an average energy $E^\text{mean,el}$ is defined for each element and $N^\text{el}$ is the number of atoms of a given element type in the molecule. 
The atomic energies $E^\text{mean,el}$ are chosen such that the loss function $\sum_J (E_J^\text{mol} - \sum_\text{elements} N_J^\text{el} \times E^\text{mean,el})^2$ over all atomisation energies $E_J^\text{mol}$ in the training set is minimized.
The prediction error of the dressed atom approach is $=20 \pm 14$~kcal/mol which is somewhat lower than for atomic energies from quantum alchemy. However, since alchemical atomic energies depend on the local environment, stability trends can in some case also be reproduced for the same stoichiometry (e.g. C$_3$H$_5$N$_3$O, see Figure~\ref{fig:local_envs_predictions}B) while molecules with the same stoichiometry have always the same energy with the dressed atom approach. In addition, the dressed atom approach is a purely statistical model while the energies $\bar{E}(Z_I, P_I)$ originate from a physics based decomposition scheme. Thus, they are less sensitive to the choice of the training set and should be more transferable.

A limitation of Eq.~\eqref{eq:estimate_lin} is that the atomisation energy of a molecule can only be calculated if all of its atomic energies $\bar{E}(Z_I, P_I)$ are known.
This restriction can be addressed by constructing a descriptor to calculate $\bar{E}(Z_I, P_I)$ for unknown local environments.
Thus, we analysed how changes in atomic energy are related to respective changes in the local environment. 

\subsection{The relation between atomic energy and nuclear electrostatic repulsion potential}

Figure~\ref{fig:local_envs_predictions}A shows that the atomic energy distributions are shifted to higher energies as the nuclear charges of the binding partners increase. For example, for a carbon atom bonded to three hydrogens and one carbon atom (C-H$_3$C) the mean atomic energy is -299~kcal/mol. By replacing one hydrogen with a carbon atom (C-H$_2$C$_2$) the atomic energy increases to -271~kcal/mol and exchanging another hydrogen for a carbon atom (C-HC$_3$) yields an atomic energy of -248~kcal/mol. Similar trends can be observed for the other elements as well (Figure~S3).

\begin{figure}
\includegraphics[width=0.48\textwidth]{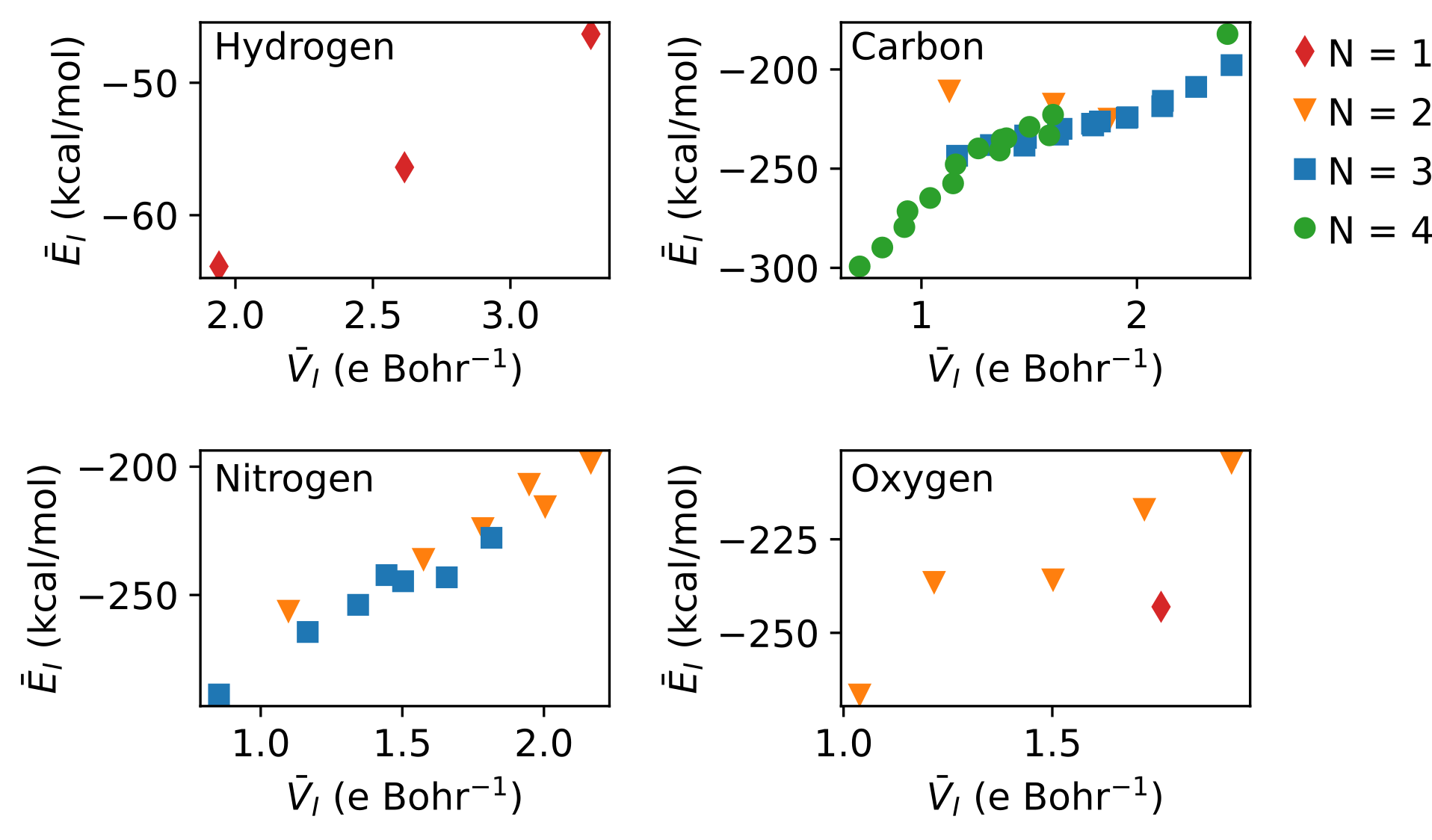}
\caption{The influence of the local environment on the respective mean atomic energy $\bar{E}_I$ (Eq.~\eqref{eq:estimate}) quantified via the relation between the average electrostatic potential of the binding partners $\bar{V}_I$ (Eq.~\eqref{eq:pot_eff}) at the position of atom $I$ for H, C, N and O and different number of binding partners $N$ which restrict bond orders.}
\label{fig:V_vs_E}
\end{figure}

The structural change can be quantified by the average electrostatic potential $V_I$ generated by the nuclei of the binding partners at the position $\vec{R}_I$ of atom $I$ as 
\begin{equation}\label{eq:pot_eff}
    V_I = \frac{1}{N} \sum_{J=1}^N \frac{Z_{J,V}}{|\vec{R}_J - \vec{R}_I|}
\end{equation}
with $N$ being the number of covalent binding partners and $Z_{J,V}$ and $\vec{R}_J$ their respective nuclear valence charges and positions. 
By averaging over the potentials $V_I$ for all atoms in the dataset with same nuclear charge and the same binding partners a potential $\bar{V} (Z_I, P_I)$ can be defined analogous to $\bar{E}(Z_I, P_I)$.
As illustrated in Figure~\ref{fig:V_vs_E} $\bar{E}(Z_I, P_I)$ and $\bar{V} (Z_I, P_I)$ correlate for the elements H, C, N and O (with $\bar{E}_I$ corresponding to $\bar{E}(Z_I, P_I)$ and $\bar{V}_I$ corresponding to $\bar{V} (Z_I, P_I)$). Since $V_I$ can easily be calculated for a given 3D-structure, it could serve as a descriptor to obtain $\bar{E}(Z_I, P_I)$ for an unknown local environment.
Furthermore, the correlation can be understood from Eq.~\eqref{eq:guido_intro}. 
Eq.~\eqref{eq:guido_intro} has an attractive and a repulsive electrostatic term that partially cancel motivating the description of the effects of the local environment by the effective potential given in Eq.~\eqref{eq:pot_eff}. We note that Eq.~\eqref{eq:guido_intro} does not include periodic boundary conditions as discussed in the Method section but it should be sufficient to explain the effect of the local environment where interactions with periodic images can be neglected. Furthermore, we choose the nuclear valence charge instead of the total nuclear charge because atomic energies are calculated using pseudopotentials. If the approach was extended to compounds containing elements like phosphorus or sulfur, with the same nuclear valence charge as nitrogen and oxygen, an effective nuclear charge\cite{slater} would be a more suitable choice to distinguish binding partners from the same group in the periodic table.

The close relationship between electrostatic potential and atomic energy emerging from the alchemical transformation, has been noted before by Politzer and Murray and motivated the following expression for the molecular energy
\begin{equation}
    E^\text{mol} = a \sum_A Z_A V_{A,0} + b
\end{equation}
with the nuclear charges $Z_A$, the electrostatic potential at position of $Z_A$ due to the other nulei and the electron density (see Eq.~\eqref{eq:elstat_politzer})
and $a$ and $b$ are fitted to reference molecular energies.\cite{Politzer2021}
Note that the fundamental importance of these relationships also motivated the definition of the
Coulomb matrix and the Bag of Bonds representation
of molecules and materials for training machine learning models of quantum properties, including energies, throughout chemical compound space\cite{Rupp2012, Montavon2013, Hansen2015}

We also investigated the influence of the bond order on the atomic energy by inspecting the energy-potential relationships for different number of binding partners $N$ for C, N and O.
For example, carbon can have a single and a triple bond or two double bonds ($ N = 2$), two single bonds and one double bond ($N = 3$) or four single bonds ($N = 4$). The relationship between energy and potential is approximately linear for fixed $N$ (Figure~\ref{fig:V_vs_E}). However, the slope changes with varying $N$ indicating that the bond order impacts the sensitivity to changes in the local environment. This should be expected as different bond orders lead to different electron density profiles in the local environment. Due to the dependence of the atomic energy on the electron density according to Eq.~\eqref{eq:guido_intro} this should result in different atomic energy trends. 
Oxygen seems to behave similarly with the atomic energy for $N = 1$ (a carbonyl group) deviating from the linear trends observed for $N = 2$ but in this case there is not enough data for a definite conclusion.
In contrast, the relation between $\bar{V}_I$ and $\bar{E}_I$ is linear for nitrogen with similar slopes for $N = 2$ and $N =3$, showing that the influence of the bond order is not the same for all elements.

In summary, this analysis reveals a linear correlation between the average nuclear electrostatic potential of the local environment (Eq.~\eqref{eq:pot_eff}) and the respective atomic energy (in case of carbon for fixed number of binding partners). 
This correlation can be rationalized by the dependence of the atomic energy on the electrostatic interactions between nuclei and electrons and could be used to estimate atomic energies for local environments not contained in the training data.

\FloatBarrier
\subsection{Comparison to other decomposition methods}
\begin{figure*}
\includegraphics[width=1.0\textwidth]{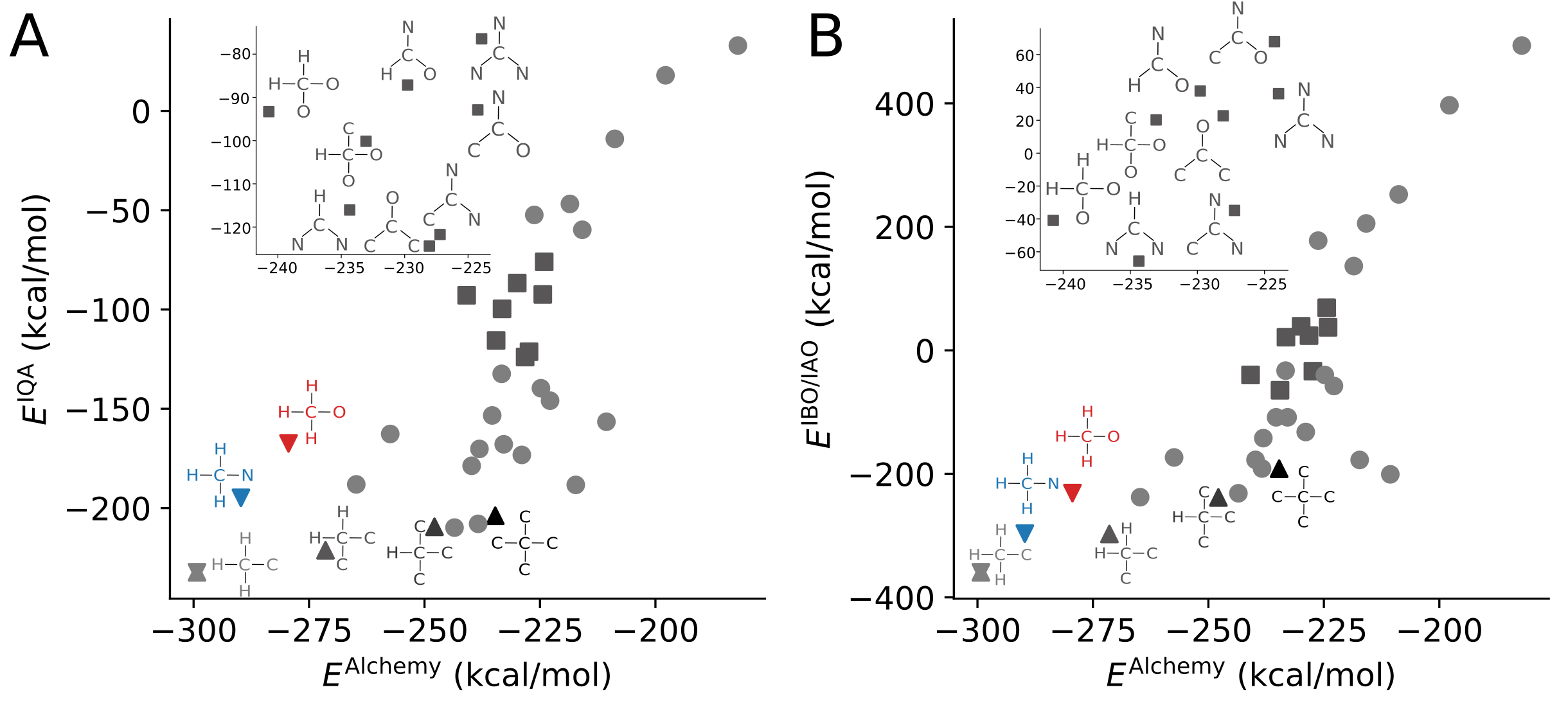}
\caption{Comparison for IQA (panel A) and IBO/IAO (panel B) partitioning schemes to alchemical partitioning in varying environments of the carbon atom. Similarities and differences are highlighted for selected series of environments. Up-pointing triangles correspond to a series of environments were hydrogens are substituted by carbons, down-pointing triangles show trends if a single heavy atom neighbor X in H$_3$C-X is varied, cubes correspond to a set of environments generating similar nuclear electrostatic potentials at the position of the central carbon atom and circles denote the remaining local environments in the dataset.}
\label{fig:compare_schemes_scatter_no_ml_modified}
\end{figure*}

We also calculated atomic energies with the interacting quantum atom approach\cite{IQA_method} and a MO/AC partitioning method. In the MO/AC decomposition different types of molecular orbitals and atomic charges can be chosen. We select intrinsic bond orbitals\cite{IBOIAO} and charges from intrinsic atomic orbitals, \cite{IBOIAO}e.g. the IBO/IAO partitioning, because it suffers less from basis set artifacts compared to other MO/AC partitionings.\cite{Eriksen2020,Kjeldal2023} The respective atomic energy distributions are displayed in Figure~\ref{fig:ae_dist_diff_methods_modified}. We note that atomic energies of 224 molecules were excluded for the interacting quantum atom approach because they contained non-nuclear attractors, likely due to basis set artifacts,\cite{Salvador2001} which prevents the partitioning of the energy into atomic contributions.

As for the alchemical decomposition, atomic energies for hydrogen are higher than for the other elements.
However, the atomic energies from IQA and IBO/IAO span wider energy ranges than energies from quantum alchemy. For example, the difference between lowest and highest atomic energy of carbon is 138~kcal/mol for alchemy, 277~kcal/mol for IQA and 863~kcal/mol for IBO/IAO.
The broader energy range is also correlated with increased error in predicting atomization energy using Eq.~\eqref{eq:estimate_lin} (MAE of 25 kcal/mol for IQA and 50 kcal/mol for IBO/IAO). Consequently, the alchemical decomposition method is the preferred choice for estimating energies solely based on average atomic energies with a specific set of binding partners.
However, atomic energies derived from the IBO/IAO decomposition scheme might be more suitable as molecular fingerprints, e.g. for machine learning applications. This preference arises from their ability to provide finer-grained differentiation between distinct local environments since the respective atomic energy distributions are more separated from each other.

Furthermore, the atomic energy distributions displayed in Figure~\ref{fig:ae_dist_diff_methods_modified}  have different shapes indicating a different relation between local environment and respective atomic energy.
We explore these differences by plotting the mean atomic energies $\bar{E}(Z_I, P_I)$ in different local environments from either IQA or IBO/IAO against the respective energies from the alchemical decomposition (Figure~\ref{fig:compare_schemes_scatter_no_ml_modified}).
This analysis reveals an overall similar dependence of the atomic energy on the local environment for the three decomposition methods.
Carbon in the H$_3$C-C fragment has always the lowest energy and the energy increases when a binding partner is replaced by a more electronegative one up to F$_3$C-C which has always the highest atomic energy.

Nevertheless, there are also distinct differences in the relative change in atomic energy from one environment to another one.
Alchemical energies change strongly if a hydrogen atom is substituted by a second row element, e.g. for the series H$_3$C-C, H$_2$C-C$_2$, HC-C$_3$ and C-C$_4$, because the electrostatic potential of the environment varies strongly. If the heavy atom is varied, e.g. for the fragments H$_3$C-C, H$_3$C-N and H$_3$C-O (see Figure~\ref{fig:compare_schemes_scatter_no_ml_modified}) the change in atomic energies with respect to the change in local environment is smaller due to the smaller variation in the electrostatic potentials $V_I$ (Eq.~\eqref{eq:pot_eff}).
IQA and IBO/IQA show the opposite behaviour with a stronger change in atomic energy for the series H$_3$C-C, H$_3$C-N and H$_3$C-O compared to the series H$_3$C-C, H$_2$C-C$_2$, HC-C$_3$ and C-C$_4$.

Furthermore, the local environments in the insets in Figure~\ref{fig:compare_schemes_scatter_no_ml_modified} have similar potentials $V_I$ leading only to a small variation of the respective alchemical atomic energies, while these energies vary more for IQA and IBO/IAO. These differences in the dependence on the local environment are responsible for the non-linear relation between $E^\text{Alchemy}$ and $E^\text{IQA}$ respectively $E^\text{IBO/IAO}$.
Similar correlations between the decomposition schemes were also found for the other elements H, N and O (see Figure S4).

The main difference between the alchemical and the other decomposition schemes in describing the relation between atomic energy and environment is the stronger dependence of the alchemical energies on the electrostatic potential of the local environment.
Thus, the alchemical decomposition might be particularly well-suited to resolve energy variations for systems with substantial electrostatic fluctuations, such as atomic diffusion in solids. Furthermore, the atomic energies derived from the alchemical decomposition could potentially be efficiently approximated using only a simple electrostatic model, while the relation between atomic energy and local environment might be more complicated for the other decomposition schemes.

\section{Conclusion}
An atomic energy partitioning based on an alchemical transformation was proposed already in 1974\cite{Politzer1974}, 
but practical applications were scarce due to arbitrary coupling and reference choices.
An important step forward, necessary to obtain the results of this work, was the introduction of the uniform electron gas as a universal reference system which can be transformed consistently into arbitrary isoelectronic compounds through linear coupling to the respective Hamiltonians.\cite{rudorff_atomic_energies} 
However, in this case periodic boundary conditions must be considered during atomic energy computation. 
Here, we addressed this problem by evaluating changes in energy due to a perturbation of the individual nuclear pseudopotentials consistent with periodic boundary conditions.

This enables the systematic and quantitative investigation of the relation between local structure and atomic energy, yielding interesting trends hinting to a general transferability of the atomic energies. 
This refined decomposition approach has been employed to compute atomic energies for a subset of 1325 molecules drawn from the QM9 dataset. Analysis of the numerical results indicates a consistent and not surprising pattern: atoms sharing the same local environment exhibit similar atomic energies across distinct molecules. Consequently, the summation of averaged atomic energies corresponding to various environments within a molecule can approximate atomization energies. For the molecules studied in this work, we find an associated mean absolute error of $23 \pm 16$~kcal/mol. This level of accuracy aligns closely with statistically determined atomic energies (utilizing the dressed atom approach)~\cite{Hansen2015}, while also potentially offering enhanced transferability. 
Naturally this approach is remarkably efficient in terms of computational complexity, being approximately 80 times faster than the semi-empirical xTB method,\cite{xtb1,xtb2}, indicating that it might be an attractive alternative for tasks such as rapid filtering of stabilities in vast chemical libraries encoding diverse chemistries. Such speed in energy estimation could accelerate research efforts that require swift assessments of molecular or material stability.
Our averaging decomposition model might also prove useful as a meaningful base-line model for $\Delta$-machine learning\cite{Ramakrishnan2015}---in analogy to recently proposed applications of Hammett's equation\cite{Hammett_ea,bragato2023occams}.

Moreover, we have also demonstrated on a fundamental level that the connection between atomic energy and local environment can be quantified using just the electrostatic potential produced by the nuclei of the binding partners at the atom's location. 
This  enables the explanation of atomic energy trends across diverse local environments by considering just the respective nuclear charges and positions.

We also compared alchemical atomic energies and those obtained by two other decomposition methods: the interacting quantum atom (IQA)\cite{IQA} and a partitioning technique employing intrinsic bond orbitals and charge distribution via intrinsic atomic orbitals (IBO/IAO).\cite{Eriksen2020}
Across these three decomposition methods, a consistent relationship emerges between atomic energy and local environment: The less electronegative the binding partners, the lower the atomic energy contribution. 
By contrast to IQA and IBO/IAO, the alchemical decomposition method displays a heightened sensitivity to substantial alterations in electrostatics within the local environment.
Consequently, the alchemical decomposition approach appears well-suited for discerning subtle energy changes tied to significant shifts in electrostatics. This capability could be particularly valuable in the modeling of scenarios such as defect formation or diffusion in solids characterized by significant variations in electrostatics between lattice and interstitial sites.
For example, the alchemical atomic energies offer the potential to assess the stability of metal or high entropy alloys, contributing to investigations in this domain based on prior studies using other decomposition schemes\cite{doi:10.1080/21663831.2022.2051763,Xu2021,Oh2019-bz,hea_defects} or could be used to explore compounds with elevated ionic conductivity, a crucial consideration in the advancement of solid-state electrolyte design.\cite{D0CS00305K,XIA2019753}

\section*{Computational Details}
Initial geometries for the 15 molecules used in the Method section were generated through the LERULI API\cite{leruli, rdkit, xtb1, xtb2}.
The geometries were further optimized with the CPMD code\cite{CPMD} (except for CH$_3$CH$_3$, CH$_3$NH$_2$, CH$_3$OH, CH$_3$F) making use of gnu parallel\cite{Tange2011a} with the PBE functional,\cite{pbe} Goedecker Teter Hutter pseudopotentials,\cite{Goedecker1,Goedecker2} a cutoff of 200~Ryd and a boxsize of $\left(N_\text{VE}/8 \times 15^3\right)^{1/3}$~\AA~with $N_\text{VE}$ being the number of valence electrons in the molecule.
For the optimized geometries, 21 single point density functional theory (DFT) calculations were carried out along the $\lambda$ path in equally spaced intervals (step size = 0.05) from $\lambda = 0$ to $\lambda = 1$ by scaling the paramters $Z_\text{V}$, $\{C_i^\text{PP}\}$ and $\{h_{ij}\}$ in the pseudopotentials by the respective $\lambda$-value. Derivatives with respect to the variation in the external potential at a specific atomic site were calculated according to Eq.~\eqref{eq:cfd} with $\Delta \lambda =  5 \times 10^{-5}$. For the perturbative approach the SCF calculations were stopped after a single SCF iteration.  For the calculation of atomic energies with Eq.~\eqref{eq:guido_intro} the valence electron density was generated with CPMD2CUBE.\cite{cpmd2cube} The numerical integration with respect to $\lambda$ was performed with the help of scipy\cite{scipy} and numpy.\cite{numpy}

In total there are 1329 molecules with 38 valence electrons in QM9. Four molecules (dsgdb9nsd\_000226, dsgdb9nsd\_000229, dsgdb9nsd\_003330, dsgdb9nsd\_003492) were excluded due to conversion problems and atomic energies were calculated for the remaining 1325 molecules. Geometries were extracted from the QM9 dataset\cite{qm9} and single points were calculated with the LDA functional, Goedecker Teter Hutter pseudopotentials, a cutoff of 200~Ryd and a boxsize of 20~\AA~for $\lambda = 8/38, 15/38, 23/38, 30/38, 1$ with the parameters $Z_\text{V}$ and $\{C_i^\text{PP}\}$ scaled by the respective $\lambda$-value. Derivatives with respect to the variation in the external potential at a specific atomic site were calculated perturbatively via Eq~\eqref{eq:pert}. The numerical integration with respect to $\lambda$ was performed with Simpson's rule using scipy and numpy. The atomic energies were rescaled using Eq.~\eqref{eq:rescaling} and atomic atomisation energies were calculated via Eq.~\eqref{eq:atomisation_energy}. The free atom energies were calculated with the same parameters as for the molecules.

The atomic energies from the interacting quantum atom approach were calculated using the AIMAll code.\cite{AIMAll} The required wavefunction files were generated with gaussian\cite{Gaussian09} from single point DFT calculations with the svwn-functional and the 6-31G(2df,p) basis set, utilizing the molecular geometries reported in the QM9 dataset. Free atom energies were calculated with the same settings with unrestricted DFT.

The atomic energies from the IBO/IAO method were calculated using the decondense code\cite{Eriksen2020} with the ibo-4 option and pyscf.\cite{pyscf} Required single point DFT calculations were performed on the geometries reported in the QM9 dataset with the LDA functional and the 6-31G(2df,p) basis set. Free atom energies were calculated with the same settings using unrestricted DFT.

Average timings for the energy evaluation with Eq.~\eqref{eq:estimate} or a single point xTB\cite{xtb1,xtb2} calculation were generated by evaluating the energy for the compounds with the QM9 IDs dsgdb9nsd\_000227, dsgdb9nsd\_001768, dsgdb9nsd\_002277, dsgdb9nsd\_002839, dsgdb9nsd\_008977. 

\section*{Supplementary Material}
Detailed derivation of Eqs.~\eqref{eq:dE_pbc} and \eqref{eq:pert}, additional data on the properties of the alchemical derivatives $\partial_\lambda E_I $, relation between local structure and alchemical atomic energies, comparison of atomic energies for the different decomposition schemes and Lewis structures of the selected QM9 compounds.

\begin{acknowledgments}
Michael J. Sahre acknowledges discussions with Giorgio Domenichini, Danish Khan, Jan Weinreich and Dominik Lehm.
ChatGPT-3.5 and GPT4 were used for rephrasing parts of the paper.
O.A.v.L. has received funding from the European Research Council (ERC) under the European Union’s Horizon 2020 research and innovation programme (grant agreement No. 772834).
This research is part of the University of Toronto’s Acceleration Consortium, which receives funding from the Canada First Research Excellence Fund (CFREF).
O.A.v.L. has received support as the Ed Clark Chair of Advanced Materials and as a Canada CIFAR AI Chair.  \end{acknowledgments}

\section*{Author Declarations}

\section*{Conflict of Interest}
The authors have no conflicts to disclose.

\section*{Author Contributions}
\textbf{M. J. Sahre:} Conceptualization (equal), Methodology (lead), Data Curation (lead), Software (lead), Writing - Original Draft (lead) , Writing - Review \& Editing (equal).
\textbf{G. F. von Rudorff:} Conceptualization (equal), Methodology (supporting), Software (supporting), Writing - Review \& Editing (equal).
\textbf{P. Marquetand:} Supervision (lead), Conceptualization (equal), Writing - Review \& Editing (equal).
\textbf{O. A. von Lilienfeld:} Supervision (supporting), Conceptualization (supporting), Writing - Review \& Editing (equal).

\section*{Data availability}
The data that support the findings of this study are available within the paper and the supplementary material. Additionally, the atomic energies of the QM9 compounds calculated with the different decomposition schemes (alchemy, IQA and IBO/IAO) and code to generate  the input-file and the rescaled pseudopotential files for the DFT calculations with the CPMD program and to calculate alchemical atomic energies from the CPMD output are openly available in Zenodo at 10.5281/zenodo.10041083. The latest version of the code can be found at https://github.com/michasahre/Alchemy\_PP.

\bibliography{ref}

\begin{thebibliography}{72}%
\makeatletter
\providecommand \@ifxundefined [1]{%
 \@ifx{#1\undefined}
}%
\providecommand \@ifnum [1]{%
 \ifnum #1\expandafter \@firstoftwo
 \else \expandafter \@secondoftwo
 \fi
}%
\providecommand \@ifx [1]{%
 \ifx #1\expandafter \@firstoftwo
 \else \expandafter \@secondoftwo
 \fi
}%
\providecommand \natexlab [1]{#1}%
\providecommand \enquote  [1]{``#1''}%
\providecommand \bibnamefont  [1]{#1}%
\providecommand \bibfnamefont [1]{#1}%
\providecommand \citenamefont [1]{#1}%
\providecommand \href@noop [0]{\@secondoftwo}%
\providecommand \href [0]{\begingroup \@sanitize@url \@href}%
\providecommand \@href[1]{\@@startlink{#1}\@@href}%
\providecommand \@@href[1]{\endgroup#1\@@endlink}%
\providecommand \@sanitize@url [0]{\catcode `\\12\catcode `\$12\catcode
  `\&12\catcode `\#12\catcode `\^12\catcode `\_12\catcode `\%12\relax}%
\providecommand \@@startlink[1]{}%
\providecommand \@@endlink[0]{}%
\providecommand \url  [0]{\begingroup\@sanitize@url \@url }%
\providecommand \@url [1]{\endgroup\@href {#1}{\urlprefix }}%
\providecommand \urlprefix  [0]{URL }%
\providecommand \Eprint [0]{\href }%
\providecommand \doibase [0]{https://doi.org/}%
\providecommand \selectlanguage [0]{\@gobble}%
\providecommand \bibinfo  [0]{\@secondoftwo}%
\providecommand \bibfield  [0]{\@secondoftwo}%
\providecommand \translation [1]{[#1]}%
\providecommand \BibitemOpen [0]{}%
\providecommand \bibitemStop [0]{}%
\providecommand \bibitemNoStop [0]{.\EOS\space}%
\providecommand \EOS [0]{\spacefactor3000\relax}%
\providecommand \BibitemShut  [1]{\csname bibitem#1\endcsname}%
\let\auto@bib@innerbib\@empty
\bibitem [{\citenamefont {Fias}\ \emph {et~al.}(2017)\citenamefont {Fias},
  \citenamefont {Heidar-Zadeh}, \citenamefont {Geerlings},\ and\ \citenamefont
  {Ayers}}]{Fias2017}%
  \BibitemOpen
  \bibfield  {author} {\bibinfo {author} {\bibfnamefont {S.}~\bibnamefont
  {Fias}}, \bibinfo {author} {\bibfnamefont {F.}~\bibnamefont {Heidar-Zadeh}},
  \bibinfo {author} {\bibfnamefont {P.}~\bibnamefont {Geerlings}},\ and\
  \bibinfo {author} {\bibfnamefont {P.~W.}\ \bibnamefont {Ayers}},\ }\bibfield
  {title} {\enquote {\bibinfo {title} {Chemical transferability of functional
  groups follows from the nearsightedness of electronic matter},}\ }\href@noop
  {} {\bibfield  {journal} {\bibinfo  {journal} {PNAS}\ }\textbf {\bibinfo
  {volume} {114}},\ \bibinfo {pages} {11633--11638} (\bibinfo {year}
  {2017})}\BibitemShut {NoStop}%
\bibitem [{\citenamefont {Herbert}(2019)}]{Herbert2019}%
  \BibitemOpen
  \bibfield  {author} {\bibinfo {author} {\bibfnamefont {J.~M.}\ \bibnamefont
  {Herbert}},\ }\bibfield  {title} {\enquote {\bibinfo {title} {Fantasy versus
  reality in fragment-based quantum chemistry},}\ }\href@noop {} {\bibfield
  {journal} {\bibinfo  {journal} {J. Chem. Phys.}\ }\textbf {\bibinfo {volume}
  {151}} (\bibinfo {year} {2019})}\BibitemShut {NoStop}%
\bibitem [{\citenamefont {Liu}\ and\ \citenamefont {He}(2020)}]{Liu2020}%
  \BibitemOpen
  \bibfield  {author} {\bibinfo {author} {\bibfnamefont {J.}~\bibnamefont
  {Liu}}\ and\ \bibinfo {author} {\bibfnamefont {X.}~\bibnamefont {He}},\
  }\bibfield  {title} {\enquote {\bibinfo {title} {Fragment-based quantum
  mechanical approach to biomolecules, molecular clusters, molecular crystals
  and liquids},}\ }\href@noop {} {\bibfield  {journal} {\bibinfo  {journal}
  {Phys. Chem. Chem. Phys.}\ }\textbf {\bibinfo {volume} {22}},\ \bibinfo
  {pages} {12341--12367} (\bibinfo {year} {2020})}\BibitemShut {NoStop}%
\bibitem [{\citenamefont {Hammett}(1935)}]{hammett_35}%
  \BibitemOpen
  \bibfield  {author} {\bibinfo {author} {\bibfnamefont {L.~P.}\ \bibnamefont
  {Hammett}},\ }\bibfield  {title} {\enquote {\bibinfo {title} {{Some Relations
  between Reaction Rates and Equilibrium Constants.}}}\ }\href@noop {}
  {\bibfield  {journal} {\bibinfo  {journal} {Chem. Rev.}\ }\textbf {\bibinfo
  {volume} {17}},\ \bibinfo {pages} {125--136} (\bibinfo {year}
  {1935})}\BibitemShut {NoStop}%
\bibitem [{\citenamefont {Hammett}(1937)}]{hammett_37}%
  \BibitemOpen
  \bibfield  {author} {\bibinfo {author} {\bibfnamefont {L.~P.}\ \bibnamefont
  {Hammett}},\ }\bibfield  {title} {\enquote {\bibinfo {title} {{The Effect of
  Structure upon the Reactions of Organic Compounds. Benzene Derivatives}},}\
  }\href@noop {} {\bibfield  {journal} {\bibinfo  {journal} {JACS}\ }\textbf
  {\bibinfo {volume} {59}},\ \bibinfo {pages} {96--103} (\bibinfo {year}
  {1937})}\BibitemShut {NoStop}%
\bibitem [{\citenamefont {von Lilienfeld}, \citenamefont {Müller},\ and\
  \citenamefont {Tkatchenko}(2020)}]{ccs_review1}%
  \BibitemOpen
  \bibfield  {author} {\bibinfo {author} {\bibfnamefont {O.~A.}\ \bibnamefont
  {von Lilienfeld}}, \bibinfo {author} {\bibfnamefont {K.-R.}\ \bibnamefont
  {Müller}},\ and\ \bibinfo {author} {\bibfnamefont {A.}~\bibnamefont
  {Tkatchenko}},\ }\bibfield  {title} {\enquote {\bibinfo {title} {Exploring
  chemical compound space with quantum-based machine learning},}\ }\href@noop
  {} {\bibfield  {journal} {\bibinfo  {journal} {Nat. Rev. Chem.}\ }\textbf
  {\bibinfo {volume} {4}},\ \bibinfo {pages} {347--358} (\bibinfo {year}
  {2020})}\BibitemShut {NoStop}%
\bibitem [{\citenamefont {Musil}\ \emph {et~al.}(2021)\citenamefont {Musil},
  \citenamefont {Grisafi}, \citenamefont {Bartók}, \citenamefont {Ortner},
  \citenamefont {Csányi},\ and\ \citenamefont {Ceriotti}}]{Musil2021}%
  \BibitemOpen
  \bibfield  {author} {\bibinfo {author} {\bibfnamefont {F.}~\bibnamefont
  {Musil}}, \bibinfo {author} {\bibfnamefont {A.}~\bibnamefont {Grisafi}},
  \bibinfo {author} {\bibfnamefont {A.~P.}\ \bibnamefont {Bartók}}, \bibinfo
  {author} {\bibfnamefont {C.}~\bibnamefont {Ortner}}, \bibinfo {author}
  {\bibfnamefont {G.}~\bibnamefont {Csányi}},\ and\ \bibinfo {author}
  {\bibfnamefont {M.}~\bibnamefont {Ceriotti}},\ }\bibfield  {title} {\enquote
  {\bibinfo {title} {{Physics-Inspired Structural Representations for Molecules
  and Materials}},}\ }\href@noop {} {\bibfield  {journal} {\bibinfo  {journal}
  {Chem. Rev.}\ }\textbf {\bibinfo {volume} {121}},\ \bibinfo {pages}
  {9759--9815} (\bibinfo {year} {2021})}\BibitemShut {NoStop}%
\bibitem [{\citenamefont {Huang}\ and\ \citenamefont {von
  Lilienfeld}(2020)}]{amons}%
  \BibitemOpen
  \bibfield  {author} {\bibinfo {author} {\bibfnamefont {B.}~\bibnamefont
  {Huang}}\ and\ \bibinfo {author} {\bibfnamefont {O.~A.}\ \bibnamefont {von
  Lilienfeld}},\ }\bibfield  {title} {\enquote {\bibinfo {title} {{Quantum
  machine learning using atom-in-molecule-based fragments selected on the
  fly}},}\ }\href@noop {} {\bibfield  {journal} {\bibinfo  {journal} {Nat.
  Chem.}\ }\textbf {\bibinfo {volume} {12}},\ \bibinfo {pages} {945--951}
  (\bibinfo {year} {2020})}\BibitemShut {NoStop}%
\bibitem [{\citenamefont {Behler}(2021)}]{behler_review}%
  \BibitemOpen
  \bibfield  {author} {\bibinfo {author} {\bibfnamefont {J.}~\bibnamefont
  {Behler}},\ }\bibfield  {title} {\enquote {\bibinfo {title} {{Four
  Generations of High-Dimensional Neural Network Potentials}},}\ }\href@noop {}
  {\bibfield  {journal} {\bibinfo  {journal} {Chem. Rev.}\ }\textbf {\bibinfo
  {volume} {121}},\ \bibinfo {pages} {10037--10072} (\bibinfo {year}
  {2021})}\BibitemShut {NoStop}%
\bibitem [{\citenamefont {Unke}\ and\ \citenamefont
  {Meuwly}(2018)}]{10.1063/1.5017898}%
  \BibitemOpen
  \bibfield  {author} {\bibinfo {author} {\bibfnamefont {O.~T.}\ \bibnamefont
  {Unke}}\ and\ \bibinfo {author} {\bibfnamefont {M.}~\bibnamefont {Meuwly}},\
  }\bibfield  {title} {\enquote {\bibinfo {title} {{A reactive, scalable, and
  transferable model for molecular energies from a neural network approach
  based on local information}},}\ }\href@noop {} {\bibfield  {journal}
  {\bibinfo  {journal} {J. Chem. Phys.}\ }\textbf {\bibinfo {volume} {148}},\
  \bibinfo {pages} {241708} (\bibinfo {year} {2018})}\BibitemShut {NoStop}%
\bibitem [{\citenamefont {Sch{\"u}tt}\ \emph {et~al.}(2017)\citenamefont
  {Sch{\"u}tt}, \citenamefont {Arbabzadah}, \citenamefont {Chmiela},
  \citenamefont {M{\"u}ller},\ and\ \citenamefont {Tkatchenko}}]{schnet}%
  \BibitemOpen
  \bibfield  {author} {\bibinfo {author} {\bibfnamefont {K.~T.}\ \bibnamefont
  {Sch{\"u}tt}}, \bibinfo {author} {\bibfnamefont {F.}~\bibnamefont
  {Arbabzadah}}, \bibinfo {author} {\bibfnamefont {S.}~\bibnamefont {Chmiela}},
  \bibinfo {author} {\bibfnamefont {K.~R.}\ \bibnamefont {M{\"u}ller}},\ and\
  \bibinfo {author} {\bibfnamefont {A.}~\bibnamefont {Tkatchenko}},\ }\bibfield
   {title} {\enquote {\bibinfo {title} {Quantum-chemical insights from deep
  tensor neural networks},}\ }\href@noop {} {\bibfield  {journal} {\bibinfo
  {journal} {Nat. Commun.}\ }\textbf {\bibinfo {volume} {8}},\ \bibinfo {pages}
  {13890} (\bibinfo {year} {2017})}\BibitemShut {NoStop}%
\bibitem [{\citenamefont {Politzer}\ and\ \citenamefont
  {Parr}(1974)}]{Politzer1974}%
  \BibitemOpen
  \bibfield  {author} {\bibinfo {author} {\bibfnamefont {P.}~\bibnamefont
  {Politzer}}\ and\ \bibinfo {author} {\bibfnamefont {R.~G.}\ \bibnamefont
  {Parr}},\ }\bibfield  {title} {\enquote {\bibinfo {title} {Some new energy
  formulas for atoms and molecules},}\ }\href@noop {} {\bibfield  {journal}
  {\bibinfo  {journal} {J. Chem. Phys.}\ }\textbf {\bibinfo {volume} {61}},\
  \bibinfo {pages} {4258--4262} (\bibinfo {year} {1974})}\BibitemShut {NoStop}%
\bibitem [{\citenamefont {Politzer}\ and\ \citenamefont
  {Murray}(2002)}]{Politzer2002}%
  \BibitemOpen
  \bibfield  {author} {\bibinfo {author} {\bibfnamefont {P.}~\bibnamefont
  {Politzer}}\ and\ \bibinfo {author} {\bibfnamefont {J.~S.}\ \bibnamefont
  {Murray}},\ }\href@noop {} {\enquote {\bibinfo {title} {The fundamental
  nature and role of the electrostatic potential in atoms and molecules},}\ }
  (\bibinfo {year} {2002})\BibitemShut {NoStop}%
\bibitem [{\citenamefont {Politzer}\ and\ \citenamefont
  {Murray}(2021)}]{Politzer2021}%
  \BibitemOpen
  \bibfield  {author} {\bibinfo {author} {\bibfnamefont {P.}~\bibnamefont
  {Politzer}}\ and\ \bibinfo {author} {\bibfnamefont {J.~S.}\ \bibnamefont
  {Murray}},\ }\bibfield  {title} {\enquote {\bibinfo {title} {Electrostatic
  potentials at the nuclei of atoms and molecules},}\ }\href@noop {} {\bibfield
   {journal} {\bibinfo  {journal} {Theor. Chem. Acc.}\ }\textbf {\bibinfo
  {volume} {140}} (\bibinfo {year} {2021})}\BibitemShut {NoStop}%
\bibitem [{\citenamefont {Wilson}(1962)}]{Wilson1962}%
  \BibitemOpen
  \bibfield  {author} {\bibinfo {author} {\bibfnamefont {E.~B.}\ \bibnamefont
  {Wilson}},\ }\bibfield  {title} {\enquote {\bibinfo {title} {Four-dimensional
  electron density function},}\ }\href@noop {} {\bibfield  {journal} {\bibinfo
  {journal} {J. Chem. Phys.}\ }\textbf {\bibinfo {volume} {36}},\ \bibinfo
  {pages} {2232--2233} (\bibinfo {year} {1962})}\BibitemShut {NoStop}%
\bibitem [{\citenamefont {Kohn}(1947)}]{Kohn}%
  \BibitemOpen
  \bibfield  {author} {\bibinfo {author} {\bibfnamefont {W.}~\bibnamefont
  {Kohn}},\ }\bibfield  {title} {\enquote {\bibinfo {title} {{Two Applications
  of the Variational Method to Quantum Mechanics}},}\ }\href
  {https://link.aps.org/doi/10.1103/PhysRev.71.635} {\bibfield  {journal}
  {\bibinfo  {journal} {Phys. Rev.}\ }\textbf {\bibinfo {volume} {71}},\
  \bibinfo {pages} {635--637} (\bibinfo {year} {1947})}\BibitemShut {NoStop}%
\bibitem [{\citenamefont {Foldy}(1951)}]{foldy}%
  \BibitemOpen
  \bibfield  {author} {\bibinfo {author} {\bibfnamefont {L.~L.}\ \bibnamefont
  {Foldy}},\ }\bibfield  {title} {\enquote {\bibinfo {title} {{A Note on Atomic
  Binding Energies}},}\ }\href
  {https://link.aps.org/doi/10.1103/PhysRev.83.397} {\bibfield  {journal}
  {\bibinfo  {journal} {Phys. Rev.}\ }\textbf {\bibinfo {volume} {83}},\
  \bibinfo {pages} {397--399} (\bibinfo {year} {1951})}\BibitemShut {NoStop}%
\bibitem [{\citenamefont {L\"owdin}(1959)}]{LOWDIN195946}%
  \BibitemOpen
  \bibfield  {author} {\bibinfo {author} {\bibfnamefont {P.-O.}\ \bibnamefont
  {L\"owdin}},\ }\bibfield  {title} {\enquote {\bibinfo {title} {Scaling
  problem, virial theorem, and connected relations in quantum mechanics},}\
  }\href {https://www.sciencedirect.com/science/article/pii/0022285259900062}
  {\bibfield  {journal} {\bibinfo  {journal} {J. Mol. Spectrosc.}\ }\textbf
  {\bibinfo {volume} {3}},\ \bibinfo {pages} {46--66} (\bibinfo {year}
  {1959})}\BibitemShut {NoStop}%
\bibitem [{\citenamefont {Hylleraas}(1930)}]{Hylleraas1930-ww}%
  \BibitemOpen
  \bibfield  {author} {\bibinfo {author} {\bibfnamefont {E.~A.}\ \bibnamefont
  {Hylleraas}},\ }\bibfield  {title} {\enquote {\bibinfo {title} {{\"Uber den
  Grundterm der Zweielektronenprobleme von H$^-$, He, Li$^+$, Be$^{++}$
  usw.}}}\ }\href@noop {} {\bibfield  {journal} {\bibinfo  {journal}
  {Zeitschrift f\"ur Physik}\ }\textbf {\bibinfo {volume} {65}},\ \bibinfo
  {pages} {209--225} (\bibinfo {year} {1930})}\BibitemShut {NoStop}%
\bibitem [{\citenamefont {Marzari}, \citenamefont {de~Gironcoli},\ and\
  \citenamefont {Baroni}(1994)}]{PhysRevLett.72.4001}%
  \BibitemOpen
  \bibfield  {author} {\bibinfo {author} {\bibfnamefont {N.}~\bibnamefont
  {Marzari}}, \bibinfo {author} {\bibfnamefont {S.}~\bibnamefont
  {de~Gironcoli}},\ and\ \bibinfo {author} {\bibfnamefont {S.}~\bibnamefont
  {Baroni}},\ }\bibfield  {title} {\enquote {\bibinfo {title} {{Structure and
  phase stability of
  ${\mathrm{Ga}}_{\mathit{x}}$${\mathrm{In}}_{1\mathrm{\ensuremath{-}}\mathit{x}}$P
  solid solutions from computational alchemy}},}\ }\href
  {https://link.aps.org/doi/10.1103/PhysRevLett.72.4001} {\bibfield  {journal}
  {\bibinfo  {journal} {Phys. Rev. Lett.}\ }\textbf {\bibinfo {volume} {72}},\
  \bibinfo {pages} {4001--4004} (\bibinfo {year} {1994})}\BibitemShut {NoStop}%
\bibitem [{\citenamefont {Lilienfeld}\ and\ \citenamefont
  {Tuckerman}(2006)}]{VonLilienfeld2006}%
  \BibitemOpen
  \bibfield  {author} {\bibinfo {author} {\bibfnamefont {O.~A.~V.}\
  \bibnamefont {Lilienfeld}}\ and\ \bibinfo {author} {\bibfnamefont {M.~E.}\
  \bibnamefont {Tuckerman}},\ }\bibfield  {title} {\enquote {\bibinfo {title}
  {Molecular grand-canonical ensemble density functional theory and exploration
  of chemical space},}\ }\href@noop {} {\bibfield  {journal} {\bibinfo
  {journal} {J. Chem. Phys.}\ }\textbf {\bibinfo {volume} {125}},\ \bibinfo
  {pages} {154104} (\bibinfo {year} {2006})}\BibitemShut {NoStop}%
\bibitem [{\citenamefont {Domenichini}, \citenamefont {von Rudorff},\ and\
  \citenamefont {von Lilienfeld}(2020)}]{giorgio}%
  \BibitemOpen
  \bibfield  {author} {\bibinfo {author} {\bibfnamefont {G.}~\bibnamefont
  {Domenichini}}, \bibinfo {author} {\bibfnamefont {G.~F.}\ \bibnamefont {von
  Rudorff}},\ and\ \bibinfo {author} {\bibfnamefont {O.~A.}\ \bibnamefont {von
  Lilienfeld}},\ }\bibfield  {title} {\enquote {\bibinfo {title} {{Effects of
  perturbation order and basis set on alchemical predictions}},}\ }\href@noop
  {} {\bibfield  {journal} {\bibinfo  {journal} {J. Chem. Phys.}\ }\textbf
  {\bibinfo {volume} {153}},\ \bibinfo {pages} {144118} (\bibinfo {year}
  {2020})}\BibitemShut {NoStop}%
\bibitem [{\citenamefont {von Rudorff}\ and\ \citenamefont {von
  Lilienfeld}(2019)}]{rudorff_atomic_energies}%
  \BibitemOpen
  \bibfield  {author} {\bibinfo {author} {\bibfnamefont {G.~F.}\ \bibnamefont
  {von Rudorff}}\ and\ \bibinfo {author} {\bibfnamefont {O.~A.}\ \bibnamefont
  {von Lilienfeld}},\ }\bibfield  {title} {\enquote {\bibinfo {title} {{Atoms
  in Molecules from Alchemical Perturbation Density Functional Theory}},}\
  }\href@noop {} {\bibfield  {journal} {\bibinfo  {journal} {J. Phys. Chem. B}\
  }\textbf {\bibinfo {volume} {123}},\ \bibinfo {pages} {10073--10082}
  (\bibinfo {year} {2019})}\BibitemShut {NoStop}%
\bibitem [{\citenamefont {Guevara-Vela}\ \emph {et~al.}(2020)\citenamefont
  {Guevara-Vela}, \citenamefont {Francisco}, \citenamefont {Rocha-Rinza},\ and\
  \citenamefont {Pend\'as}}]{IQA}%
  \BibitemOpen
  \bibfield  {author} {\bibinfo {author} {\bibfnamefont {J.~M.}\ \bibnamefont
  {Guevara-Vela}}, \bibinfo {author} {\bibfnamefont {E.}~\bibnamefont
  {Francisco}}, \bibinfo {author} {\bibfnamefont {T.}~\bibnamefont
  {Rocha-Rinza}},\ and\ \bibinfo {author} {\bibfnamefont {A.~M.}\ \bibnamefont
  {Pend\'as}},\ }\bibfield  {title} {\enquote {\bibinfo {title} {{Interacting
  Quantum Atoms - A Review}},}\ }\href@noop {} {\bibfield  {journal} {\bibinfo
  {journal} {Molecules}\ }\textbf {\bibinfo {volume} {25}},\ \bibinfo {pages}
  {1--35} (\bibinfo {year} {2020})}\BibitemShut {NoStop}%
\bibitem [{\citenamefont {Francisco}, \citenamefont {Martín~Pendás},\ and\
  \citenamefont {Blanco}(2006)}]{IQA_method}%
  \BibitemOpen
  \bibfield  {author} {\bibinfo {author} {\bibfnamefont {E.}~\bibnamefont
  {Francisco}}, \bibinfo {author} {\bibfnamefont {A.}~\bibnamefont
  {Martín~Pendás}},\ and\ \bibinfo {author} {\bibfnamefont {M.~A.}\
  \bibnamefont {Blanco}},\ }\bibfield  {title} {\enquote {\bibinfo {title} {{A
  Molecular Energy Decomposition Scheme for Atoms in Molecules}},}\ }\href@noop
  {} {\bibfield  {journal} {\bibinfo  {journal} {J. Chem. Theory Comput.}\
  }\textbf {\bibinfo {volume} {2}},\ \bibinfo {pages} {90--102} (\bibinfo
  {year} {2006})}\BibitemShut {NoStop}%
\bibitem [{\citenamefont {Bader}(1991)}]{bader}%
  \BibitemOpen
  \bibfield  {author} {\bibinfo {author} {\bibfnamefont {R.~F.~W.}\
  \bibnamefont {Bader}},\ }\bibfield  {title} {\enquote {\bibinfo {title} {A
  quantum theory of molecular structure and its applications},}\ }\href@noop {}
  {\bibfield  {journal} {\bibinfo  {journal} {Chem. Rev.}\ }\textbf {\bibinfo
  {volume} {91}},\ \bibinfo {pages} {893--928} (\bibinfo {year}
  {1991})}\BibitemShut {NoStop}%
\bibitem [{\citenamefont {Eriksen}(2020)}]{Eriksen2020}%
  \BibitemOpen
  \bibfield  {author} {\bibinfo {author} {\bibfnamefont {J.~J.}\ \bibnamefont
  {Eriksen}},\ }\bibfield  {title} {\enquote {\bibinfo {title} {Mean-field
  density matrix decompositions},}\ }\href {https://doi.org/10.1063/5.0030764}
  {\bibfield  {journal} {\bibinfo  {journal} {J. Chem. Phys.}\ }\textbf
  {\bibinfo {volume} {153}} (\bibinfo {year} {2020})}\BibitemShut {NoStop}%
\bibitem [{\citenamefont {Martín~Pendás}, \citenamefont {Francisco},\ and\
  \citenamefont {Blanco}(2006)}]{binding_IQA}%
  \BibitemOpen
  \bibfield  {author} {\bibinfo {author} {\bibfnamefont {A.}~\bibnamefont
  {Martín~Pendás}}, \bibinfo {author} {\bibfnamefont {E.}~\bibnamefont
  {Francisco}},\ and\ \bibinfo {author} {\bibfnamefont {M.~A.}\ \bibnamefont
  {Blanco}},\ }\bibfield  {title} {\enquote {\bibinfo {title} {{Binding
  Energies of First Row Diatomics in the Light of the Interacting Quantum Atoms
  Approach}},}\ }\href@noop {} {\bibfield  {journal} {\bibinfo  {journal} {J.
  Phys. Chem. A}\ }\textbf {\bibinfo {volume} {110}},\ \bibinfo {pages}
  {12864--12869} (\bibinfo {year} {2006})}\BibitemShut {NoStop}%
\bibitem [{\citenamefont {Darley}\ and\ \citenamefont
  {Popelier}(2008)}]{Darley2008}%
  \BibitemOpen
  \bibfield  {author} {\bibinfo {author} {\bibfnamefont {M.~G.}\ \bibnamefont
  {Darley}}\ and\ \bibinfo {author} {\bibfnamefont {P.~L.}\ \bibnamefont
  {Popelier}},\ }\bibfield  {title} {\enquote {\bibinfo {title} {Role of
  short-range electrostatics in torsional potentials},}\ }\href@noop {}
  {\bibfield  {journal} {\bibinfo  {journal} {J. Phys. Chem. A}\ }\textbf
  {\bibinfo {volume} {112}},\ \bibinfo {pages} {12954--12965} (\bibinfo {year}
  {2008})}\BibitemShut {NoStop}%
\bibitem [{\citenamefont {Popelier}(2015)}]{Popelier2015}%
  \BibitemOpen
  \bibfield  {author} {\bibinfo {author} {\bibfnamefont {P.~L.}\ \bibnamefont
  {Popelier}},\ }\bibfield  {title} {\enquote {\bibinfo {title} {{QCTFF: On the
  construction of a novel protein force field}},}\ }\href@noop {} {\bibfield
  {journal} {\bibinfo  {journal} {Int. J. Quantum Chem}\ }\textbf {\bibinfo
  {volume} {115}},\ \bibinfo {pages} {1005--1011} (\bibinfo {year}
  {2015})}\BibitemShut {NoStop}%
\bibitem [{\citenamefont {Burn}\ and\ \citenamefont {Popelier}(2020)}]{fflux}%
  \BibitemOpen
  \bibfield  {author} {\bibinfo {author} {\bibfnamefont {M.~J.}\ \bibnamefont
  {Burn}}\ and\ \bibinfo {author} {\bibfnamefont {P.~L.}\ \bibnamefont
  {Popelier}},\ }\bibfield  {title} {\enquote {\bibinfo {title} {{Creating
  Gaussian process regression models for molecular simulations using adaptive
  sampling}},}\ }\href@noop {} {\bibfield  {journal} {\bibinfo  {journal} {J.
  Chem. Phys.}\ }\textbf {\bibinfo {volume} {153}} (\bibinfo {year}
  {2020})}\BibitemShut {NoStop}%
\bibitem [{\citenamefont {Eriksen}(2022)}]{Eriksen2022}%
  \BibitemOpen
  \bibfield  {author} {\bibinfo {author} {\bibfnamefont {J.~J.}\ \bibnamefont
  {Eriksen}},\ }\bibfield  {title} {\enquote {\bibinfo {title} {Electronic
  excitations through the prism of mean-field decomposition techniques},}\
  }\href {https://doi.org/10.1063/5.0082938} {\bibfield  {journal} {\bibinfo
  {journal} {J. Chem. Phys.}\ }\textbf {\bibinfo {volume} {156}} (\bibinfo
  {year} {2022})}\BibitemShut {NoStop}%
\bibitem [{\citenamefont {Kjeldal}\ and\ \citenamefont
  {Eriksen}(2023)}]{Kjeldal2023}%
  \BibitemOpen
  \bibfield  {author} {\bibinfo {author} {\bibfnamefont {F.~{\O}.}\
  \bibnamefont {Kjeldal}}\ and\ \bibinfo {author} {\bibfnamefont {J.~J.}\
  \bibnamefont {Eriksen}},\ }\bibfield  {title} {\enquote {\bibinfo {title}
  {{Decomposing Chemical Space: Applications to the Machine Learning of Atomic
  Energies}},}\ }\href@noop {} {\bibfield  {journal} {\bibinfo  {journal} {J.
  Chem. Theory Comput.}\ } (\bibinfo {year} {2023})}\BibitemShut {NoStop}%
\bibitem [{\citenamefont {Sahre}, \citenamefont {von Rudorff},\ and\
  \citenamefont {von Lilienfeld}(2023)}]{qaa_bde}%
  \BibitemOpen
  \bibfield  {author} {\bibinfo {author} {\bibfnamefont {M.~J.}\ \bibnamefont
  {Sahre}}, \bibinfo {author} {\bibfnamefont {G.~F.}\ \bibnamefont {von
  Rudorff}},\ and\ \bibinfo {author} {\bibfnamefont {O.~A.}\ \bibnamefont {von
  Lilienfeld}},\ }\bibfield  {title} {\enquote {\bibinfo {title} {{Quantum
  Alchemy Based Bonding Trends and Their Link to Hammett’s Equation and
  Pauling’s Electronegativity Model}},}\ }\href@noop {} {\bibfield  {journal}
  {\bibinfo  {journal} {JACS}\ }\textbf {\bibinfo {volume} {145}},\ \bibinfo
  {pages} {5899--5908} (\bibinfo {year} {2023})}\BibitemShut {NoStop}%
\bibitem [{\citenamefont {Hartwigsen}, \citenamefont {Goedecker},\ and\
  \citenamefont {Hutter}(1998{\natexlab{a}})}]{Hartwigsen1998}%
  \BibitemOpen
  \bibfield  {author} {\bibinfo {author} {\bibfnamefont {C.}~\bibnamefont
  {Hartwigsen}}, \bibinfo {author} {\bibfnamefont {S.}~\bibnamefont
  {Goedecker}},\ and\ \bibinfo {author} {\bibfnamefont {J.}~\bibnamefont
  {Hutter}},\ }\bibfield  {title} {\enquote {\bibinfo {title} {{Relativistic
  separable dual-space Gaussian pseudopotentials from H to Rn}},}\ }\href@noop
  {} {\bibfield  {journal} {\bibinfo  {journal} {Phys. Rev. B}\ }\textbf
  {\bibinfo {volume} {58}},\ \bibinfo {pages} {3641--3662} (\bibinfo {year}
  {1998}{\natexlab{a}})}\BibitemShut {NoStop}%
\bibitem [{\citenamefont {Krack}(2005)}]{Krack2005}%
  \BibitemOpen
  \bibfield  {author} {\bibinfo {author} {\bibfnamefont {M.}~\bibnamefont
  {Krack}},\ }\bibfield  {title} {\enquote {\bibinfo {title} {{Pseudopotentials
  for H to Kr optimized for gradient-corrected exchange-correlation
  functionals}},}\ }\href@noop {} {\bibfield  {journal} {\bibinfo  {journal}
  {Theor. Chem. Acc.}\ }\textbf {\bibinfo {volume} {114}},\ \bibinfo {pages}
  {145--152} (\bibinfo {year} {2005})}\BibitemShut {NoStop}%
\bibitem [{\citenamefont {Kumagai}\ and\ \citenamefont
  {Oba}(2014)}]{PhysRevB.89.195205}%
  \BibitemOpen
  \bibfield  {author} {\bibinfo {author} {\bibfnamefont {Y.}~\bibnamefont
  {Kumagai}}\ and\ \bibinfo {author} {\bibfnamefont {F.}~\bibnamefont {Oba}},\
  }\bibfield  {title} {\enquote {\bibinfo {title} {Electrostatics-based
  finite-size corrections for first-principles point defect calculations},}\
  }\href@noop {} {\bibfield  {journal} {\bibinfo  {journal} {Phys. Rev. B}\
  }\textbf {\bibinfo {volume} {89}},\ \bibinfo {pages} {195205} (\bibinfo
  {year} {2014})}\BibitemShut {NoStop}%
\bibitem [{\citenamefont {Lilienfeld}(2009)}]{lili_2009}%
  \BibitemOpen
  \bibfield  {author} {\bibinfo {author} {\bibfnamefont {O.~A.~V.}\
  \bibnamefont {Lilienfeld}},\ }\bibfield  {title} {\enquote {\bibinfo {title}
  {Accurate ab initio energy gradients in chemical compound space},}\
  }\href@noop {} {\bibfield  {journal} {\bibinfo  {journal} {J. Chem. Phys.}\
  }\textbf {\bibinfo {volume} {131}} (\bibinfo {year} {2009})}\BibitemShut
  {NoStop}%
\bibitem [{\citenamefont {von Rudorff}\ and\ \citenamefont {von
  Lilienfeld}(2020)}]{VonRudorff2020}%
  \BibitemOpen
  \bibfield  {author} {\bibinfo {author} {\bibfnamefont {G.~F.}\ \bibnamefont
  {von Rudorff}}\ and\ \bibinfo {author} {\bibfnamefont {O.~A.}\ \bibnamefont
  {von Lilienfeld}},\ }\bibfield  {title} {\enquote {\bibinfo {title}
  {Alchemical perturbation density functional theory},}\ }\href
  {https://doi.org/10.1103/PhysRevResearch.2.023220} {\bibfield  {journal}
  {\bibinfo  {journal} {Phys. Rev. Res.}\ }\textbf {\bibinfo {volume} {2}},\
  \bibinfo {pages} {23220} (\bibinfo {year} {2020})}\BibitemShut {NoStop}%
\bibitem [{\citenamefont {Lilienfeld}, \citenamefont {Lins},\ and\
  \citenamefont {Rothlisberger}(2005)}]{VonLilienfeld2005}%
  \BibitemOpen
  \bibfield  {author} {\bibinfo {author} {\bibfnamefont {O.~A.~V.}\
  \bibnamefont {Lilienfeld}}, \bibinfo {author} {\bibfnamefont {R.~D.}\
  \bibnamefont {Lins}},\ and\ \bibinfo {author} {\bibfnamefont
  {U.}~\bibnamefont {Rothlisberger}},\ }\bibfield  {title} {\enquote {\bibinfo
  {title} {Variational particle number approach for rational compound
  design},}\ }\href@noop {} {\bibfield  {journal} {\bibinfo  {journal} {Phys.
  Rev. Lett.}\ }\textbf {\bibinfo {volume} {95}},\ \bibinfo {pages} {1--4}
  (\bibinfo {year} {2005})}\BibitemShut {NoStop}%
\bibitem [{\citenamefont {Chang}\ and\ \citenamefont
  {Lilienfeld}(2018)}]{Chang2018}%
  \BibitemOpen
  \bibfield  {author} {\bibinfo {author} {\bibfnamefont {K.~Y.}\ \bibnamefont
  {Chang}}\ and\ \bibinfo {author} {\bibfnamefont {O.~A.~V.}\ \bibnamefont
  {Lilienfeld}},\ }\bibfield  {title} {\enquote {\bibinfo {title} {{Al$_x$
  Ga$_{1-x}$As crystals with direct 2 eV band gaps from computational
  alchemy}},}\ }\href@noop {} {\bibfield  {journal} {\bibinfo  {journal} {Phys.
  Rev. Mater.}\ }\textbf {\bibinfo {volume} {2}},\ \bibinfo {pages} {1--9}
  (\bibinfo {year} {2018})}\BibitemShut {NoStop}%
\bibitem [{\citenamefont {{Copyright IBM Corp. 1990-2023 and Copyright
  1994-2001 by Max Planck Institute, Stuttgart.}}()}]{CPMD}%
  \BibitemOpen
  \bibfield  {author} {\bibinfo {author} {\bibnamefont {{Copyright IBM Corp.
  1990-2023 and Copyright 1994-2001 by Max Planck Institute, Stuttgart.}}},\
  }\href@noop {} {\enquote {\bibinfo {title} {{CPMD 4.3}},}\ }\bibinfo {note}
  {{https://github.com/CPMD-code/CPMD. (date of access: 13:42
  30.01.2022)}}\BibitemShut {NoStop}%
\bibitem [{\citenamefont {Ramakrishnan}\ \emph {et~al.}(2014)\citenamefont
  {Ramakrishnan}, \citenamefont {Dral}, \citenamefont {Rupp},\ and\
  \citenamefont {Lilienfeld}}]{qm9}%
  \BibitemOpen
  \bibfield  {author} {\bibinfo {author} {\bibfnamefont {R.}~\bibnamefont
  {Ramakrishnan}}, \bibinfo {author} {\bibfnamefont {P.~O.}\ \bibnamefont
  {Dral}}, \bibinfo {author} {\bibfnamefont {M.}~\bibnamefont {Rupp}},\ and\
  \bibinfo {author} {\bibfnamefont {O.~A.~V.}\ \bibnamefont {Lilienfeld}},\
  }\bibfield  {title} {\enquote {\bibinfo {title} {Quantum chemistry structures
  and properties of 134 kilo molecules},}\ }\href@noop {} {\bibfield  {journal}
  {\bibinfo  {journal} {Sci. Data}\ }\textbf {\bibinfo {volume} {1}},\ \bibinfo
  {pages} {1--7} (\bibinfo {year} {2014})}\BibitemShut {NoStop}%
\bibitem [{\citenamefont {Rupp}\ \emph {et~al.}(2012)\citenamefont {Rupp},
  \citenamefont {Tkatchenko}, \citenamefont {Müller},\ and\ \citenamefont
  {Lilienfeld}}]{Rupp2012}%
  \BibitemOpen
  \bibfield  {author} {\bibinfo {author} {\bibfnamefont {M.}~\bibnamefont
  {Rupp}}, \bibinfo {author} {\bibfnamefont {A.}~\bibnamefont {Tkatchenko}},
  \bibinfo {author} {\bibfnamefont {K.~R.}\ \bibnamefont {Müller}},\ and\
  \bibinfo {author} {\bibfnamefont {O.~A.~V.}\ \bibnamefont {Lilienfeld}},\
  }\bibfield  {title} {\enquote {\bibinfo {title} {Fast and accurate modeling
  of molecular atomization energies with machine learning},}\ }\href@noop {}
  {\bibfield  {journal} {\bibinfo  {journal} {Phys. Rev. Lett.}\ }\textbf
  {\bibinfo {volume} {108}},\ \bibinfo {pages} {1--5} (\bibinfo {year}
  {2012})}\BibitemShut {NoStop}%
\bibitem [{\citenamefont {Hansen}\ \emph {et~al.}(2015)\citenamefont {Hansen},
  \citenamefont {Biegler}, \citenamefont {Ramakrishnan}, \citenamefont
  {Pronobis}, \citenamefont {Lilienfeld}, \citenamefont {Müller},\ and\
  \citenamefont {Tkatchenko}}]{Hansen2015}%
  \BibitemOpen
  \bibfield  {author} {\bibinfo {author} {\bibfnamefont {K.}~\bibnamefont
  {Hansen}}, \bibinfo {author} {\bibfnamefont {F.}~\bibnamefont {Biegler}},
  \bibinfo {author} {\bibfnamefont {R.}~\bibnamefont {Ramakrishnan}}, \bibinfo
  {author} {\bibfnamefont {W.}~\bibnamefont {Pronobis}}, \bibinfo {author}
  {\bibfnamefont {O.~A.~V.}\ \bibnamefont {Lilienfeld}}, \bibinfo {author}
  {\bibfnamefont {K.~R.}\ \bibnamefont {Müller}},\ and\ \bibinfo {author}
  {\bibfnamefont {A.}~\bibnamefont {Tkatchenko}},\ }\bibfield  {title}
  {\enquote {\bibinfo {title} {{Machine learning predictions of molecular
  properties: Accurate many-body potentials and nonlocality in chemical
  space}},}\ }\href@noop {} {\bibfield  {journal} {\bibinfo  {journal} {J.
  Phys. Chem. Lett.}\ }\textbf {\bibinfo {volume} {6}},\ \bibinfo {pages}
  {2326--2331} (\bibinfo {year} {2015})}\BibitemShut {NoStop}%
\bibitem [{\citenamefont {Slater}(1930)}]{slater}%
  \BibitemOpen
  \bibfield  {author} {\bibinfo {author} {\bibfnamefont {J.~C.}\ \bibnamefont
  {Slater}},\ }\bibfield  {title} {\enquote {\bibinfo {title} {{Atomic
  Shielding Constants}},}\ }\href
  {https://link.aps.org/doi/10.1103/PhysRev.36.57} {\bibfield  {journal}
  {\bibinfo  {journal} {Phys. Rev.}\ }\textbf {\bibinfo {volume} {36}},\
  \bibinfo {pages} {57--64} (\bibinfo {year} {1930})}\BibitemShut {NoStop}%
\bibitem [{\citenamefont {Montavon}\ \emph {et~al.}(2013)\citenamefont
  {Montavon}, \citenamefont {Rupp}, \citenamefont {Gobre}, \citenamefont
  {Vazquez-Mayagoitia}, \citenamefont {Hansen}, \citenamefont {Tkatchenko},
  \citenamefont {M\"uller},\ and\ \citenamefont {von
  Lilienfeld}}]{Montavon2013}%
  \BibitemOpen
  \bibfield  {author} {\bibinfo {author} {\bibfnamefont {G.}~\bibnamefont
  {Montavon}}, \bibinfo {author} {\bibfnamefont {M.}~\bibnamefont {Rupp}},
  \bibinfo {author} {\bibfnamefont {V.}~\bibnamefont {Gobre}}, \bibinfo
  {author} {\bibfnamefont {A.}~\bibnamefont {Vazquez-Mayagoitia}}, \bibinfo
  {author} {\bibfnamefont {K.}~\bibnamefont {Hansen}}, \bibinfo {author}
  {\bibfnamefont {A.}~\bibnamefont {Tkatchenko}}, \bibinfo {author}
  {\bibfnamefont {K.-R.}\ \bibnamefont {M\"uller}},\ and\ \bibinfo {author}
  {\bibfnamefont {O.~A.}\ \bibnamefont {von Lilienfeld}},\ }\bibfield  {title}
  {\enquote {\bibinfo {title} {Machine learning of molecular electronic
  properties in chemical compound space},}\ }\href
  {http://stacks.iop.org/1367-2630/15/i=9/a=095003} {\bibfield  {journal}
  {\bibinfo  {journal} {New Journal of Physics}\ }\textbf {\bibinfo {volume}
  {15}},\ \bibinfo {pages} {095003} (\bibinfo {year} {2013})}\BibitemShut
  {NoStop}%
\bibitem [{\citenamefont {Knizia}(2013)}]{IBOIAO}%
  \BibitemOpen
  \bibfield  {author} {\bibinfo {author} {\bibfnamefont {G.}~\bibnamefont
  {Knizia}},\ }\bibfield  {title} {\enquote {\bibinfo {title} {{Intrinsic
  Atomic Orbitals: An Unbiased Bridge between Quantum Theory and Chemical
  Concepts}},}\ }\href@noop {} {\bibfield  {journal} {\bibinfo  {journal} {J.
  Chem. Theory Comput.}\ }\textbf {\bibinfo {volume} {9}},\ \bibinfo {pages}
  {4834--4843} (\bibinfo {year} {2013})}\BibitemShut {NoStop}%
\bibitem [{\citenamefont {Salvador}, \citenamefont {Duran},\ and\ \citenamefont
  {Mayer}(2001)}]{Salvador2001}%
  \BibitemOpen
  \bibfield  {author} {\bibinfo {author} {\bibfnamefont {P.}~\bibnamefont
  {Salvador}}, \bibinfo {author} {\bibfnamefont {M.}~\bibnamefont {Duran}},\
  and\ \bibinfo {author} {\bibfnamefont {I.}~\bibnamefont {Mayer}},\ }\bibfield
   {title} {\enquote {\bibinfo {title} {One- and two-center energy components
  in the atoms in molecules theory},}\ }\href@noop {} {\bibfield  {journal}
  {\bibinfo  {journal} {J. Chem. Phys.}\ }\textbf {\bibinfo {volume} {115}},\
  \bibinfo {pages} {1153--1157} (\bibinfo {year} {2001})}\BibitemShut {NoStop}%
\bibitem [{\citenamefont {Bannwarth}, \citenamefont {Ehlert},\ and\
  \citenamefont {Grimme}(2019)}]{xtb1}%
  \BibitemOpen
  \bibfield  {author} {\bibinfo {author} {\bibfnamefont {C.}~\bibnamefont
  {Bannwarth}}, \bibinfo {author} {\bibfnamefont {S.}~\bibnamefont {Ehlert}},\
  and\ \bibinfo {author} {\bibfnamefont {S.}~\bibnamefont {Grimme}},\
  }\bibfield  {title} {\enquote {\bibinfo {title} {{GFN2-xTB-An Accurate and
  Broadly Parametrized Self-Consistent Tight-Binding Quantum Chemical Method
  with Multipole Electrostatics and Density-Dependent Dispersion
  Contributions}},}\ }\href {https://doi.org/10.1021/acs.jctc.8b01176}
  {\bibfield  {journal} {\bibinfo  {journal} {J. Chem. Theory Comput.}\
  }\textbf {\bibinfo {volume} {15}},\ \bibinfo {pages} {1652--1671} (\bibinfo
  {year} {2019})}\BibitemShut {NoStop}%
\bibitem [{\citenamefont {Bannwarth}\ \emph {et~al.}(2021)\citenamefont
  {Bannwarth}, \citenamefont {Caldeweyher}, \citenamefont {Ehlert},
  \citenamefont {Hansen}, \citenamefont {Pracht}, \citenamefont {Seibert},
  \citenamefont {Spicher},\ and\ \citenamefont {Grimme}}]{xtb2}%
  \BibitemOpen
  \bibfield  {author} {\bibinfo {author} {\bibfnamefont {C.}~\bibnamefont
  {Bannwarth}}, \bibinfo {author} {\bibfnamefont {E.}~\bibnamefont
  {Caldeweyher}}, \bibinfo {author} {\bibfnamefont {S.}~\bibnamefont {Ehlert}},
  \bibinfo {author} {\bibfnamefont {A.}~\bibnamefont {Hansen}}, \bibinfo
  {author} {\bibfnamefont {P.}~\bibnamefont {Pracht}}, \bibinfo {author}
  {\bibfnamefont {J.}~\bibnamefont {Seibert}}, \bibinfo {author} {\bibfnamefont
  {S.}~\bibnamefont {Spicher}},\ and\ \bibinfo {author} {\bibfnamefont
  {S.}~\bibnamefont {Grimme}},\ }\bibfield  {title} {\enquote {\bibinfo {title}
  {{Extended tight-binding quantum chemistry methods}},}\ }\href@noop {}
  {\bibfield  {journal} {\bibinfo  {journal} {WIREs Comput. Mol. Sci.}\
  }\textbf {\bibinfo {volume} {11}},\ \bibinfo {pages} {e1493} (\bibinfo {year}
  {2021})}\BibitemShut {NoStop}%
\bibitem [{\citenamefont {Ramakrishnan}\ \emph {et~al.}(2015)\citenamefont
  {Ramakrishnan}, \citenamefont {Dral}, \citenamefont {Rupp},\ and\
  \citenamefont {Lilienfeld}}]{Ramakrishnan2015}%
  \BibitemOpen
  \bibfield  {author} {\bibinfo {author} {\bibfnamefont {R.}~\bibnamefont
  {Ramakrishnan}}, \bibinfo {author} {\bibfnamefont {P.~O.}\ \bibnamefont
  {Dral}}, \bibinfo {author} {\bibfnamefont {M.}~\bibnamefont {Rupp}},\ and\
  \bibinfo {author} {\bibfnamefont {O.~A.~V.}\ \bibnamefont {Lilienfeld}},\
  }\bibfield  {title} {\enquote {\bibinfo {title} {{Big data meets quantum
  chemistry approximations: The $\Delta$-machine learning approach}},}\
  }\href@noop {} {\bibfield  {journal} {\bibinfo  {journal} {J. Chem. Theory
  Comput.}\ }\textbf {\bibinfo {volume} {11}},\ \bibinfo {pages} {2087--2096}
  (\bibinfo {year} {2015})}\BibitemShut {NoStop}%
\bibitem [{\citenamefont {Bragato}, \citenamefont {{Von Rudorff}},\ and\
  \citenamefont {{Von Lilienfeld}}(2020)}]{Hammett_ea}%
  \BibitemOpen
  \bibfield  {author} {\bibinfo {author} {\bibfnamefont {M.}~\bibnamefont
  {Bragato}}, \bibinfo {author} {\bibfnamefont {G.~F.}\ \bibnamefont {{Von
  Rudorff}}},\ and\ \bibinfo {author} {\bibfnamefont {O.~A.}\ \bibnamefont
  {{Von Lilienfeld}}},\ }\bibfield  {title} {\enquote {\bibinfo {title} {{Data
  enhanced Hammett-equation: Reaction barriers in chemical space}},}\
  }\href@noop {} {\bibfield  {journal} {\bibinfo  {journal} {Chem. Sci.}\
  }\textbf {\bibinfo {volume} {11}},\ \bibinfo {pages} {11859--11868} (\bibinfo
  {year} {2020})}\BibitemShut {NoStop}%
\bibitem [{\citenamefont {Bragato}, \citenamefont {von Rudorff},\ and\
  \citenamefont {von Lilienfeld}()}]{bragato2023occams}%
  \BibitemOpen
  \bibfield  {author} {\bibinfo {author} {\bibfnamefont {M.}~\bibnamefont
  {Bragato}}, \bibinfo {author} {\bibfnamefont {G.~F.}\ \bibnamefont {von
  Rudorff}},\ and\ \bibinfo {author} {\bibfnamefont {O.~A.}\ \bibnamefont {von
  Lilienfeld}},\ }\href@noop {} {\enquote {\bibinfo {title} {{Occam's razor for
  AI: Coarse-graining Hammett Inspired Product Ansatz in Chemical Space, arXiv
  (Chemical Physics). DOI: arXiv:2305.07010. (accessed 2023-11-07)}},}\
  }\BibitemShut {NoStop}%
\bibitem [{\citenamefont {Dan}\ and\ \citenamefont
  {Trinkle}(2022)}]{doi:10.1080/21663831.2022.2051763}%
  \BibitemOpen
  \bibfield  {author} {\bibinfo {author} {\bibfnamefont {Y.}~\bibnamefont
  {Dan}}\ and\ \bibinfo {author} {\bibfnamefont {D.~R.}\ \bibnamefont
  {Trinkle}},\ }\bibfield  {title} {\enquote {\bibinfo {title}
  {First-principles core energies of isolated basal and prism screw
  dislocations in magnesium},}\ }\href
  {https://doi.org/10.1080/21663831.2022.2051763} {\bibfield  {journal}
  {\bibinfo  {journal} {Mater. Res. Lett.}\ }\textbf {\bibinfo {volume} {10}},\
  \bibinfo {pages} {360--368} (\bibinfo {year} {2022})},\ \Eprint
  {https://arxiv.org/abs/https://doi.org/10.1080/21663831.2022.2051763}
  {https://doi.org/10.1080/21663831.2022.2051763} \BibitemShut {NoStop}%
\bibitem [{\citenamefont {Xu}, \citenamefont {Tanaka},\ and\ \citenamefont
  {Kohyama}(2021)}]{Xu2021}%
  \BibitemOpen
  \bibfield  {author} {\bibinfo {author} {\bibfnamefont {Z.}~\bibnamefont
  {Xu}}, \bibinfo {author} {\bibfnamefont {S.}~\bibnamefont {Tanaka}},\ and\
  \bibinfo {author} {\bibfnamefont {M.}~\bibnamefont {Kohyama}},\ }\bibfield
  {title} {\enquote {\bibinfo {title} {{Atomic configurations and energies of
  Mg symmetric tilt grain boundaries: Ab initio local analysis}},}\ }\href@noop
  {} {\bibfield  {journal} {\bibinfo  {journal} {Modell. Simul. Mater. Sci.
  Eng.}\ }\textbf {\bibinfo {volume} {29}} (\bibinfo {year}
  {2021})}\BibitemShut {NoStop}%
\bibitem [{\citenamefont {Oh}\ \emph {et~al.}(2019)\citenamefont {Oh},
  \citenamefont {Kim}, \citenamefont {Odbadrakh}, \citenamefont {Ryu},
  \citenamefont {Yoon}, \citenamefont {Mu}, \citenamefont {K{\"o}rmann},
  \citenamefont {Ikeda}, \citenamefont {Tasan}, \citenamefont {Raabe},
  \citenamefont {Egami},\ and\ \citenamefont {Park}}]{Oh2019-bz}%
  \BibitemOpen
  \bibfield  {author} {\bibinfo {author} {\bibfnamefont {H.~S.}\ \bibnamefont
  {Oh}}, \bibinfo {author} {\bibfnamefont {S.~J.}\ \bibnamefont {Kim}},
  \bibinfo {author} {\bibfnamefont {K.}~\bibnamefont {Odbadrakh}}, \bibinfo
  {author} {\bibfnamefont {W.~H.}\ \bibnamefont {Ryu}}, \bibinfo {author}
  {\bibfnamefont {K.~N.}\ \bibnamefont {Yoon}}, \bibinfo {author}
  {\bibfnamefont {S.}~\bibnamefont {Mu}}, \bibinfo {author} {\bibfnamefont
  {F.}~\bibnamefont {K{\"o}rmann}}, \bibinfo {author} {\bibfnamefont
  {Y.}~\bibnamefont {Ikeda}}, \bibinfo {author} {\bibfnamefont {C.~C.}\
  \bibnamefont {Tasan}}, \bibinfo {author} {\bibfnamefont {D.}~\bibnamefont
  {Raabe}}, \bibinfo {author} {\bibfnamefont {T.}~\bibnamefont {Egami}},\ and\
  \bibinfo {author} {\bibfnamefont {E.~S.}\ \bibnamefont {Park}},\ }\bibfield
  {title} {\enquote {\bibinfo {title} {Engineering atomic-level complexity in
  high-entropy and complex concentrated alloys},}\ }\href@noop {} {\bibfield
  {journal} {\bibinfo  {journal} {Nat. Commun.}\ }\textbf {\bibinfo {volume}
  {10}},\ \bibinfo {pages} {2090} (\bibinfo {year} {2019})}\BibitemShut
  {NoStop}%
\bibitem [{\citenamefont {Li}\ \emph {et~al.}(2019)\citenamefont {Li},
  \citenamefont {Yin}, \citenamefont {Odbadrakh}, \citenamefont {Sales},
  \citenamefont {Zinkle}, \citenamefont {Stocks},\ and\ \citenamefont
  {Wirth}}]{hea_defects}%
  \BibitemOpen
  \bibfield  {author} {\bibinfo {author} {\bibfnamefont {C.}~\bibnamefont
  {Li}}, \bibinfo {author} {\bibfnamefont {J.}~\bibnamefont {Yin}}, \bibinfo
  {author} {\bibfnamefont {K.}~\bibnamefont {Odbadrakh}}, \bibinfo {author}
  {\bibfnamefont {B.~C.}\ \bibnamefont {Sales}}, \bibinfo {author}
  {\bibfnamefont {S.~J.}\ \bibnamefont {Zinkle}}, \bibinfo {author}
  {\bibfnamefont {G.~M.}\ \bibnamefont {Stocks}},\ and\ \bibinfo {author}
  {\bibfnamefont {B.~D.}\ \bibnamefont {Wirth}},\ }\bibfield  {title} {\enquote
  {\bibinfo {title} {{First principle study of magnetism and vacancy energetics
  in a near equimolar NiFeMnCr high entropy alloy}},}\ }\href@noop {}
  {\bibfield  {journal} {\bibinfo  {journal} {J. Appl. Phys.}\ }\textbf
  {\bibinfo {volume} {125}},\ \bibinfo {pages} {155103} (\bibinfo {year}
  {2019})}\BibitemShut {NoStop}%
\bibitem [{\citenamefont {Zheng}\ \emph {et~al.}(2020)\citenamefont {Zheng},
  \citenamefont {Yao}, \citenamefont {Ou}, \citenamefont {Li}, \citenamefont
  {Luo}, \citenamefont {Dou}, \citenamefont {Li}, \citenamefont {Amine},
  \citenamefont {Yu},\ and\ \citenamefont {Chen}}]{D0CS00305K}%
  \BibitemOpen
  \bibfield  {author} {\bibinfo {author} {\bibfnamefont {Y.}~\bibnamefont
  {Zheng}}, \bibinfo {author} {\bibfnamefont {Y.}~\bibnamefont {Yao}}, \bibinfo
  {author} {\bibfnamefont {J.}~\bibnamefont {Ou}}, \bibinfo {author}
  {\bibfnamefont {M.}~\bibnamefont {Li}}, \bibinfo {author} {\bibfnamefont
  {D.}~\bibnamefont {Luo}}, \bibinfo {author} {\bibfnamefont {H.}~\bibnamefont
  {Dou}}, \bibinfo {author} {\bibfnamefont {Z.}~\bibnamefont {Li}}, \bibinfo
  {author} {\bibfnamefont {K.}~\bibnamefont {Amine}}, \bibinfo {author}
  {\bibfnamefont {A.}~\bibnamefont {Yu}},\ and\ \bibinfo {author}
  {\bibfnamefont {Z.}~\bibnamefont {Chen}},\ }\bibfield  {title} {\enquote
  {\bibinfo {title} {A review of composite solid-state electrolytes for lithium
  batteries: fundamentals{,} key materials and advanced structures},}\ }\href
  {https://doi.org/10.1039/D0CS00305K} {\bibfield  {journal} {\bibinfo
  {journal} {Chem. Soc. Rev.}\ }\textbf {\bibinfo {volume} {49}},\ \bibinfo
  {pages} {8790--8839} (\bibinfo {year} {2020})}\BibitemShut {NoStop}%
\bibitem [{\citenamefont {Xia}\ \emph {et~al.}(2019)\citenamefont {Xia},
  \citenamefont {Wu}, \citenamefont {Zhang}, \citenamefont {Cui},\ and\
  \citenamefont {Liu}}]{XIA2019753}%
  \BibitemOpen
  \bibfield  {author} {\bibinfo {author} {\bibfnamefont {S.}~\bibnamefont
  {Xia}}, \bibinfo {author} {\bibfnamefont {X.}~\bibnamefont {Wu}}, \bibinfo
  {author} {\bibfnamefont {Z.}~\bibnamefont {Zhang}}, \bibinfo {author}
  {\bibfnamefont {Y.}~\bibnamefont {Cui}},\ and\ \bibinfo {author}
  {\bibfnamefont {W.}~\bibnamefont {Liu}},\ }\bibfield  {title} {\enquote
  {\bibinfo {title} {{Practical Challenges and Future Perspectives of
  All-Solid-State Lithium-Metal Batteries}},}\ }\href
  {https://www.sciencedirect.com/science/article/pii/S2451929418305308}
  {\bibfield  {journal} {\bibinfo  {journal} {Chem}\ }\textbf {\bibinfo
  {volume} {5}},\ \bibinfo {pages} {753--785} (\bibinfo {year}
  {2019})}\BibitemShut {NoStop}%
\bibitem [{\citenamefont {Lemm}, \citenamefont {von Rudorff},\ and\
  \citenamefont {von Lilienfeld}(2021)}]{leruli}%
  \BibitemOpen
  \bibfield  {author} {\bibinfo {author} {\bibfnamefont {D.}~\bibnamefont
  {Lemm}}, \bibinfo {author} {\bibfnamefont {G.~F.}\ \bibnamefont {von
  Rudorff}},\ and\ \bibinfo {author} {\bibfnamefont {A.}~\bibnamefont {von
  Lilienfeld}},\ }\href@noop {} {\enquote {\bibinfo {title} {{LERULI.com,
  Online molecular property predictions in real time and for free}},}\
  }\bibinfo {howpublished} {\url{www.leruli.com}} (\bibinfo {year} {2021}),\
  \bibinfo {note} {(date of access: 13:46 30.01.2022)}\BibitemShut {NoStop}%
\bibitem [{\citenamefont {Landrum}(2010)}]{rdkit}%
  \BibitemOpen
  \bibfield  {author} {\bibinfo {author} {\bibfnamefont {G.}~\bibnamefont
  {Landrum}},\ }\href {https://www.rdkit.org/} {\enquote {\bibinfo {title}
  {{RDKit}},}\ } (\bibinfo {year} {2010}),\ \bibinfo {note} {(date of access:
  13:45 30.01.2022)}\BibitemShut {NoStop}%
\bibitem [{\citenamefont {Tange}(2011)}]{Tange2011a}%
  \BibitemOpen
  \bibfield  {author} {\bibinfo {author} {\bibfnamefont {O.}~\bibnamefont
  {Tange}},\ }\bibfield  {title} {\enquote {\bibinfo {title} {{GNU Parallel -
  The Command-Line Power Tool}},}\ }\href {http://www.gnu.org/s/parallel}
  {\bibfield  {journal} {\bibinfo  {journal} {;login: The USENIX Magazine}\
  }\textbf {\bibinfo {volume} {36}},\ \bibinfo {pages} {42--47} (\bibinfo
  {year} {2011})}\BibitemShut {NoStop}%
\bibitem [{\citenamefont {Perdew}, \citenamefont {Burke},\ and\ \citenamefont
  {Ernzerhof}(1996)}]{pbe}%
  \BibitemOpen
  \bibfield  {author} {\bibinfo {author} {\bibfnamefont {J.~P.}\ \bibnamefont
  {Perdew}}, \bibinfo {author} {\bibfnamefont {K.}~\bibnamefont {Burke}},\ and\
  \bibinfo {author} {\bibfnamefont {M.}~\bibnamefont {Ernzerhof}},\ }\bibfield
  {title} {\enquote {\bibinfo {title} {{Generalized Gradient Approximation Made
  Simple}},}\ }\href {https://link.aps.org/doi/10.1103/PhysRevLett.77.3865}
  {\bibfield  {journal} {\bibinfo  {journal} {Phys. Rev. Lett.}\ }\textbf
  {\bibinfo {volume} {77}},\ \bibinfo {pages} {3865--3868} (\bibinfo {year}
  {1996})}\BibitemShut {NoStop}%
\bibitem [{\citenamefont {Goedecker}, \citenamefont {Teter},\ and\
  \citenamefont {Hutter}(1996)}]{Goedecker1}%
  \BibitemOpen
  \bibfield  {author} {\bibinfo {author} {\bibfnamefont {S.}~\bibnamefont
  {Goedecker}}, \bibinfo {author} {\bibfnamefont {M.}~\bibnamefont {Teter}},\
  and\ \bibinfo {author} {\bibfnamefont {J.}~\bibnamefont {Hutter}},\
  }\bibfield  {title} {\enquote {\bibinfo {title} {{Separable dual-space
  Gaussian pseudopotentials}},}\ }\href
  {https://link.aps.org/doi/10.1103/PhysRevB.54.1703} {\bibfield  {journal}
  {\bibinfo  {journal} {Phys. Rev. B}\ }\textbf {\bibinfo {volume} {54}},\
  \bibinfo {pages} {1703--1710} (\bibinfo {year} {1996})}\BibitemShut {NoStop}%
\bibitem [{\citenamefont {Hartwigsen}, \citenamefont {Goedecker},\ and\
  \citenamefont {Hutter}(1998{\natexlab{b}})}]{Goedecker2}%
  \BibitemOpen
  \bibfield  {author} {\bibinfo {author} {\bibfnamefont {C.}~\bibnamefont
  {Hartwigsen}}, \bibinfo {author} {\bibfnamefont {S.}~\bibnamefont
  {Goedecker}},\ and\ \bibinfo {author} {\bibfnamefont {J.}~\bibnamefont
  {Hutter}},\ }\bibfield  {title} {\enquote {\bibinfo {title} {{Relativistic
  separable dual-space Gaussian pseudopotentials from H to Rn}},}\ }\href@noop
  {} {\bibfield  {journal} {\bibinfo  {journal} {Phys. Rev. B}\ }\textbf
  {\bibinfo {volume} {58}},\ \bibinfo {pages} {3641--3662} (\bibinfo {year}
  {1998}{\natexlab{b}})}\BibitemShut {NoStop}%
\bibitem [{\citenamefont {{Silvestri, Walter and Seitsonen, Ari P. and Boero,
  Mauro and Kohlmeyer, Axel}}()}]{cpmd2cube}%
  \BibitemOpen
  \bibfield  {author} {\bibinfo {author} {\bibnamefont {{Silvestri, Walter and
  Seitsonen, Ari P. and Boero, Mauro and Kohlmeyer, Axel}}},\ }\href@noop {}
  {\enquote {\bibinfo {title} {{CPMD2CUBE}},}\ }\bibinfo {note}
  {{https://github.com/CPMD-code/Addons. (date of access: 13:39 30.01.2022)
  }}\BibitemShut {NoStop}%
\bibitem [{\citenamefont {Virtanen}\ \emph {et~al.}(2020)\citenamefont
  {Virtanen}, \citenamefont {Gommers}, \citenamefont {Oliphant}, \citenamefont
  {Haberland}, \citenamefont {Reddy}, \citenamefont {Cournapeau}, \citenamefont
  {Burovski}, \citenamefont {Peterson}, \citenamefont {Weckesser},
  \citenamefont {Bright}, \citenamefont {{van der Walt}}, \citenamefont
  {Brett}, \citenamefont {Wilson}, \citenamefont {Millman}, \citenamefont
  {Mayorov}, \citenamefont {Nelson}, \citenamefont {Jones}, \citenamefont
  {Kern}, \citenamefont {Larson}, \citenamefont {Carey}, \citenamefont {Polat},
  \citenamefont {Feng}, \citenamefont {Moore}, \citenamefont {{VanderPlas}},
  \citenamefont {Laxalde}, \citenamefont {Perktold}, \citenamefont {Cimrman},
  \citenamefont {Henriksen}, \citenamefont {Quintero}, \citenamefont {Harris},
  \citenamefont {Archibald}, \citenamefont {Ribeiro}, \citenamefont
  {Pedregosa}, \citenamefont {{van Mulbregt}},\ and\ \citenamefont {{SciPy 1.0
  Contributors}}}]{scipy}%
  \BibitemOpen
  \bibfield  {author} {\bibinfo {author} {\bibfnamefont {P.}~\bibnamefont
  {Virtanen}}, \bibinfo {author} {\bibfnamefont {R.}~\bibnamefont {Gommers}},
  \bibinfo {author} {\bibfnamefont {T.~E.}\ \bibnamefont {Oliphant}}, \bibinfo
  {author} {\bibfnamefont {M.}~\bibnamefont {Haberland}}, \bibinfo {author}
  {\bibfnamefont {T.}~\bibnamefont {Reddy}}, \bibinfo {author} {\bibfnamefont
  {D.}~\bibnamefont {Cournapeau}}, \bibinfo {author} {\bibfnamefont
  {E.}~\bibnamefont {Burovski}}, \bibinfo {author} {\bibfnamefont
  {P.}~\bibnamefont {Peterson}}, \bibinfo {author} {\bibfnamefont
  {W.}~\bibnamefont {Weckesser}}, \bibinfo {author} {\bibfnamefont
  {J.}~\bibnamefont {Bright}}, \bibinfo {author} {\bibfnamefont {S.~J.}\
  \bibnamefont {{van der Walt}}}, \bibinfo {author} {\bibfnamefont
  {M.}~\bibnamefont {Brett}}, \bibinfo {author} {\bibfnamefont
  {J.}~\bibnamefont {Wilson}}, \bibinfo {author} {\bibfnamefont {K.~J.}\
  \bibnamefont {Millman}}, \bibinfo {author} {\bibfnamefont {N.}~\bibnamefont
  {Mayorov}}, \bibinfo {author} {\bibfnamefont {A.~R.~J.}\ \bibnamefont
  {Nelson}}, \bibinfo {author} {\bibfnamefont {E.}~\bibnamefont {Jones}},
  \bibinfo {author} {\bibfnamefont {R.}~\bibnamefont {Kern}}, \bibinfo {author}
  {\bibfnamefont {E.}~\bibnamefont {Larson}}, \bibinfo {author} {\bibfnamefont
  {C.~J.}\ \bibnamefont {Carey}}, \bibinfo {author} {\bibfnamefont
  {{\.I}.}~\bibnamefont {Polat}}, \bibinfo {author} {\bibfnamefont
  {Y.}~\bibnamefont {Feng}}, \bibinfo {author} {\bibfnamefont {E.~W.}\
  \bibnamefont {Moore}}, \bibinfo {author} {\bibfnamefont {J.}~\bibnamefont
  {{VanderPlas}}}, \bibinfo {author} {\bibfnamefont {D.}~\bibnamefont
  {Laxalde}}, \bibinfo {author} {\bibfnamefont {J.}~\bibnamefont {Perktold}},
  \bibinfo {author} {\bibfnamefont {R.}~\bibnamefont {Cimrman}}, \bibinfo
  {author} {\bibfnamefont {I.}~\bibnamefont {Henriksen}}, \bibinfo {author}
  {\bibfnamefont {E.~A.}\ \bibnamefont {Quintero}}, \bibinfo {author}
  {\bibfnamefont {C.~R.}\ \bibnamefont {Harris}}, \bibinfo {author}
  {\bibfnamefont {A.~M.}\ \bibnamefont {Archibald}}, \bibinfo {author}
  {\bibfnamefont {A.~H.}\ \bibnamefont {Ribeiro}}, \bibinfo {author}
  {\bibfnamefont {F.}~\bibnamefont {Pedregosa}}, \bibinfo {author}
  {\bibfnamefont {P.}~\bibnamefont {{van Mulbregt}}},\ and\ \bibinfo {author}
  {\bibnamefont {{SciPy 1.0 Contributors}}},\ }\bibfield  {title} {\enquote
  {\bibinfo {title} {{{SciPy} 1.0: Fundamental Algorithms for Scientific
  Computing in Python}},}\ }\href@noop {} {\bibfield  {journal} {\bibinfo
  {journal} {Nat. Methods}\ }\textbf {\bibinfo {volume} {17}},\ \bibinfo
  {pages} {261--272} (\bibinfo {year} {2020})}\BibitemShut {NoStop}%
\bibitem [{\citenamefont {Harris}\ \emph {et~al.}(2020)\citenamefont {Harris},
  \citenamefont {Millman}, \citenamefont {van~der Walt}, \citenamefont
  {Gommers}, \citenamefont {Virtanen}, \citenamefont {Cournapeau},
  \citenamefont {Wieser}, \citenamefont {Taylor}, \citenamefont {Berg},
  \citenamefont {Smith}, \citenamefont {Kern}, \citenamefont {Picus},
  \citenamefont {Hoyer}, \citenamefont {van Kerkwijk}, \citenamefont {Brett},
  \citenamefont {Haldane}, \citenamefont {del R{\'{i}}o}, \citenamefont
  {Wiebe}, \citenamefont {Peterson}, \citenamefont {G{\'{e}}rard-Marchant},
  \citenamefont {Sheppard}, \citenamefont {Reddy}, \citenamefont {Weckesser},
  \citenamefont {Abbasi}, \citenamefont {Gohlke},\ and\ \citenamefont
  {Oliphant}}]{numpy}%
  \BibitemOpen
  \bibfield  {author} {\bibinfo {author} {\bibfnamefont {C.~R.}\ \bibnamefont
  {Harris}}, \bibinfo {author} {\bibfnamefont {K.~J.}\ \bibnamefont {Millman}},
  \bibinfo {author} {\bibfnamefont {S.~J.}\ \bibnamefont {van~der Walt}},
  \bibinfo {author} {\bibfnamefont {R.}~\bibnamefont {Gommers}}, \bibinfo
  {author} {\bibfnamefont {P.}~\bibnamefont {Virtanen}}, \bibinfo {author}
  {\bibfnamefont {D.}~\bibnamefont {Cournapeau}}, \bibinfo {author}
  {\bibfnamefont {E.}~\bibnamefont {Wieser}}, \bibinfo {author} {\bibfnamefont
  {J.}~\bibnamefont {Taylor}}, \bibinfo {author} {\bibfnamefont
  {S.}~\bibnamefont {Berg}}, \bibinfo {author} {\bibfnamefont {N.~J.}\
  \bibnamefont {Smith}}, \bibinfo {author} {\bibfnamefont {R.}~\bibnamefont
  {Kern}}, \bibinfo {author} {\bibfnamefont {M.}~\bibnamefont {Picus}},
  \bibinfo {author} {\bibfnamefont {S.}~\bibnamefont {Hoyer}}, \bibinfo
  {author} {\bibfnamefont {M.~H.}\ \bibnamefont {van Kerkwijk}}, \bibinfo
  {author} {\bibfnamefont {M.}~\bibnamefont {Brett}}, \bibinfo {author}
  {\bibfnamefont {A.}~\bibnamefont {Haldane}}, \bibinfo {author} {\bibfnamefont
  {J.~F.}\ \bibnamefont {del R{\'{i}}o}}, \bibinfo {author} {\bibfnamefont
  {M.}~\bibnamefont {Wiebe}}, \bibinfo {author} {\bibfnamefont
  {P.}~\bibnamefont {Peterson}}, \bibinfo {author} {\bibfnamefont
  {P.}~\bibnamefont {G{\'{e}}rard-Marchant}}, \bibinfo {author} {\bibfnamefont
  {K.}~\bibnamefont {Sheppard}}, \bibinfo {author} {\bibfnamefont
  {T.}~\bibnamefont {Reddy}}, \bibinfo {author} {\bibfnamefont
  {W.}~\bibnamefont {Weckesser}}, \bibinfo {author} {\bibfnamefont
  {H.}~\bibnamefont {Abbasi}}, \bibinfo {author} {\bibfnamefont
  {C.}~\bibnamefont {Gohlke}},\ and\ \bibinfo {author} {\bibfnamefont {T.~E.}\
  \bibnamefont {Oliphant}},\ }\bibfield  {title} {\enquote {\bibinfo {title}
  {Array programming with {NumPy}},}\ }\href@noop {} {\bibfield  {journal}
  {\bibinfo  {journal} {Nature}\ }\textbf {\bibinfo {volume} {585}},\ \bibinfo
  {pages} {357--362} (\bibinfo {year} {2020})}\BibitemShut {NoStop}%
\bibitem [{\citenamefont {Keith}(2019)}]{AIMAll}%
  \BibitemOpen
  \bibfield  {author} {\bibinfo {author} {\bibfnamefont {T.~A.}\ \bibnamefont
  {Keith}},\ }\href@noop {} {\enquote {\bibinfo {title} {{AIMAll} (version
  19.10.12)},}\ }\bibinfo {howpublished} {\url{aim.tkgristmill.com}} (\bibinfo
  {year} {2019}),\ \bibinfo {note} {{TK Gristmill Software, Overland Park KS,
  USA}}\BibitemShut {NoStop}%
\bibitem [{\citenamefont {Frisch}\ \emph {et~al.}(2009)\citenamefont {Frisch},
  \citenamefont {Trucks}, \citenamefont {Schlegel}, \citenamefont {Scuseria},
  \citenamefont {Robb}, \citenamefont {Cheeseman}, \citenamefont {Scalmani},
  \citenamefont {Barone}, \citenamefont {Mennucci}, \citenamefont {Petersson},
  \citenamefont {Nakatsuji}, \citenamefont {Caricato}, \citenamefont {Li},
  \citenamefont {Hratchian}, \citenamefont {Izmaylov}, \citenamefont {Bloino},
  \citenamefont {Zheng}, \citenamefont {Sonnenberg}, \citenamefont {Hada},
  \citenamefont {Ehara}, \citenamefont {Toyota}, \citenamefont {Fukuda},
  \citenamefont {Hasegawa}, \citenamefont {Ishida}, \citenamefont {Nakajima},
  \citenamefont {Honda}, \citenamefont {Kitao}, \citenamefont {Nakai},
  \citenamefont {Vreven}, \citenamefont {Montgomery}, \citenamefont {Peralta},
  \citenamefont {Ogliaro}, \citenamefont {Bearpark}, \citenamefont {Heyd},
  \citenamefont {Brothers}, \citenamefont {Kudin}, \citenamefont {Staroverov},
  \citenamefont {Kobayashi}, \citenamefont {Normand}, \citenamefont
  {Raghavachari}, \citenamefont {Rendell}, \citenamefont {Burant},
  \citenamefont {Iyengar}, \citenamefont {Tomasi}, \citenamefont {Cossi},
  \citenamefont {Rega}, \citenamefont {Millam}, \citenamefont {Klene},
  \citenamefont {Knox}, \citenamefont {Cross}, \citenamefont {Bakken},
  \citenamefont {Adamo}, \citenamefont {Jaramillo}, \citenamefont {Gomperts},
  \citenamefont {Stratmann}, \citenamefont {Yazyev}, \citenamefont {Austin},
  \citenamefont {Cammi}, \citenamefont {Pomelli}, \citenamefont {Ochterski},
  \citenamefont {Martin}, \citenamefont {Morokuma}, \citenamefont {Zakrzewski},
  \citenamefont {Voth}, \citenamefont {Salvador}, \citenamefont {Dannenberg},
  \citenamefont {Dapprich}, \citenamefont {Daniels}, \citenamefont {Farkas},
  \citenamefont {Foresman}, \citenamefont {Ortiz}, \citenamefont {Cioslowski},\
  and\ \citenamefont {Fox}}]{Gaussian09}%
  \BibitemOpen
  \bibfield  {author} {\bibinfo {author} {\bibfnamefont {M.~J.}\ \bibnamefont
  {Frisch}}, \bibinfo {author} {\bibfnamefont {G.~W.}\ \bibnamefont {Trucks}},
  \bibinfo {author} {\bibfnamefont {H.~B.}\ \bibnamefont {Schlegel}}, \bibinfo
  {author} {\bibfnamefont {G.~E.}\ \bibnamefont {Scuseria}}, \bibinfo {author}
  {\bibfnamefont {M.~A.}\ \bibnamefont {Robb}}, \bibinfo {author}
  {\bibfnamefont {J.~R.}\ \bibnamefont {Cheeseman}}, \bibinfo {author}
  {\bibfnamefont {G.}~\bibnamefont {Scalmani}}, \bibinfo {author}
  {\bibfnamefont {V.}~\bibnamefont {Barone}}, \bibinfo {author} {\bibfnamefont
  {B.}~\bibnamefont {Mennucci}}, \bibinfo {author} {\bibfnamefont {G.~A.}\
  \bibnamefont {Petersson}}, \bibinfo {author} {\bibfnamefont {H.}~\bibnamefont
  {Nakatsuji}}, \bibinfo {author} {\bibfnamefont {M.}~\bibnamefont {Caricato}},
  \bibinfo {author} {\bibfnamefont {X.}~\bibnamefont {Li}}, \bibinfo {author}
  {\bibfnamefont {H.~P.}\ \bibnamefont {Hratchian}}, \bibinfo {author}
  {\bibfnamefont {A.~F.}\ \bibnamefont {Izmaylov}}, \bibinfo {author}
  {\bibfnamefont {J.}~\bibnamefont {Bloino}}, \bibinfo {author} {\bibfnamefont
  {G.}~\bibnamefont {Zheng}}, \bibinfo {author} {\bibfnamefont {J.~L.}\
  \bibnamefont {Sonnenberg}}, \bibinfo {author} {\bibfnamefont
  {M.}~\bibnamefont {Hada}}, \bibinfo {author} {\bibfnamefont {M.}~\bibnamefont
  {Ehara}}, \bibinfo {author} {\bibfnamefont {K.}~\bibnamefont {Toyota}},
  \bibinfo {author} {\bibfnamefont {R.}~\bibnamefont {Fukuda}}, \bibinfo
  {author} {\bibfnamefont {J.}~\bibnamefont {Hasegawa}}, \bibinfo {author}
  {\bibfnamefont {M.}~\bibnamefont {Ishida}}, \bibinfo {author} {\bibfnamefont
  {T.}~\bibnamefont {Nakajima}}, \bibinfo {author} {\bibfnamefont
  {Y.}~\bibnamefont {Honda}}, \bibinfo {author} {\bibfnamefont
  {O.}~\bibnamefont {Kitao}}, \bibinfo {author} {\bibfnamefont
  {H.}~\bibnamefont {Nakai}}, \bibinfo {author} {\bibfnamefont
  {T.}~\bibnamefont {Vreven}}, \bibinfo {author} {\bibfnamefont {J.~A.}\
  \bibnamefont {Montgomery}}, \bibinfo {author} {\bibfnamefont {J.~E.}\
  \bibnamefont {Peralta}}, \bibinfo {author} {\bibfnamefont {F.}~\bibnamefont
  {Ogliaro}}, \bibinfo {author} {\bibfnamefont {M.}~\bibnamefont {Bearpark}},
  \bibinfo {author} {\bibfnamefont {J.~J.}\ \bibnamefont {Heyd}}, \bibinfo
  {author} {\bibfnamefont {E.}~\bibnamefont {Brothers}}, \bibinfo {author}
  {\bibfnamefont {K.~N.}\ \bibnamefont {Kudin}}, \bibinfo {author}
  {\bibfnamefont {V.~N.}\ \bibnamefont {Staroverov}}, \bibinfo {author}
  {\bibfnamefont {R.}~\bibnamefont {Kobayashi}}, \bibinfo {author}
  {\bibfnamefont {J.}~\bibnamefont {Normand}}, \bibinfo {author} {\bibfnamefont
  {K.}~\bibnamefont {Raghavachari}}, \bibinfo {author} {\bibfnamefont
  {A.}~\bibnamefont {Rendell}}, \bibinfo {author} {\bibfnamefont {J.~C.}\
  \bibnamefont {Burant}}, \bibinfo {author} {\bibfnamefont {S.~S.}\
  \bibnamefont {Iyengar}}, \bibinfo {author} {\bibfnamefont {J.}~\bibnamefont
  {Tomasi}}, \bibinfo {author} {\bibfnamefont {M.}~\bibnamefont {Cossi}},
  \bibinfo {author} {\bibfnamefont {N.}~\bibnamefont {Rega}}, \bibinfo {author}
  {\bibfnamefont {J.~M.}\ \bibnamefont {Millam}}, \bibinfo {author}
  {\bibfnamefont {M.}~\bibnamefont {Klene}}, \bibinfo {author} {\bibfnamefont
  {J.~E.}\ \bibnamefont {Knox}}, \bibinfo {author} {\bibfnamefont {J.~B.}\
  \bibnamefont {Cross}}, \bibinfo {author} {\bibfnamefont {V.}~\bibnamefont
  {Bakken}}, \bibinfo {author} {\bibfnamefont {C.}~\bibnamefont {Adamo}},
  \bibinfo {author} {\bibfnamefont {J.}~\bibnamefont {Jaramillo}}, \bibinfo
  {author} {\bibfnamefont {R.}~\bibnamefont {Gomperts}}, \bibinfo {author}
  {\bibfnamefont {R.~E.}\ \bibnamefont {Stratmann}}, \bibinfo {author}
  {\bibfnamefont {O.}~\bibnamefont {Yazyev}}, \bibinfo {author} {\bibfnamefont
  {A.~J.}\ \bibnamefont {Austin}}, \bibinfo {author} {\bibfnamefont
  {R.}~\bibnamefont {Cammi}}, \bibinfo {author} {\bibfnamefont
  {C.}~\bibnamefont {Pomelli}}, \bibinfo {author} {\bibfnamefont {J.~W.}\
  \bibnamefont {Ochterski}}, \bibinfo {author} {\bibfnamefont {R.~L.}\
  \bibnamefont {Martin}}, \bibinfo {author} {\bibfnamefont {K.}~\bibnamefont
  {Morokuma}}, \bibinfo {author} {\bibfnamefont {V.~G.}\ \bibnamefont
  {Zakrzewski}}, \bibinfo {author} {\bibfnamefont {G.~A.}\ \bibnamefont
  {Voth}}, \bibinfo {author} {\bibfnamefont {P.}~\bibnamefont {Salvador}},
  \bibinfo {author} {\bibfnamefont {J.~J.}\ \bibnamefont {Dannenberg}},
  \bibinfo {author} {\bibfnamefont {S.}~\bibnamefont {Dapprich}}, \bibinfo
  {author} {\bibfnamefont {A.~D.}\ \bibnamefont {Daniels}}, \bibinfo {author}
  {\bibfnamefont {{\"{O}}.}~\bibnamefont {Farkas}}, \bibinfo {author}
  {\bibfnamefont {J.~B.}\ \bibnamefont {Foresman}}, \bibinfo {author}
  {\bibfnamefont {J.~V.}\ \bibnamefont {Ortiz}}, \bibinfo {author}
  {\bibfnamefont {J.}~\bibnamefont {Cioslowski}},\ and\ \bibinfo {author}
  {\bibfnamefont {D.~J.}\ \bibnamefont {Fox}},\ }\href@noop {} {\enquote
  {\bibinfo {title} {{Gaussian 09 Revision D.01}},}\ } (\bibinfo {year}
  {2009})\BibitemShut {NoStop}%
\bibitem [{\citenamefont {Sun}\ \emph {et~al.}(2020)\citenamefont {Sun},
  \citenamefont {Zhang}, \citenamefont {Banerjee}, \citenamefont {Bao},
  \citenamefont {Barbry}, \citenamefont {Blunt}, \citenamefont {Bogdanov},
  \citenamefont {Booth}, \citenamefont {Chen}, \citenamefont {Cui},
  \citenamefont {Eriksen}, \citenamefont {Gao}, \citenamefont {Guo},
  \citenamefont {Hermann}, \citenamefont {Hermes}, \citenamefont {Koh},
  \citenamefont {Koval}, \citenamefont {Lehtola}, \citenamefont {Li},
  \citenamefont {Liu}, \citenamefont {Mardirossian}, \citenamefont {McClain},
  \citenamefont {Motta}, \citenamefont {Mussard}, \citenamefont {Pham},
  \citenamefont {Pulkin}, \citenamefont {Purwanto}, \citenamefont {Robinson},
  \citenamefont {Ronca}, \citenamefont {Sayfutyarova}, \citenamefont
  {Scheurer}, \citenamefont {Schurkus}, \citenamefont {Smith}, \citenamefont
  {Sun}, \citenamefont {Sun}, \citenamefont {Upadhyay}, \citenamefont {Wagner},
  \citenamefont {Wang}, \citenamefont {White}, \citenamefont {Whitfield},
  \citenamefont {Williamson}, \citenamefont {Wouters}, \citenamefont {Yang},
  \citenamefont {Yu}, \citenamefont {Zhu}, \citenamefont {Berkelbach},
  \citenamefont {Sharma}, \citenamefont {Sokolov},\ and\ \citenamefont
  {Chan}}]{pyscf}%
  \BibitemOpen
  \bibfield  {author} {\bibinfo {author} {\bibfnamefont {Q.}~\bibnamefont
  {Sun}}, \bibinfo {author} {\bibfnamefont {X.}~\bibnamefont {Zhang}}, \bibinfo
  {author} {\bibfnamefont {S.}~\bibnamefont {Banerjee}}, \bibinfo {author}
  {\bibfnamefont {P.}~\bibnamefont {Bao}}, \bibinfo {author} {\bibfnamefont
  {M.}~\bibnamefont {Barbry}}, \bibinfo {author} {\bibfnamefont {N.~S.}\
  \bibnamefont {Blunt}}, \bibinfo {author} {\bibfnamefont {N.~A.}\ \bibnamefont
  {Bogdanov}}, \bibinfo {author} {\bibfnamefont {G.~H.}\ \bibnamefont {Booth}},
  \bibinfo {author} {\bibfnamefont {J.}~\bibnamefont {Chen}}, \bibinfo {author}
  {\bibfnamefont {Z.-H.}\ \bibnamefont {Cui}}, \bibinfo {author} {\bibfnamefont
  {J.~J.}\ \bibnamefont {Eriksen}}, \bibinfo {author} {\bibfnamefont
  {Y.}~\bibnamefont {Gao}}, \bibinfo {author} {\bibfnamefont {S.}~\bibnamefont
  {Guo}}, \bibinfo {author} {\bibfnamefont {J.}~\bibnamefont {Hermann}},
  \bibinfo {author} {\bibfnamefont {M.~R.}\ \bibnamefont {Hermes}}, \bibinfo
  {author} {\bibfnamefont {K.}~\bibnamefont {Koh}}, \bibinfo {author}
  {\bibfnamefont {P.}~\bibnamefont {Koval}}, \bibinfo {author} {\bibfnamefont
  {S.}~\bibnamefont {Lehtola}}, \bibinfo {author} {\bibfnamefont
  {Z.}~\bibnamefont {Li}}, \bibinfo {author} {\bibfnamefont {J.}~\bibnamefont
  {Liu}}, \bibinfo {author} {\bibfnamefont {N.}~\bibnamefont {Mardirossian}},
  \bibinfo {author} {\bibfnamefont {J.~D.}\ \bibnamefont {McClain}}, \bibinfo
  {author} {\bibfnamefont {M.}~\bibnamefont {Motta}}, \bibinfo {author}
  {\bibfnamefont {B.}~\bibnamefont {Mussard}}, \bibinfo {author} {\bibfnamefont
  {H.~Q.}\ \bibnamefont {Pham}}, \bibinfo {author} {\bibfnamefont
  {A.}~\bibnamefont {Pulkin}}, \bibinfo {author} {\bibfnamefont
  {W.}~\bibnamefont {Purwanto}}, \bibinfo {author} {\bibfnamefont {P.~J.}\
  \bibnamefont {Robinson}}, \bibinfo {author} {\bibfnamefont {E.}~\bibnamefont
  {Ronca}}, \bibinfo {author} {\bibfnamefont {E.~R.}\ \bibnamefont
  {Sayfutyarova}}, \bibinfo {author} {\bibfnamefont {M.}~\bibnamefont
  {Scheurer}}, \bibinfo {author} {\bibfnamefont {H.~F.}\ \bibnamefont
  {Schurkus}}, \bibinfo {author} {\bibfnamefont {J.~E.~T.}\ \bibnamefont
  {Smith}}, \bibinfo {author} {\bibfnamefont {C.}~\bibnamefont {Sun}}, \bibinfo
  {author} {\bibfnamefont {S.-N.}\ \bibnamefont {Sun}}, \bibinfo {author}
  {\bibfnamefont {S.}~\bibnamefont {Upadhyay}}, \bibinfo {author}
  {\bibfnamefont {L.~K.}\ \bibnamefont {Wagner}}, \bibinfo {author}
  {\bibfnamefont {X.}~\bibnamefont {Wang}}, \bibinfo {author} {\bibfnamefont
  {A.}~\bibnamefont {White}}, \bibinfo {author} {\bibfnamefont {J.~D.}\
  \bibnamefont {Whitfield}}, \bibinfo {author} {\bibfnamefont {M.~J.}\
  \bibnamefont {Williamson}}, \bibinfo {author} {\bibfnamefont
  {S.}~\bibnamefont {Wouters}}, \bibinfo {author} {\bibfnamefont
  {J.}~\bibnamefont {Yang}}, \bibinfo {author} {\bibfnamefont {J.~M.}\
  \bibnamefont {Yu}}, \bibinfo {author} {\bibfnamefont {T.}~\bibnamefont
  {Zhu}}, \bibinfo {author} {\bibfnamefont {T.~C.}\ \bibnamefont {Berkelbach}},
  \bibinfo {author} {\bibfnamefont {S.}~\bibnamefont {Sharma}}, \bibinfo
  {author} {\bibfnamefont {A.~Y.}\ \bibnamefont {Sokolov}},\ and\ \bibinfo
  {author} {\bibfnamefont {G.~K.-L.}\ \bibnamefont {Chan}},\ }\bibfield
  {title} {\enquote {\bibinfo {title} {{Recent developments in the PySCF
  program package}},}\ }\href {https://doi.org/10.1063/5.0006074} {\bibfield
  {journal} {\bibinfo  {journal} {J. Chem. Phys.}\ }\textbf {\bibinfo {volume}
  {153}},\ \bibinfo {pages} {024109} (\bibinfo {year} {2020})}\BibitemShut
  {NoStop}%
\end{thebibliography}%

\end{document}


\title{Supplementary Material: Transferability of atomic energies from alchemical decomposition}
 
\author{Michael J. Sahre}

\affiliation{Vienna Doctoral School in Chemistry (DoSChem) and Institute  of  Theoretical  Chemistry and Faculty of Physics, University of Vienna, 1090 Vienna, Austria}

\author{Guido Falk von Rudorff}
\affiliation{University Kassel, Department of Chemistry, Heinrich-Plett-Str.40, 34132 Kassel, Germany}
\affiliation{Center for Interdisciplinary Nanostructure Science and Technology (CINSaT), Heinrich-Plett-Straße 40, 34132 Kassel}

\author{Philipp Marquetand}
\affiliation{Institute  of  Theoretical  Chemistry, Faculty of Chemistry, University of Vienna, W\"ahringer Str. 17, 1090 Vienna, Austria}

\author{O. Anatole von Lilienfeld}
\affiliation{Chemical Physics Theory Group, Department of Chemistry, University of Toronto, St. George Campus, Toronto, ON, Canada}
\affiliation{Department of Materials Science and Engineering, University of Toronto, St. George Campus, Toronto, ON, Canada}
\affiliation{Vector Institute for Artificial Intelligence, Toronto, ON, M5S 1M1, Canada}
\affiliation{ML Group, Technische Universit\"at Berlin and Institute for the Foundations of Learning and Data, 10587 Berlin, Germany}
\affiliation{Berlin Institute for the Foundations of Learning and Data, 10587 Berlin, Germany}
\affiliation{Department of Physics, University of Toronto, St. George Campus, Toronto, ON, Canada}
\affiliation{Acceleration Consortium, University of Toronto, Toronto, ON}

\maketitle

\section{Derivation of Eq.(15)}
According to the Hellmann-Feynman theorem, the derivative of the energy with respect to $\lambda$, $\pdv{E}{\lambda}$, is given by
\begin{equation}\label{eq:hf_theorem}
    \pdv{E}{\lambda} = \pdv{}{\lambda} \expval{H(\lambda)}{\Psi} = \expval{\pdv{H}{\lambda}}{\Psi}.
\end{equation}
For a compound isoelectronic to the uniform electron gas, only the external potential and the nuclear repulsion depend explicitly on $\lambda$ so that
\begin{equation}
    \pdv{H}{\lambda} = \pdv{V_\text{ext}^\text{pbc}}{\lambda} + \pdv{E^\text{nuc}}{\lambda}.
\end{equation}
From the definition of $V_\text{ext}^\text{pbc}$ (see Eqs.~(5)-(10) in the main text) follows by application of the chain rule
\begin{equation}
    \pdv{V_\text{ext}^\text{pbc}}{\lambda} = \sum_I \sum_\vec{T} \left(\sum_K \pdv{V_I^\text{PP}}{V_{IK}} \pdv{V_{IK}}{\lambda}\right) + 
    \pdv{\rho^\text{bg}}{Z_{I,V}}\pdv{Z_{I,V}}{\lambda} V^\text{bg}
\end{equation}
where $V_{IK}$ is the $K$-th $\lambda$-dependent pseudopotential parameter of nucleus $I$. The pseudopotentials are linearly dependent on the parameters $V_{IK}$ (see Eqs.~(7,8) in the main text). Moreover, and as defined in the main text, the parameters $V_{IK}$ are themselves linearly dependent on $\lambda$, e.g. $V_{IK}(\lambda) = \bar{V}_{IK} \lambda$ with $\bar{V}_{IK}$ being the optimized $K$-th pseudopotential parameter for element $I$. Hence, the directional derivative $\pdv{V_I}{\lambda}$ is equal to the respective pseudopotential at $\lambda = 1$
\begin{equation}\label{eq:deltaVI}
    \pdv{V_I}{\lambda} = \sum_K \pdv{V_I^\text{PP}}{V_{IK}} \pdv{V_{IK}}{\lambda} = V_I(\lambda = 1).
\end{equation}

Furthermore from Eq.~(9) in the main text follows,
\begin{equation}
    \pdv{\rho^\text{bg}}{Z_{I,V}}\pdv{Z_{I,V}}{\lambda} = -\frac{\bar{Z}_{I,V}}{\Omega_\text{cell}}
\end{equation}
such that
\begin{equation}\label{eq:dVI}
    \pdv{V_\text{ext}^\text{pbc}}{\lambda} = \sum_I \sum_\vec{T} V_I(\lambda=1,\vec{r}-\vec{R}_I-\vec{T}) -\frac{\bar{Z}_{I,V}}{\Omega_\text{cell}} V^\text{bg} = \sum_I \Delta V_I.
\end{equation}

The derivative of the nuclear repulsion energy (see Eqs.~(11)-(14) in the main text) is given by
\begin{equation}
    \pdv{E^\text{nuc}}{\lambda} = \pdv{E^\text{NN}}{\lambda} + \pdv{E^\text{Nbg}}{\lambda} + \pdv{E^\text{bg}}{\lambda}
\end{equation}
with 
\begin{equation}
    \pdv{E^\text{NN}}{\lambda} = \sum_I \pdv{E^\text{NN}}{Z_{I,V}} \pdv{Z_{I,V}}{\lambda} = \sum_I \sum_{J'} \sum_\vec{T}^{'} \frac{\bar{Z}_{I,V} Z_{J,V}(\lambda)}{|\vec{R}_I-\vec{R}_J-\vec{T}|}
\end{equation}
and
\begin{equation}
    \pdv{E^\text{Nbg}}{\lambda} = \sum_I \pdv{E^\text{Nbg}}{Z_{I,V}}\pdv{Z_{I,V}}{\lambda} = \sum_{I,\vec{T}} \bar{Z}_{I,V} \left( \frac{\sum_J Z_{J,V}(\lambda)}{\Omega_\text{cell}} - \rho^\text{bg} \right) V^\text{bg}(\vec{R}_I-\vec{T})
\end{equation}
and
\begin{equation}
    \pdv{E^\text{bg}}{\lambda} = \sum_I \pdv{E^\text{bg}}{Z_{I,V}}\pdv{Z_{I,V}}{\lambda} = \sum_{I,\vec{T}} \bar{Z}_{I,V} \frac{\rho^\text{bg}}{\Omega_\text{cell}} \int_{\Omega_\text{cell}} d\vec{r} V^\text{bg}(\vec{r}-\vec{T}).
\end{equation}
Consequently, $\pdv{E^\text{nuc}}{\lambda}$ can be decomposed into atomic contributions as
\begin{equation}
    \pdv{E^\text{nuc}}{\lambda} = \sum_I \pdv{E_I^\text{nuc}}{Z_{I,V}} \pdv{Z_{I,V}}{\lambda}
\end{equation}
with
\begin{equation}\label{eq:EInuc}
\begin{split}
    \pdv{E_I^\text{nuc}}{Z_{I,V}}\pdv{Z_{I,V}}{\lambda} &= \sum_\vec{T}^{'} \sum_{J'} \frac{\bar{Z}_{I,V} Z_{J,V}(\lambda)}{|\vec{R}_I-\vec{R}_J-\vec{T}|} \\
    &+ \sum_\vec{T} \bar{Z}_{I,V} \left( \frac{\sum_J Z_{J,V}(\lambda)}{\Omega_\text{cell}} - \rho^\text{bg} \right) V^\text{bg}(\vec{R}_I-\vec{T}) \\
    &+ \sum_\vec{T} \bar{Z}_{I,V} \frac{\rho^\text{bg}}{\Omega_\text{cell}} \int_{\Omega_\text{cell}} d\vec{r} V^\text{bg}(\vec{r}-\vec{T}).
\end{split}
\end{equation}
Hence, $\pdv{E}{\lambda}$ can be expressed as
\begin{equation}
\begin{split}
    \pdv{E}{\lambda} &= \sum_I \expval{\Delta V_I + \pdv{E_I^\text{nuc}}{Z_{I,V}}\pdv{Z_{I,V}}{\lambda}}{\Psi} \\
    &= \sum_I \int d\vec{r} \Delta V_I \rho_\lambda(\vec{r}) + \pdv{E_I^\text{nuc}}{Z_{I,V}}\pdv{Z_{I,V}}{\lambda}
\end{split}
\end{equation}
with the definitions of $\Delta V_I$ and $\pdv{E_I^\text{nuc}}{\lambda}$ in Eq.~\eqref{eq:dVI} and Eq.~\eqref{eq:EInuc}, respectively.
The atomic energy decomposition of the difference between uniform electron gas and compound is finally
\begin{equation}
    E^\text{mol} - E^\text{UEG} = \sum_I  \int_0^1 d\lambda \left( \pdv{E_I^\text{nuc}}{Z_{I,V}}\pdv{Z_{I,V}}{\lambda} + \int d\vec{r} \Delta V_I \rho_\lambda(\vec{r}) \right).
\end{equation}

\section{Perturbative derivatives}
\begin{equation}
\begin{split}
    \langle V_\text{ext}^\pm \rangle &= \sum_\vec{T} \int_{\Omega_\text{cell}} d\vec{r} \left( V_I^\text{PP}(\lambda \pm \Delta \lambda ) + \sum_{J \neq I} V_J^\text{PP}(\lambda) + \frac{N_e-\sum_J Z_{J,V}(\lambda) \mp \Delta \lambda \bar{Z}_{I,V}}{\Omega_\text{cell}} V^\text{bg} \right) \rho_\lambda(\vec{r}) \\
    &= \sum_\vec{T} \int_{\Omega_\text{cell}} d\vec{r} \left( V_I^\text{PP}(\lambda) \pm \Delta \lambda V_I^\text{PP}(\lambda=1) + \sum_{J \neq I} V_J^\text{PP}(\lambda) + \frac{N_e-\sum_J Z_{J,V}(\lambda) \mp \Delta \lambda \bar{Z}_{I,V}}{\Omega_\text{cell}} V^\text{bg} \right) \rho_\lambda(\vec{r})
\end{split}
\end{equation}

\begin{equation}
    \langle \frac{V_\text{ext}^+ \rangle - \langle V_\text{ext}^- \rangle}{2 \Delta  \lambda} = \sum_\vec{T} \int_{\Omega_\text{cell}} d\vec{r} \left( V_I^\text{PP}(\lambda) - \frac{\bar{Z}_{I,V}}{\Omega_\text{cell}} V^\text{bg} \right) \rho_\lambda(\vec{r}) = \int_{\Omega_\text{cell}} d\vec{r} \Delta V_I \rho_\lambda(\vec{r})
\end{equation}

\begin{equation}
\begin{split}
    E_I^{\text{NN} \pm} &= \frac{1}{2} \sum_\vec{T}^{'} \frac{(Z_{I,V}(\lambda)\pm\Delta \bar{Z}_{I,V})^2}{|\vec{R}_{IIT}|} + \frac{1}{2} \sum_{J \neq I} \sum_\vec{T} \frac{(Z_{I,V}(\lambda) \pm \Delta \bar{Z}_{I,V}) Z_{J,V}(\lambda)}{|\vec{R}_{IJT}|} \\ &+ \frac{1}{2} \sum_{K \neq I} \sum_\vec{T} \frac{Z_K(\lambda) (Z_{I,V}(\lambda) \pm \Delta \bar{Z}_{I,V}) }{|\vec{R}_{KIT}|} 
    +  \frac{1}{2} \sum_{K\neq I} \sum_{J' \neq I} \sum_\vec{T}^{'} \frac{Z_K(\lambda) Z_{J,V}(\lambda)}{|\vec{R}_{KJT}|} \\
    &= \frac{1}{2} \sum_\vec{T}^{'} \frac{(Z_{I,V}(\lambda)\pm\Delta \bar{Z}_{I,V})^2}{|\vec{R}_{IIT}|} + \sum_{J \neq I} \sum_\vec{T} \frac{(Z_{I,V}(\lambda) \pm \Delta \bar{Z}_{I,V}) Z_{J,V}(\lambda)}{|\vec{R}_{IJT}|} + \frac{1}{2} \sum_{K\neq I} \sum_{J' \neq I} \sum_\vec{T}^{'} \frac{Z_K(\lambda) Z_{J,V}(\lambda)}{|\vec{R}_{KJT}|}
\end{split}
\end{equation}
with $\vec{R}_{IJT} = \vec{R}_I - \vec{R}_J - \vec{R}_T$.

\begin{equation}
    \frac{E_I^{\text{NN} +} - E_I^{\text{NN} -}}{2 \Delta \lambda} = \bar{Z}_{I,V} \sum_{J'} \sum_\vec{T}^{'} \frac{Z_{J,V}(\lambda)}{|\vec{R}_{IJT}|} = \pdv{E^\text{NN}}{Z_{I,V}}\pdv{Z_{I,V}}{\lambda}
\end{equation}

\begin{equation}
\begin{split}
    E_I^{\text{Nbg} \pm} &= \sum_\vec{T} -V^\text{bg} \left[ 
    (Z_{I,V}(\lambda) \pm \Delta \lambda \bar{Z}_{I,V}) \left(\frac{N_e-\sum_J Z_{J,V}(\lambda) \mp \Delta \lambda \bar{Z}_{I,V}}{\Omega_\text{cell}} \right) \right. \\
    &\left. + \sum_{J \neq I} Z_{J,V}(\lambda) \left(\frac{N_e-\sum_J Z_{J,V}(\lambda) \mp \Delta \lambda \bar{Z}_{I,V}}{\Omega_\text{cell}} \right)
    \right]
\end{split}
\end{equation}

\begin{equation}
\frac{E_I^{\text{Nbg}+} - E_I^{\text{Nbg}-}}{2 \Delta \lambda} = \sum_\vec{T} \bar{Z}_{I,V} \left( \frac{\sum_J Z_{J,V}(\lambda) }{\Omega_\text{cell}} -\rho^\text{bg}\right) V^\text{bg} (\vec{R}_I-\vec{T}) = \pdv{E^\text{Nbg}}{Z_{I,V}}\pdv{Z_{I,V}}{\lambda}
\end{equation}

\begin{equation}
    E^{\text{bg} \pm} = \frac{1}{2} \left(\frac{N_e-\sum_K Z_K(\lambda) \mp \Delta \lambda \bar{Z}_{I,V}}{\Omega_\text{cell}}\right)^2 \sum_\vec{T} - \int_{\Omega_\text{cell}} d\vec{r} V^\text{bg}(\vec{r}-\vec{T})
\end{equation}

\begin{equation}
    \frac{E^\text{bg+}-E^\text{bg-}}{2 \Delta \lambda} = \bar{Z}_{I,V} \frac{\rho^\text{bg}}{\Omega_\text{cell}} \sum_\vec{T} \int_{\Omega_\text{cell}} d\vec{r} V^\text{bg}(\vec{r}-\vec{T}) = \pdv{E^\text{bg}}{Z_{I,V}} \pdv{Z_{I,V}}{\lambda}
\end{equation}

\FloatBarrier

\begin{figure}
\includegraphics[width=0.7\textwidth]{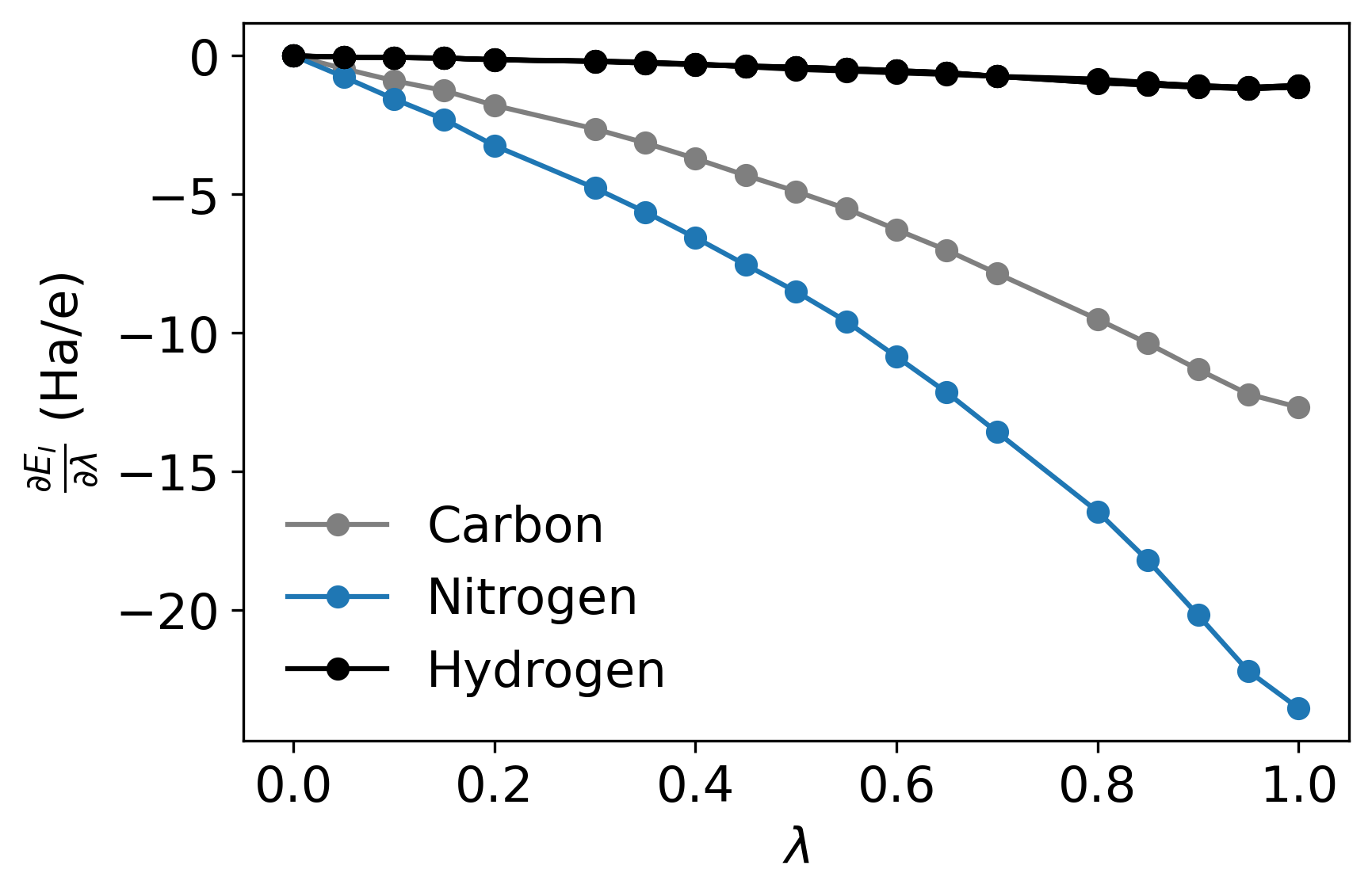}
\caption{The partial derivatives $\pdv{E_I}{\lambda}$ of the energy with respect to a change in the pseuodopotential of a specific atom as a function of $\lambda$ for the atoms in methylamine.}
\label{fig:deri_example}
\end{figure}

\begin{figure}
\includegraphics[width=0.7\textwidth]{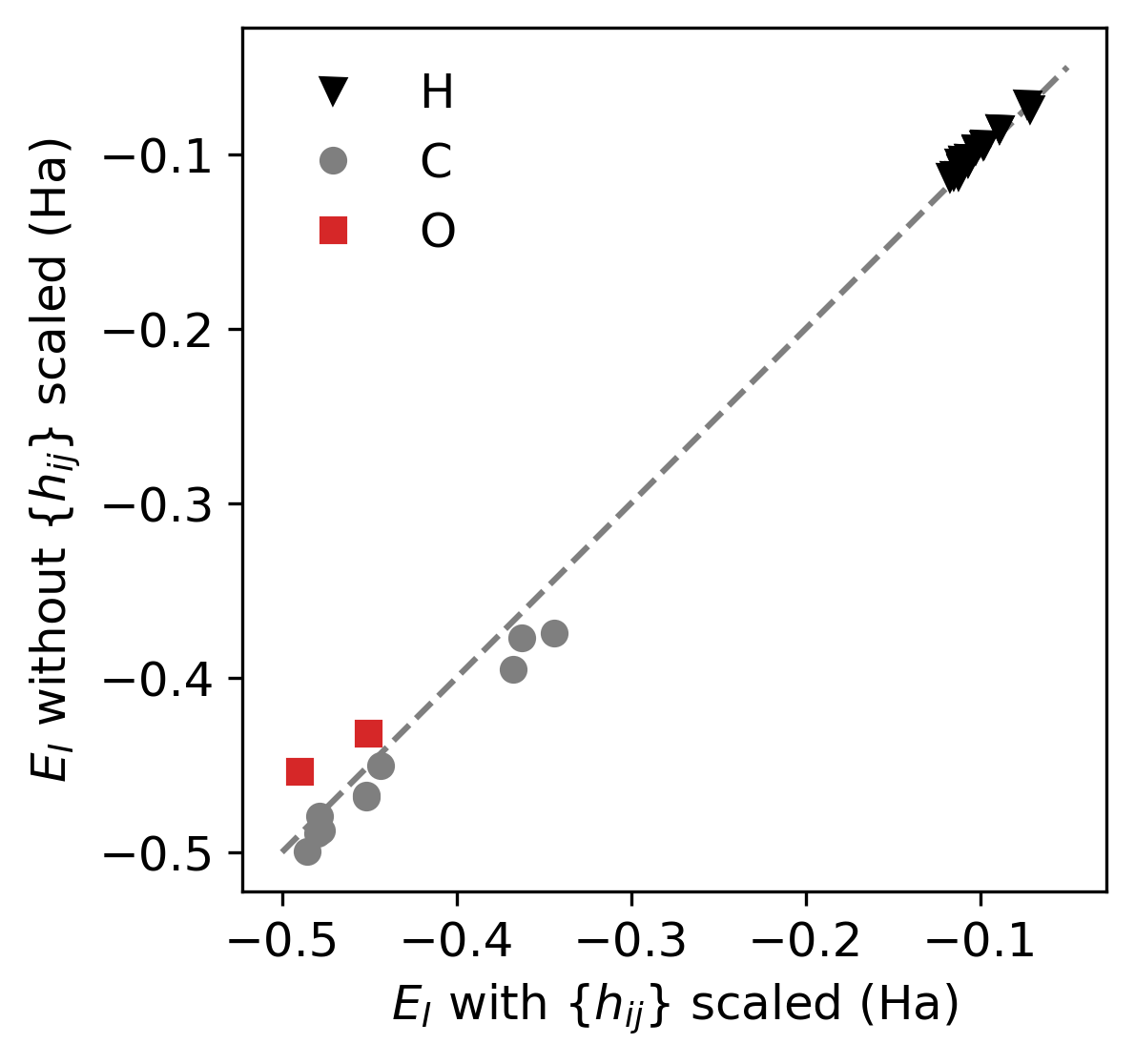}
\caption{Impact of the scaling of the non-local parameters of the pseudopotential $\{h_{ij}\}$ on the atomic energies exemplified for two molecules of QM9 (IDs dsgdb9nsd\_000227 and dsgdb9nsd\_000228).}
\label{fig:v_non_loc}
\end{figure}

\begin{figure}
\includegraphics[width=1.0\textwidth]{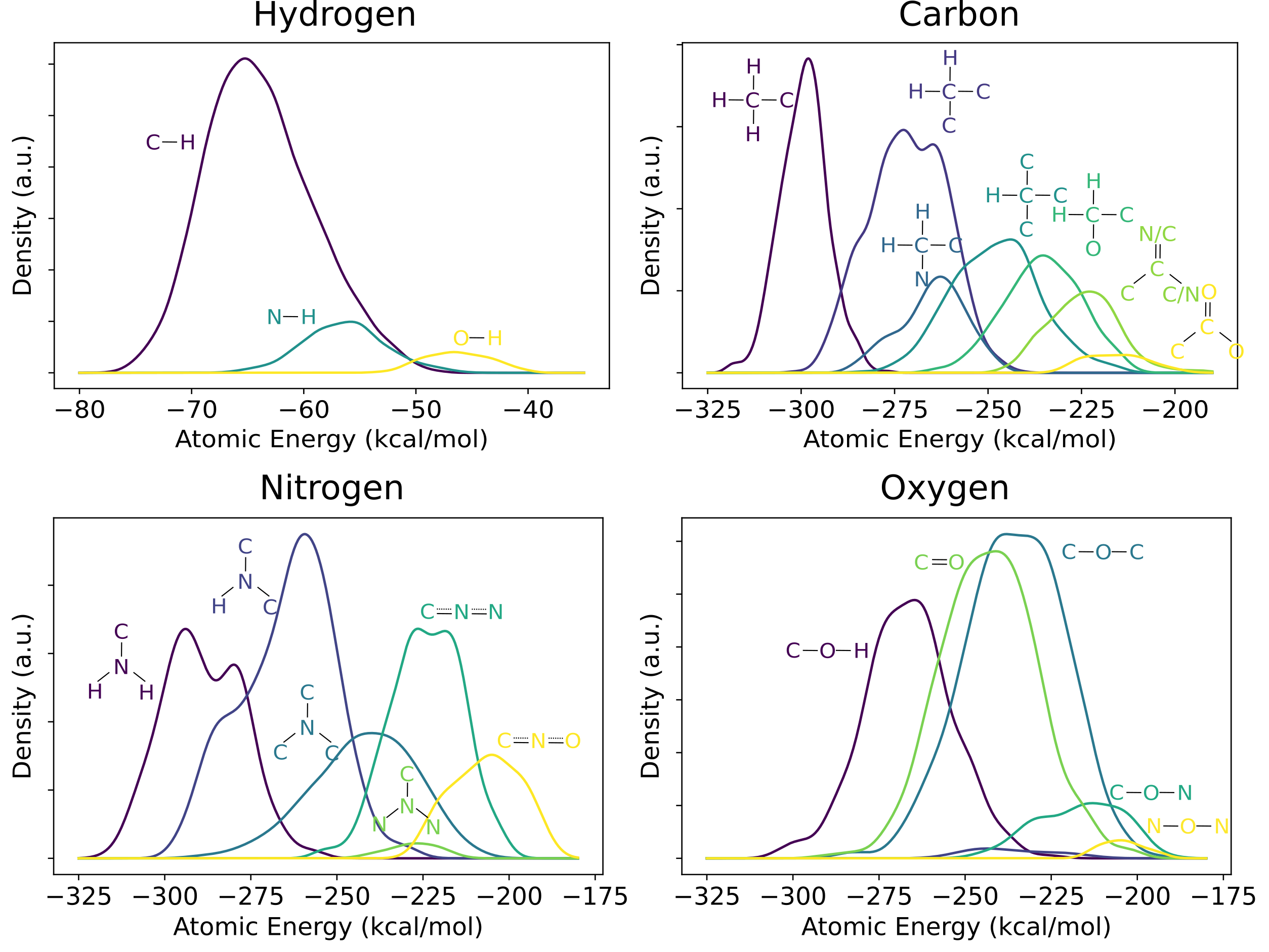}
\caption{Distribution of alchemical atomic energies for selected local environments for H, C, N, O.}
\label{fig:SI_ae_local_envs}
\end{figure}

\begin{figure}
\includegraphics[width=1.0\textwidth]{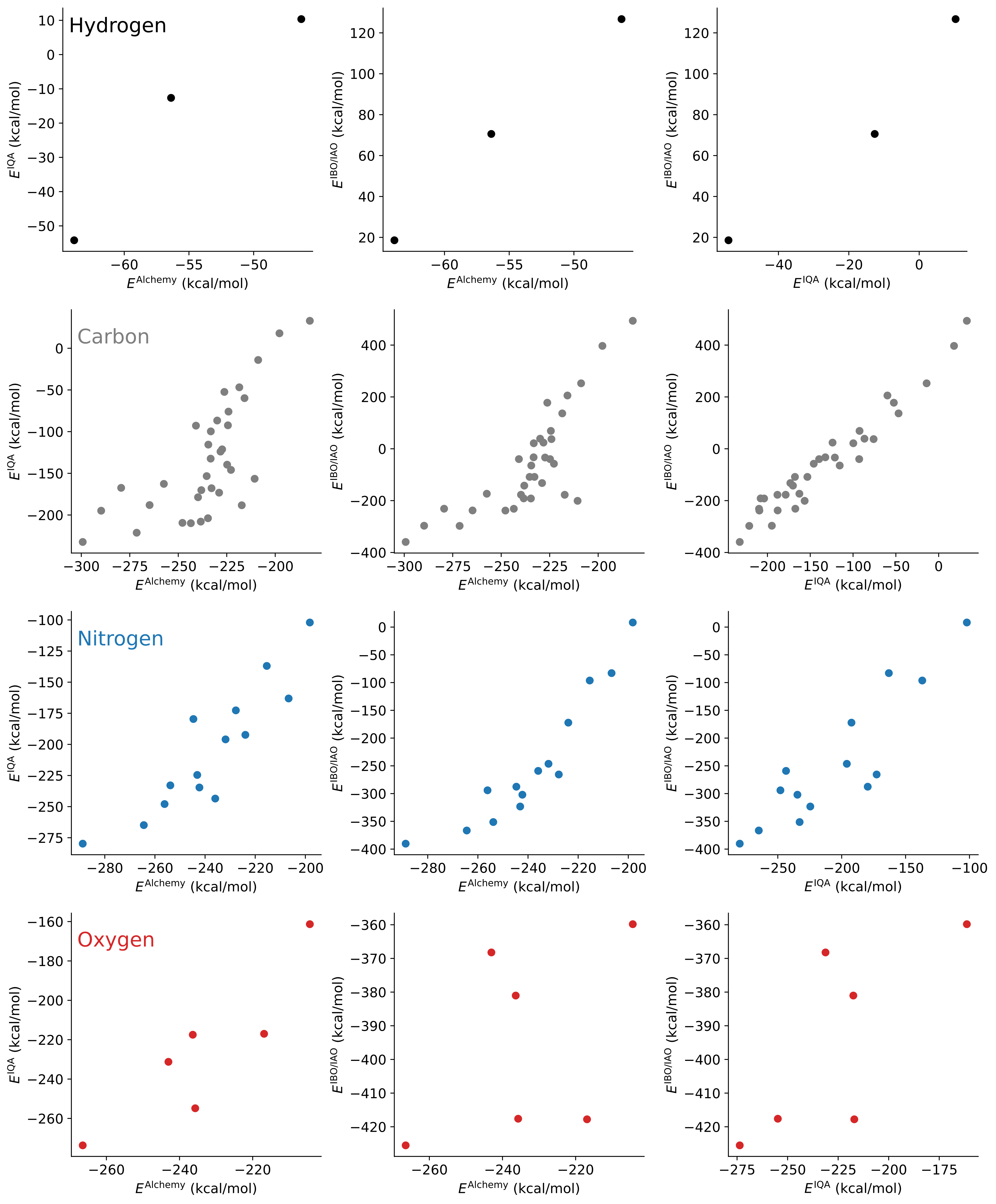}
\caption{The relation between atomic energy and the local environment compared for different decompositon schemes (from left to right alchemy vs IQA, alchemy vs IBO/IAO and IQA vs IBO/IAO) for the elements H, C, N and O.}
\label{fig:SI_compare_decompositions}
\end{figure}

\begin{figure}
\includegraphics[width=0.99\textwidth]{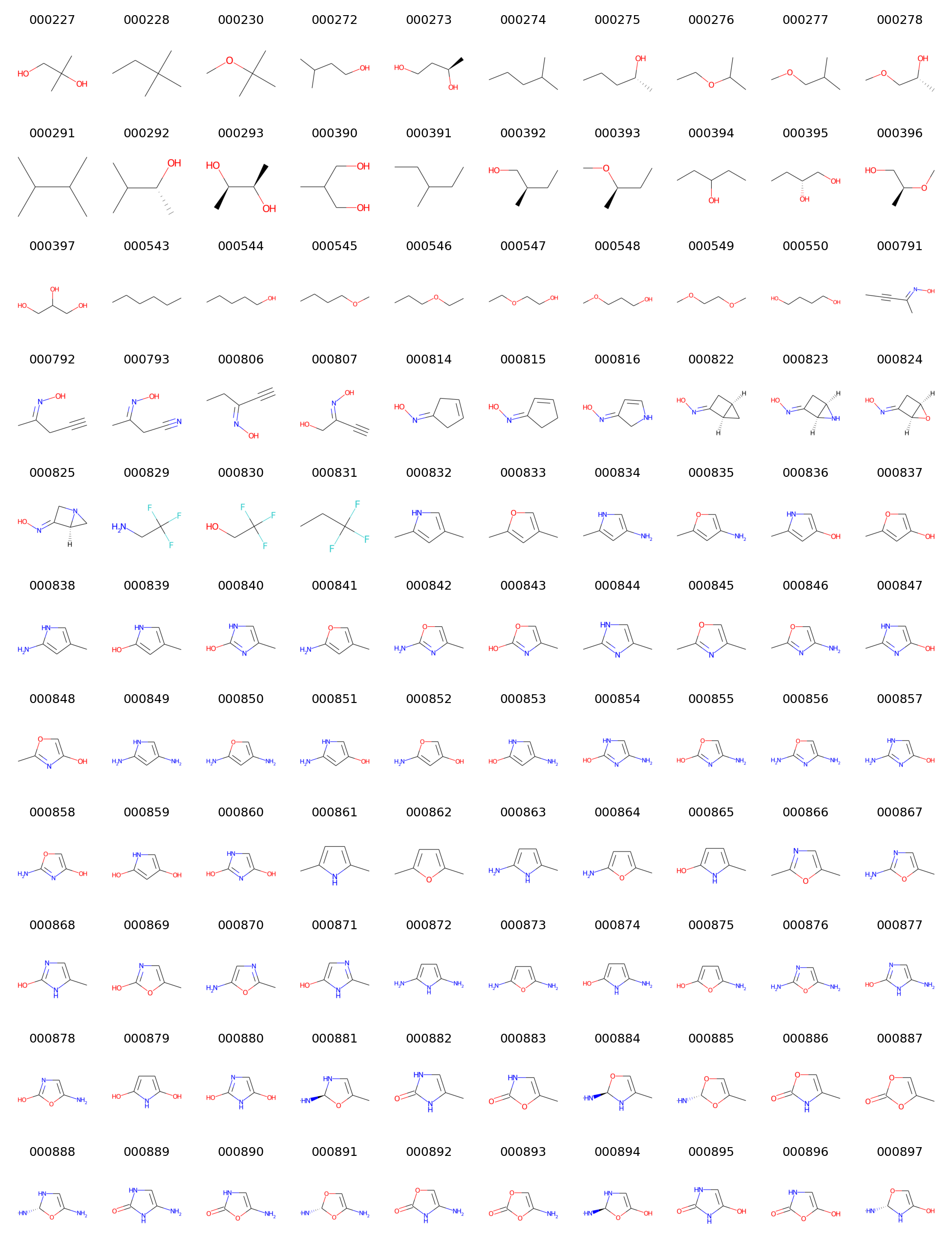}
\caption{The structures of the molecules for which atomic energies where calculated in this study. The numbers correspond to the respective ID in the QM9 dataset.}
\label{fig:ae_dist_diff_methods_modified0}
\end{figure}
\begin{figure}
\includegraphics[width=0.99\textwidth]{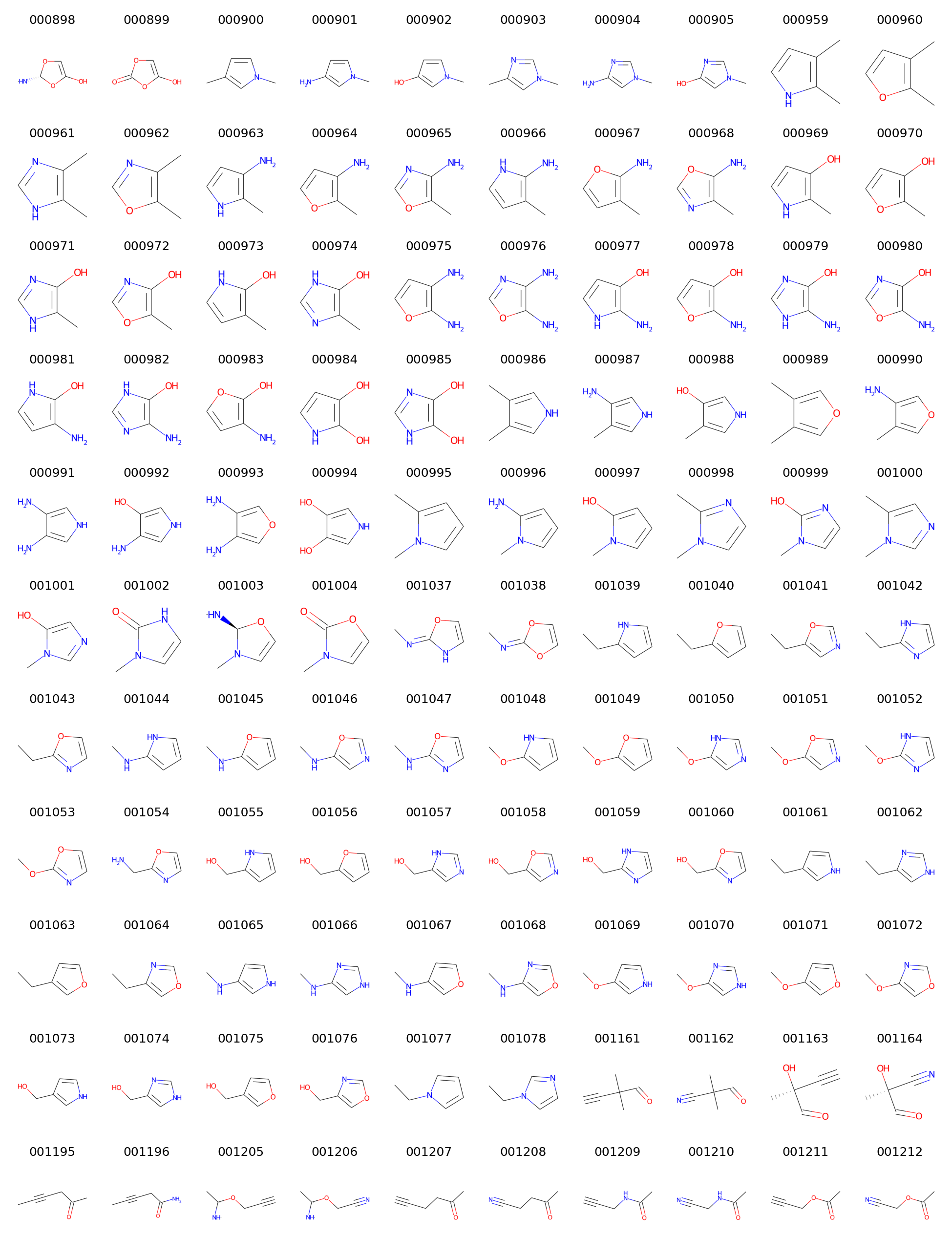}
\caption{The structures of the molecules for which atomic energies where calculated in this study. The numbers correspond to the respective ID in the QM9 dataset.}
\label{fig:ae_dist_diff_methods_modified1}
\end{figure}
\begin{figure}
\includegraphics[width=0.99\textwidth]{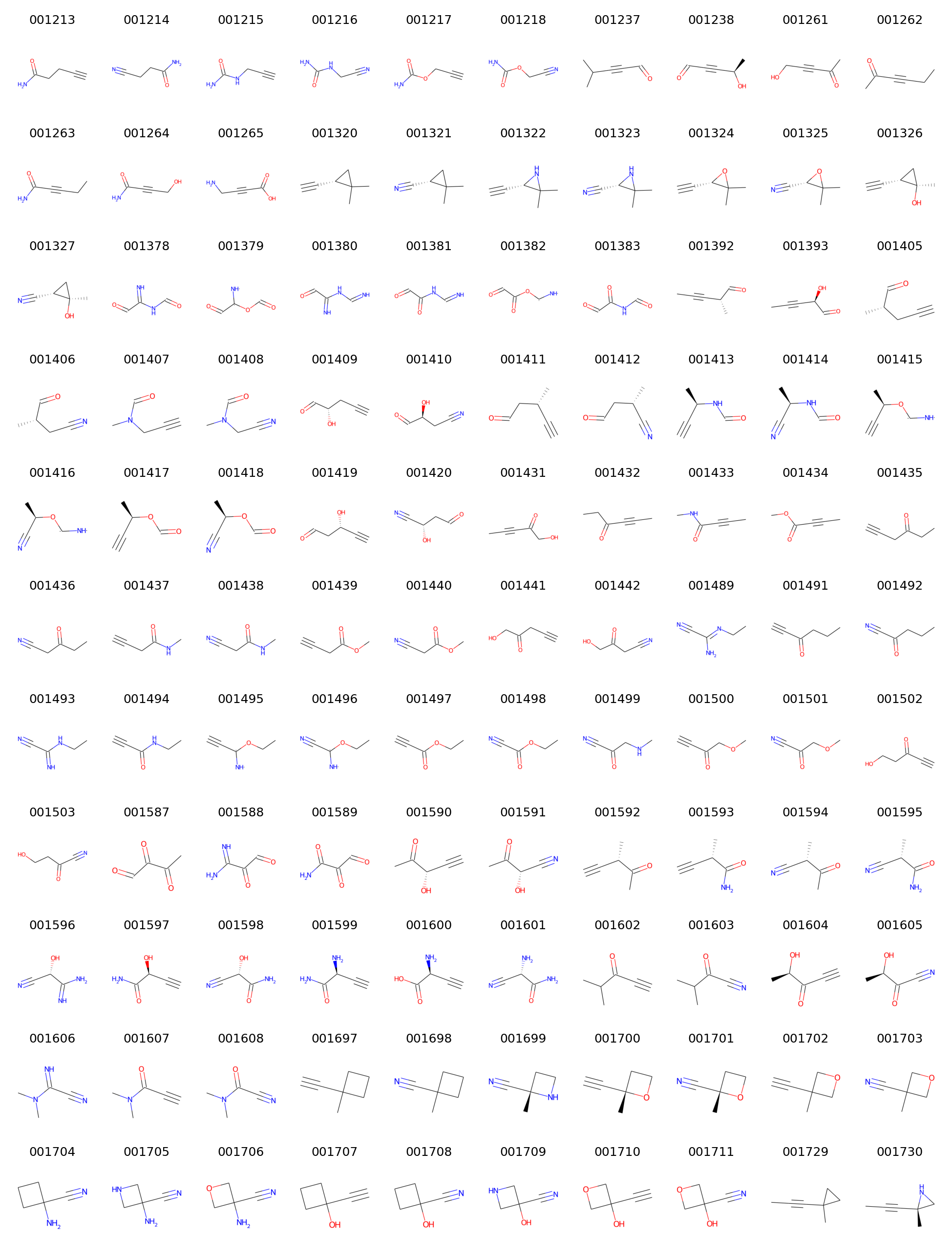}
\caption{The structures of the molecules for which atomic energies where calculated in this study. The numbers correspond to the respective ID in the QM9 dataset.}
\label{fig:ae_dist_diff_methods_modified}2
\end{figure}
\begin{figure}
\includegraphics[width=0.99\textwidth]{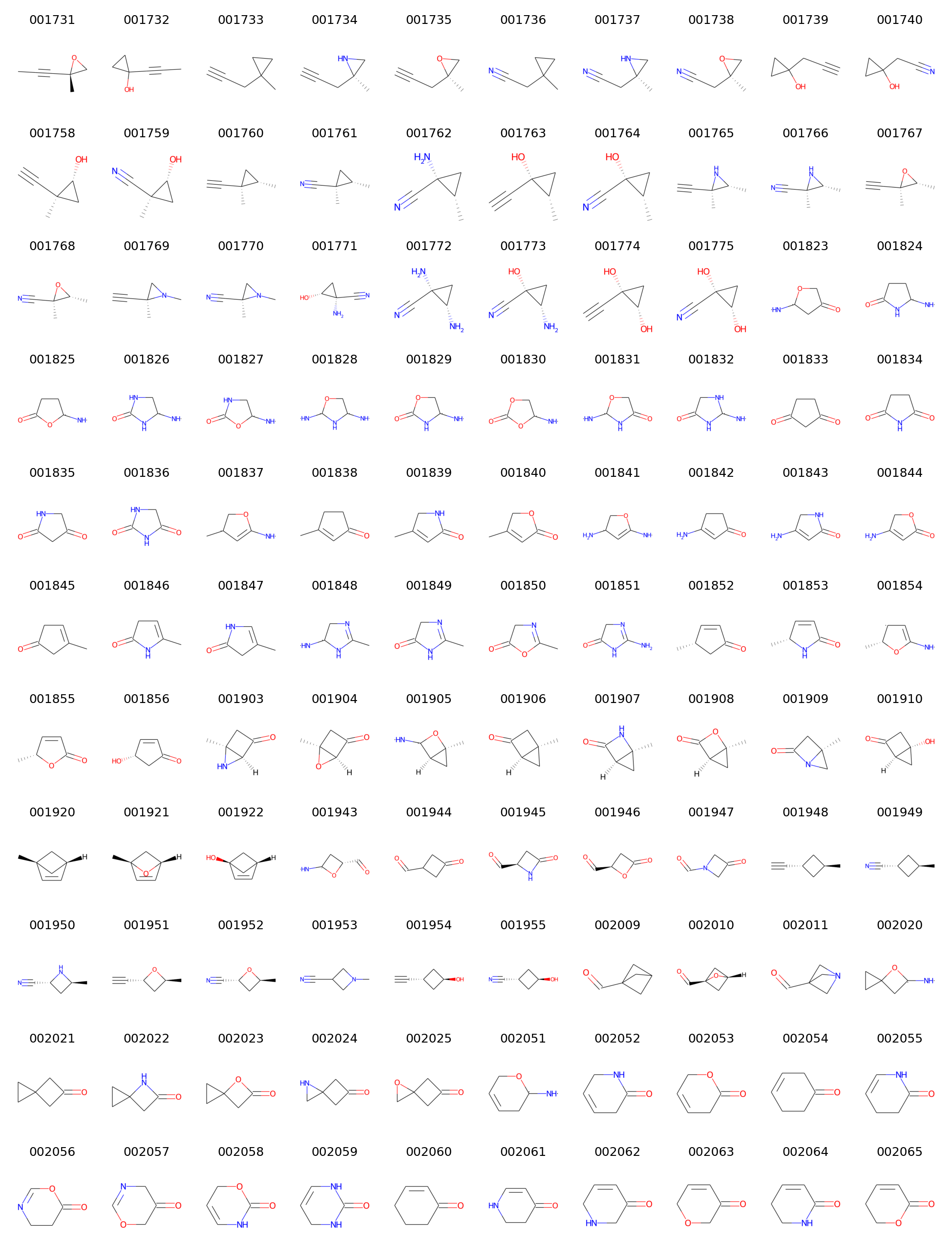}
\caption{The structures of the molecules for which atomic energies where calculated in this study. The numbers correspond to the respective ID in the QM9 dataset.}
\label{fig:ae_dist_diff_methods_modified3}
\end{figure}
\begin{figure}
\includegraphics[width=0.99\textwidth]{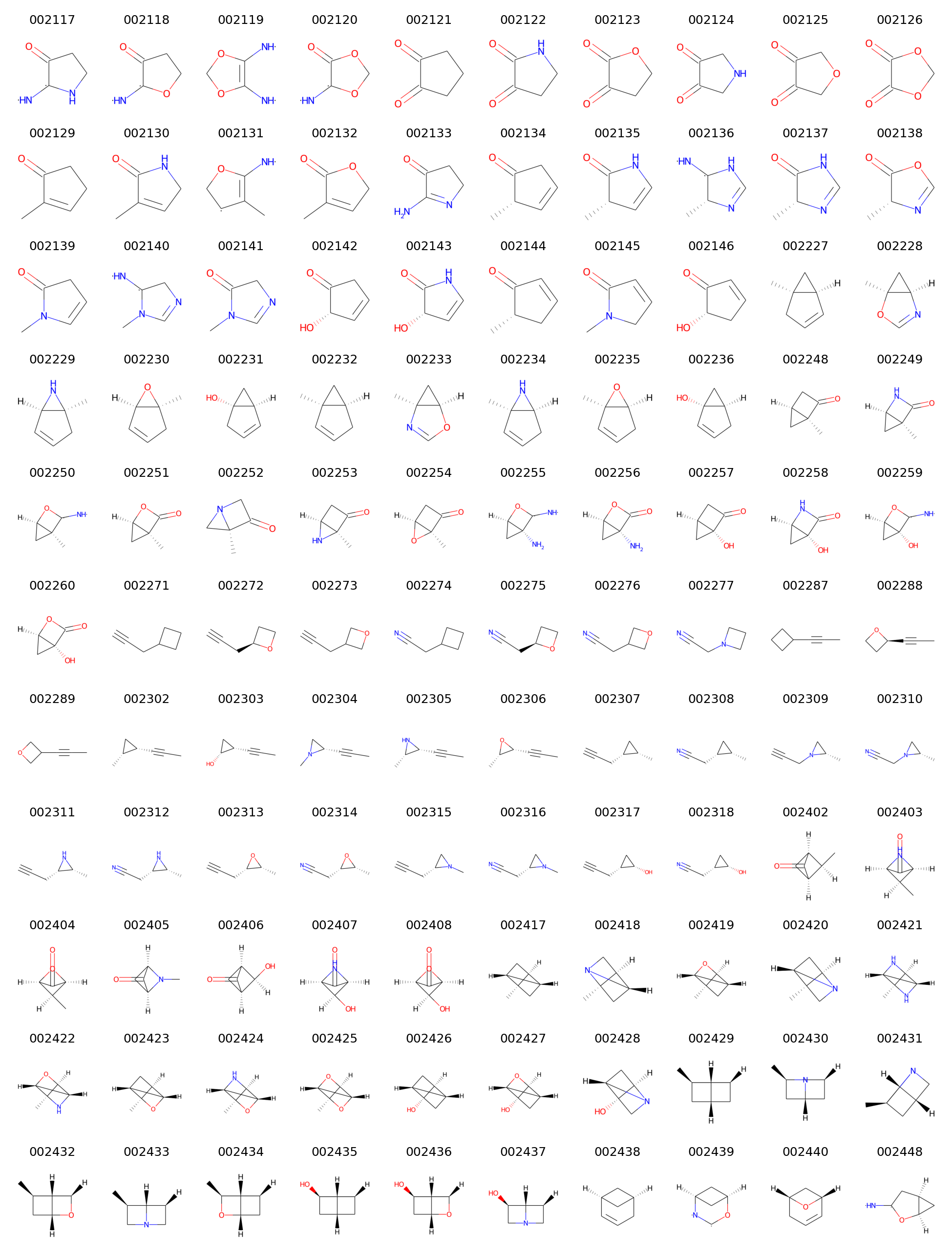}
\caption{The structures of the molecules for which atomic energies where calculated in this study. The numbers correspond to the respective ID in the QM9 dataset.}
\label{fig:ae_dist_diff_methods_modified4}
\end{figure}
\begin{figure}
\includegraphics[width=0.99\textwidth]{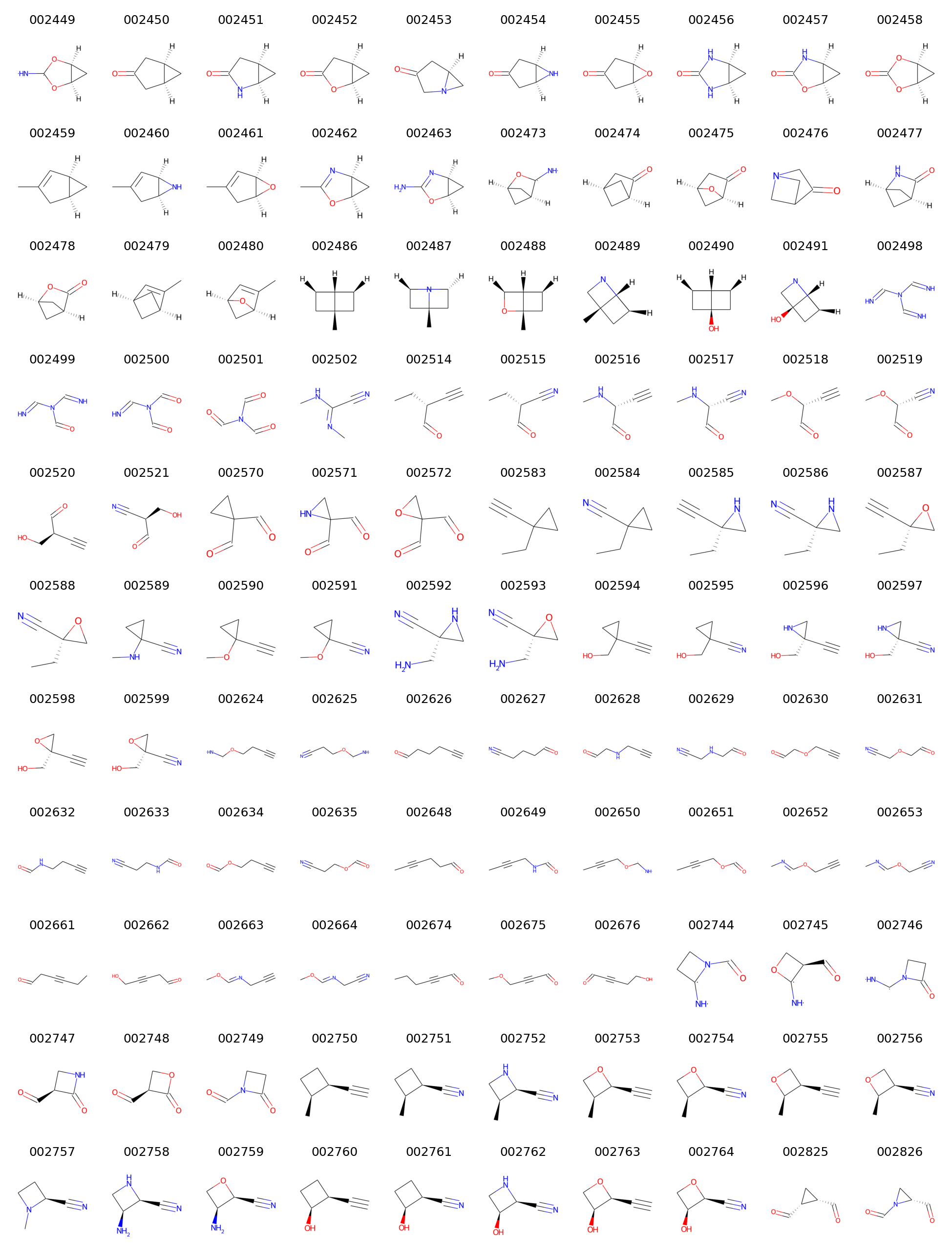}
\caption{The structures of the molecules for which atomic energies where calculated in this study. The numbers correspond to the respective ID in the QM9 dataset.}
\label{fig:ae_dist_diff_methods_modified5}
\end{figure}
\begin{figure}
\includegraphics[width=0.99\textwidth]{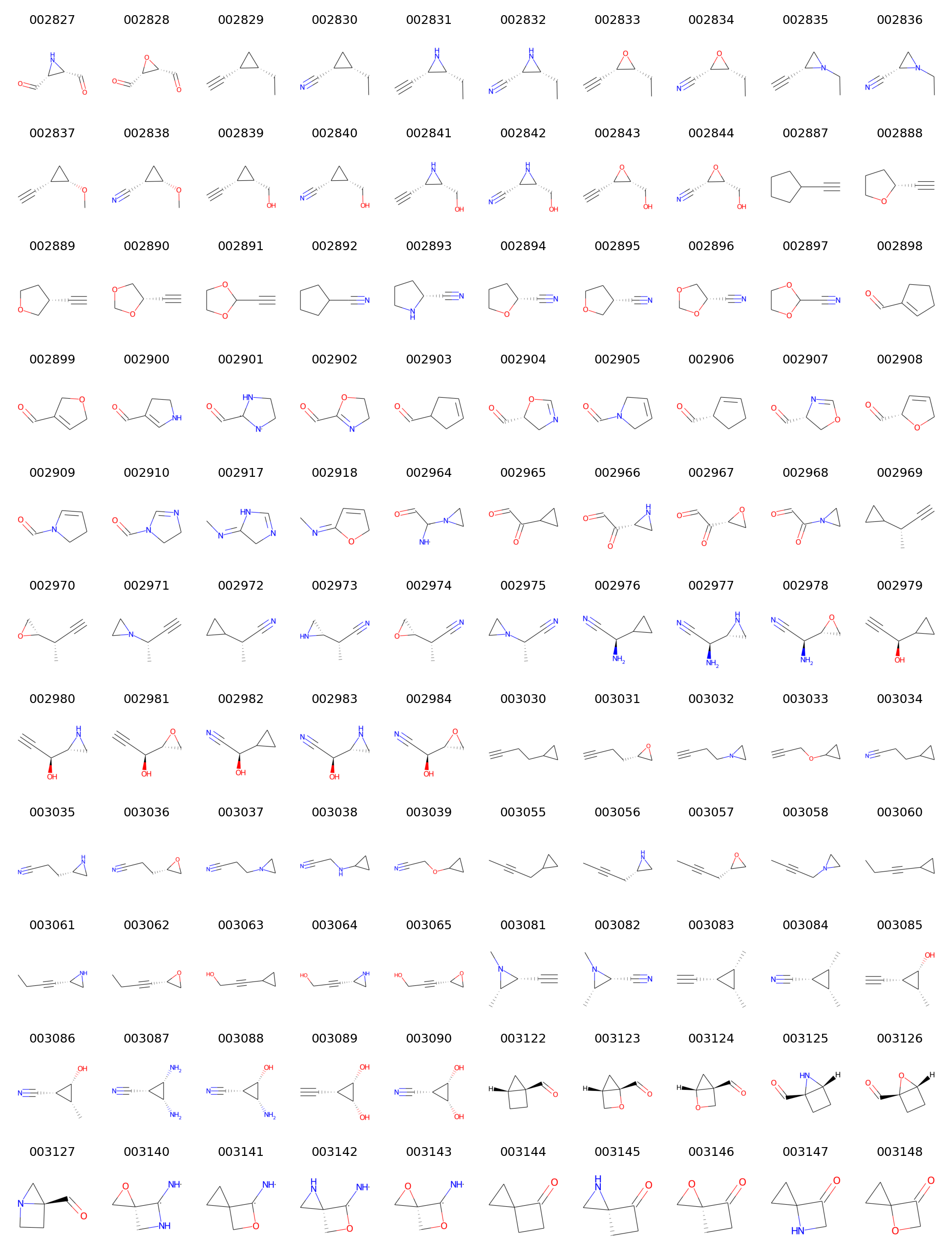}
\caption{The structures of the molecules for which atomic energies where calculated in this study. The numbers correspond to the respective ID in the QM9 dataset.}
\label{fig:ae_dist_diff_methods_modified6}
\end{figure}
\begin{figure}
\includegraphics[width=0.99\textwidth]{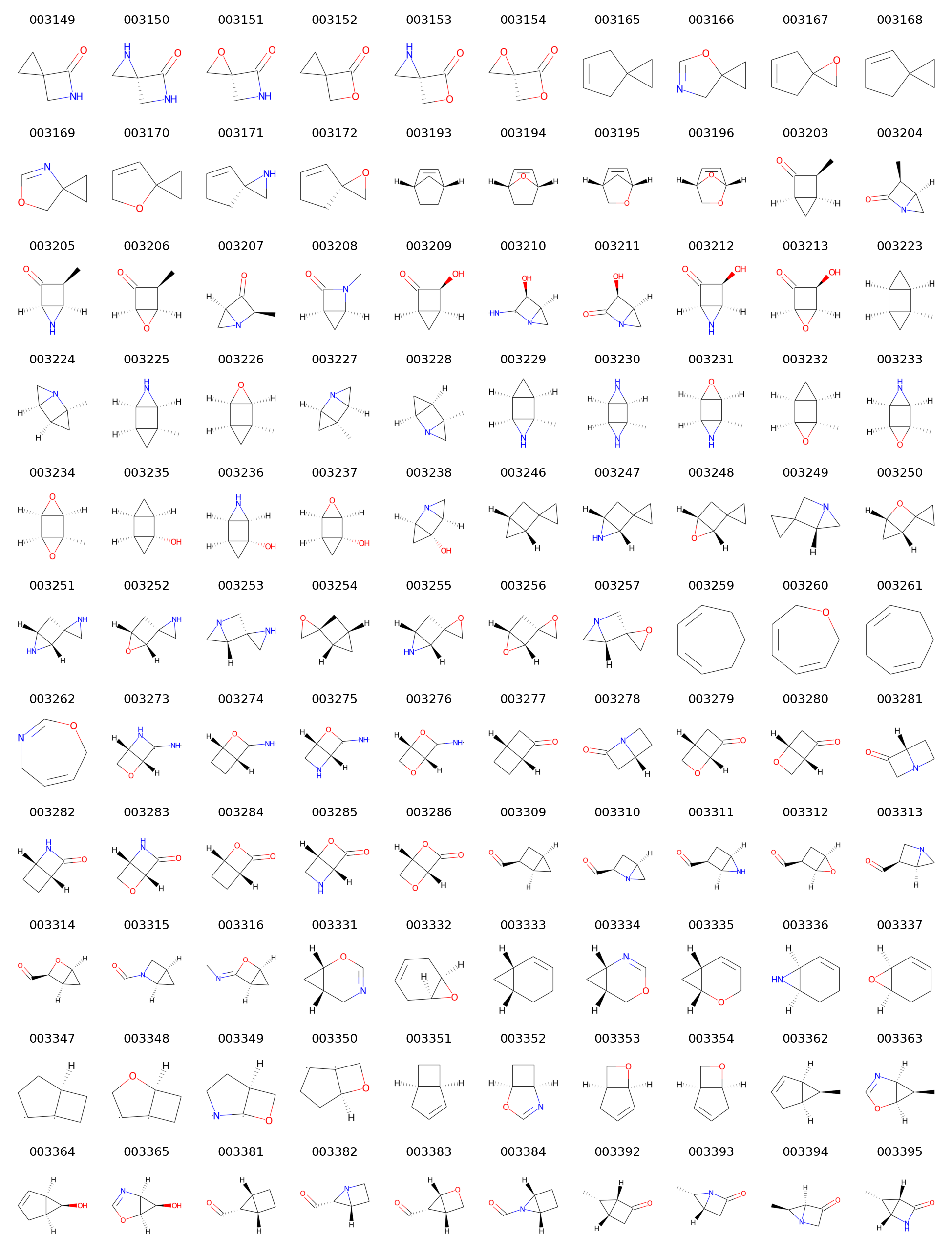}
\caption{The structures of the molecules for which atomic energies where calculated in this study. The numbers correspond to the respective ID in the QM9 dataset.}
\label{fig:ae_dist_diff_methods_modified7}
\end{figure}
\begin{figure}
\includegraphics[width=0.99\textwidth]{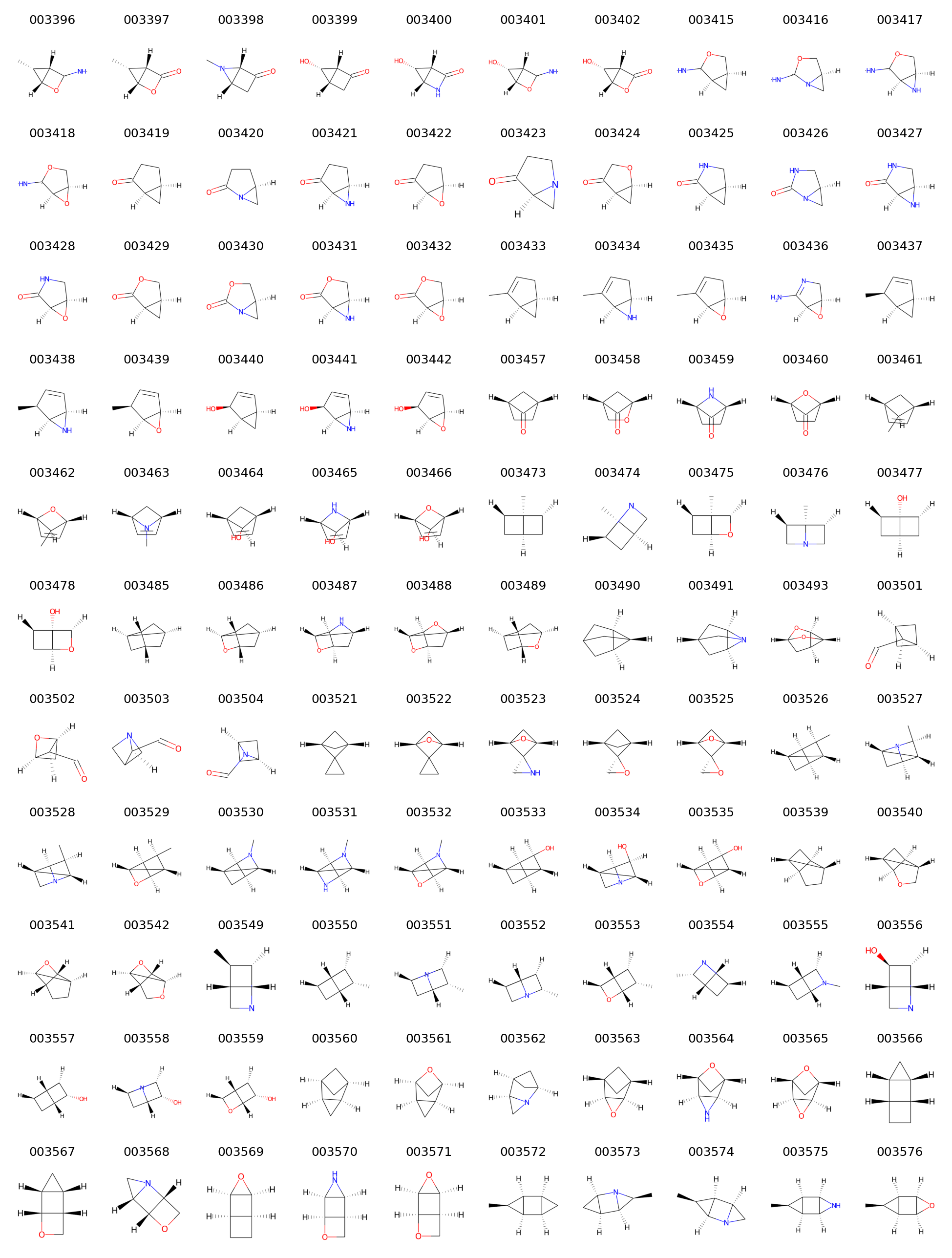}
\caption{The structures of the molecules for which atomic energies where calculated in this study. The numbers correspond to the respective ID in the QM9 dataset.}
\label{fig:ae_dist_diff_methods_modified8}
\end{figure}
\begin{figure}
\includegraphics[width=0.99\textwidth]{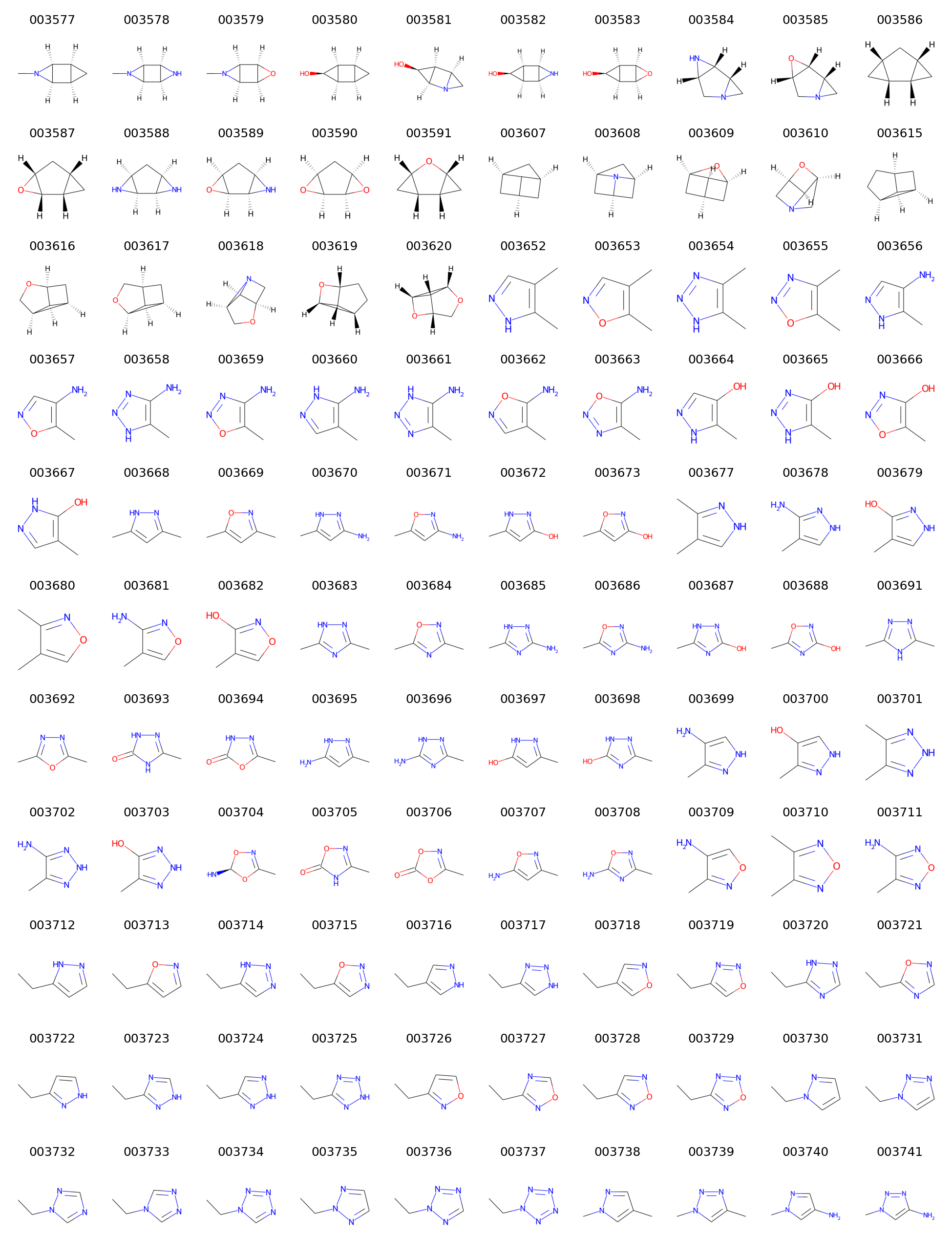}
\caption{The structures of the molecules for which atomic energies where calculated in this study. The numbers correspond to the respective ID in the QM9 dataset.}
\label{fig:ae_dist_diff_methods_modified9}
\end{figure}
\begin{figure}
\includegraphics[width=0.99\textwidth]{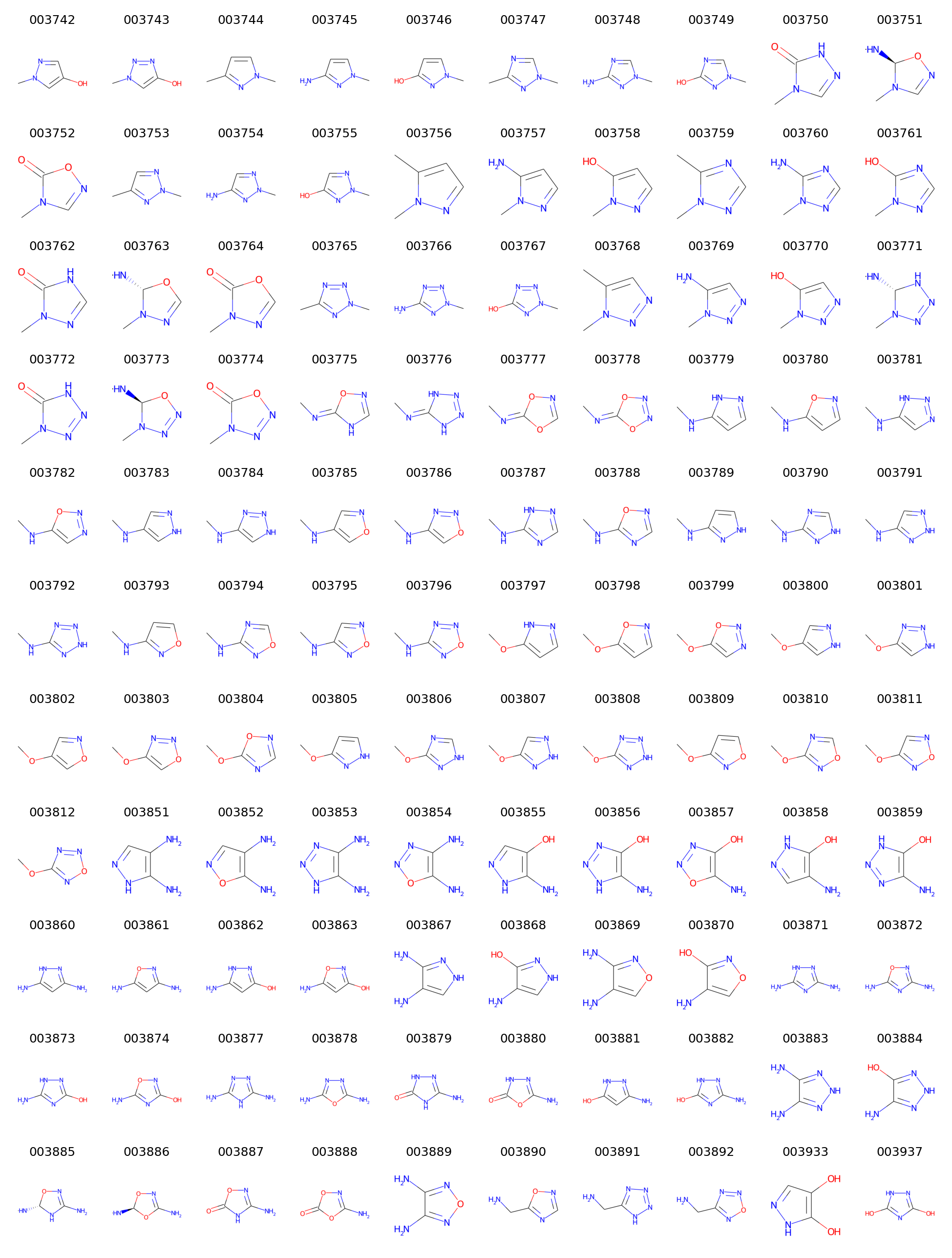}
\caption{The structures of the molecules for which atomic energies where calculated in this study. The numbers correspond to the respective ID in the QM9 dataset.}
\label{fig:ae_dist_diff_methods_modified10}
\end{figure}
\begin{figure}
\includegraphics[width=0.99\textwidth]{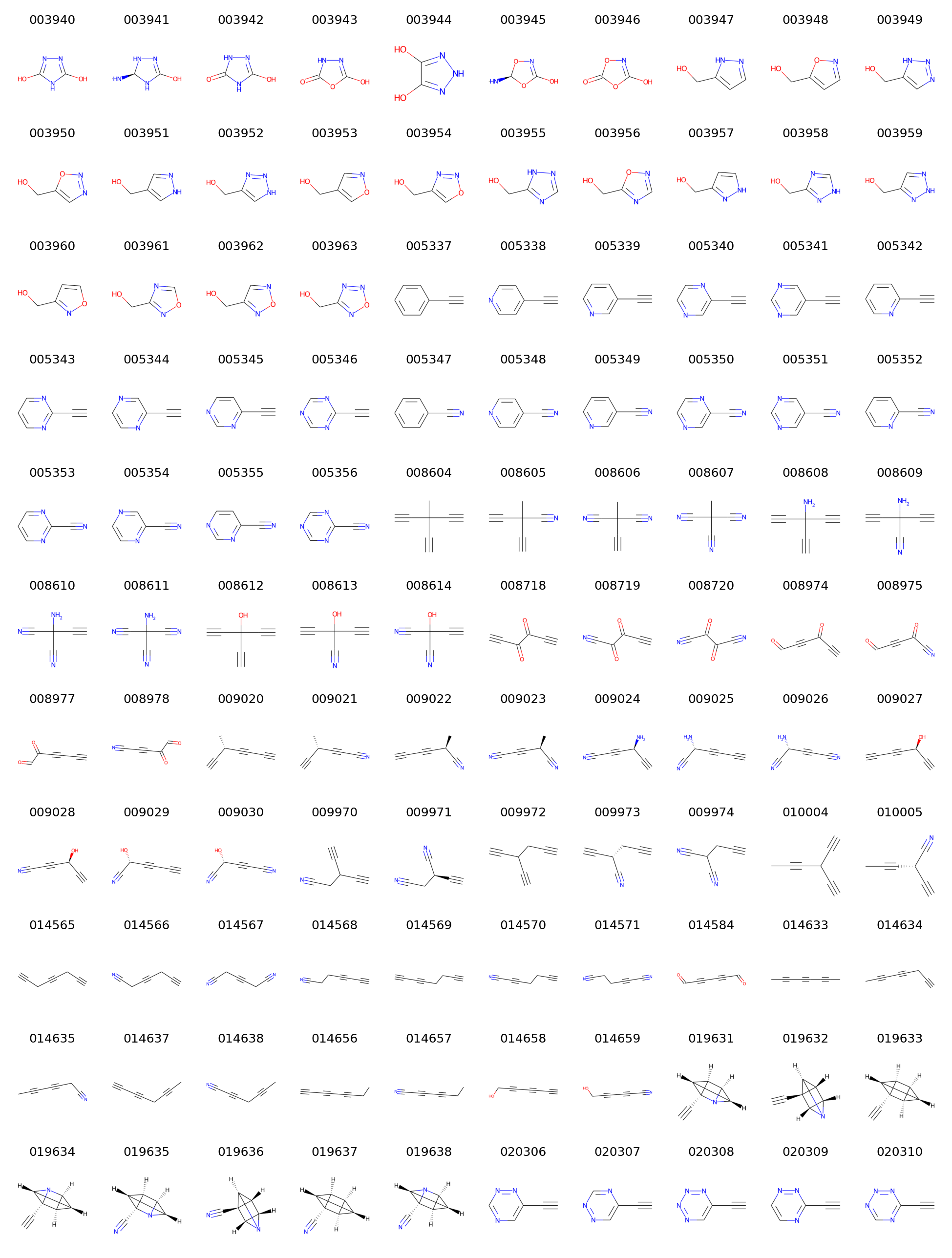}
\caption{The structures of the molecules for which atomic energies where calculated in this study. The numbers correspond to the respective ID in the QM9 dataset.}
\label{fig:ae_dist_diff_methods_modified11}
\end{figure}
\begin{figure}
\includegraphics{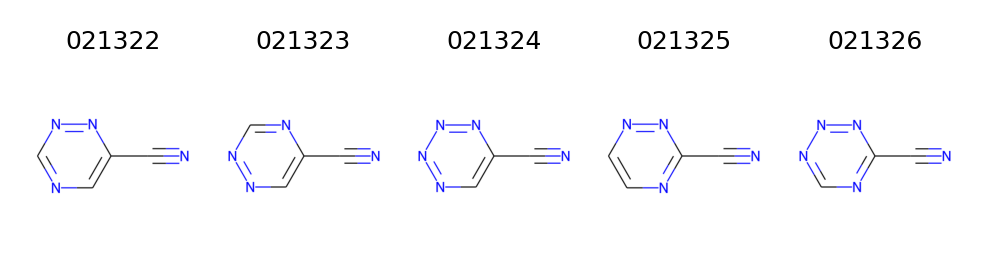}
\caption{The structures of the molecules for which atomic energies where calculated in this study. The numbers correspond to the respective ID in the QM9 dataset.}
\label{fig:ae_dist_diff_methods_modified12}
\end{figure}
